\documentclass[12pt]{article}
\usepackage[dvips]{graphics}
\def\reff#1{(\ref{#1})}

\newtheorem{Thm}{Theorem}

\newtheorem{Lem}[Thm]{Lemma}
\newtheorem{Prop}[Thm]{Proposition}

\newtheorem{Cor}[Thm]{Corollary}
\newenvironment{proof}{\noindent {\bf Proof: }}{\QED}
\newcounter{masection} 
\newcommand{\masection}[1]{\setcounter{equation}{0}
\refstepcounter{masection} \vspace{10pt plus 10pt minus 3pt} \noindent
{\large\bf \arabic{masection} #1}\par\vspace{5pt}}

\renewcommand{\theequation}{\mbox{\arabic{masection}.\arabic{equation}}}

\newcounter{masubsection}[masection] 
\newcommand{\masubsection}[1]{ \refstepcounter{masubsection} \vspace{5pt
plus 5pt minus 2pt} \renewcommand{\themasubsection}{%
\arabic{masection}.\arabic{masubsection}} \noindent {\bf
\arabic{masection}.\arabic{masubsection} #1} \par\vspace{5pt}}

\newcounter{saveeqn}

\newcommand{\eq}[1]{(\ref{#1})}

\newcommand{\ba}[1]{\begin{array}{#1}}
\newcommand{\ea}{\end{array}}
\newcommand{\arr}[2]{\ba{#1} #2 \ea}
\newcommand{\be}{\begin{equation}}
\newcommand{\ee}{\end{equation}}
\newcommand{\bea}{\begin{eqnarray}}
\newcommand{\eea}{\end{eqnarray}}
\newcommand{\beann}{\begin{eqnarray*}}
\newcommand{\eeann}{\end{eqnarray*}}
\newcommand{\p}{{\varrho}}
\newcommand{\E}{{\cal E}}
\newcommand{\Ea}{{\cal E}_1}
\newcommand{\Eb}{{\cal E}_2}
\newcommand{\ho}{{\rm h.o.}}           
\newcommand{\bare}{{\rm bare}}  
\newcommand{\dec}{{\rm dec}}
\newcommand{\vol}{\Lambda}             

\newcommand{\vo}{\Lambda}
   
\newcommand{\tcalI}{\widetilde{{\cal I}}} 
\newcommand{\tE}{\widetilde{E}}
\newcommand{\tI}{{\tilde{I}}}
\newcommand{\tPhi}{{\widetilde{\Phi}}}

\newcommand{\pp}{{\prime\prime}}
\newcommand{\Zpol}{{Z}_{\mbox{\tiny pol}}}
\newcommand{\zed}{\cal{Z}}
\newcommand{\Proj}{{\cal P}}
 
\newcommand{\supp}{\mbox{\rm supp}}
\newcommand{\prob}{\mbox{\rm Prob}}

\newcommand{\xx}{x_0^*}
\newcommand{\ovup}{\overline{\Upsilon}}
\newcommand{\spi}{\zeta}
\newcommand{\uphi}[1]{\Phi_{{\underline{B}}_{#1}}}
\newcommand{\sphiq}[1]{\Phi^{\rm q}_{{\underline{B}}_{#1}}}
\newcommand{\cphi}[1]{\Phi_{{\underline{C}}_{#1}}}

\newcommand{\llD}{{\underline{D}}}
\newcommand{\llC}{{\underline{C}}}
\newcommand{\llB}{{\underline{B}}}

\def\idty{{\leavevmode{\rm 1\ifmmode\mkern -5.4mu\else
\kern -.3em\fi I}}} \def\Ibb #1{ {\rm I\ifmmode\mkern -3.6mu\else\kern
-.2em\fi#1}} \def\Ird{{\hbox{\kern2pt\vbox{\hrule height0pt depth.4pt
width5.7pt \hbox{\kern-1pt\sevensy\char"36\kern2pt\char"36} \vskip-.2pt
\hrule height.4pt depth0pt width6pt}}}}
\def\Irs{{\hbox{\kern2pt\vbox{\hrule height0pt depth.34pt width5pt
\hbox{\kern-1pt\fivesy\char"36\kern1.6pt\char"36} \vskip -.1pt \hrule
height .34 pt depth 0pt width 5.1 pt}}}}
\def\reff#1{(\ref{#1})} 
\def\trunc{{\rm T}}
\def\QED{{\hspace*{\fill}{\vrule height 1.8ex width 1.8ex }\quad} \vskip
0pt plus20pt}  
 \def\A1n{A_1\otimes\cdots\otimes A_n} \def\Bar{\overline}

\def\phi{\varphi}            
\def\epsilon{\varepsilon}    

\def\A{{\cal A}} \def\B{{\cal B}} \def\C{{\cal C}_\Lambda} \def\D{{\cal D}}
 \def\H{{\cal H}}  
 \def\R{{\cal R}} \def\T{{\cal T}} \def\Z{{\bf Z}}
 \def\V{{\cal V}} \def\I{{\cal I}}
\def\x{{x_1}} \def\y{{x_2}} \def\z{{x_3}}

   \def\P{{\Ibb
P}}

\newcommand{\condmat}[1]{archived as {\tt cond-mat/#1}}
\def\bbbc{{\mathchoice {\setbox0=\hbox{$\displaystyle    C$}\hbox{\hbox
to0pt{\kern0.4\wd0\vrule height0.9\ht0\hss}\box0}}
{\setbox0=\hbox{$\textstyle    C$}\hbox{\hbox
to0pt{\kern0.4\wd0\vrule height0.9\ht0\hss}\box0}}
{\setbox0=\hbox{$\scriptstyle    C$}\hbox{\hbox
to0pt{\kern0.4\wd0\vrule height0.9\ht0\hss}\box0}}
{\setbox0=\hbox{$\scriptscriptstyle    C$}\hbox{\hbox
to0pt{\kern0.4\wd0\vrule height0.9\ht0\hss}\box0}}}}
\def\bbbz{{\mathchoice
{\hbox{$\sf\textstyle Z\kern-0.4em Z$}} {\hbox{$\sf\textstyle
Z\kern-0.4em Z$}} {\hbox{$\sf\scriptstyle Z\kern-0.3em Z$}}
{\hbox{$\sf\scriptscriptstyle Z\kern-0.2em Z$}}}}
\def\ibb
#1{\leavevmode\hbox{\kern.3em\vrule height 1.5ex depth -.1ex width
.2pt\kern-.3em\rm#1}}

\def\Rl{{\Ibb R}}

\let\szed=\bbbz 

\newcommand{\neighbours}[2]{<\! #1#2 \! >}
\def\intsum{\mathop{\hbox to 0pt{$\sum$}\int}}
\def\llB{{\underline{B}}}
\def\indic{\mathop{\rm I}\nolimits}
\newcommand{\heff}{H_{\rm{eff}}}
\def\bydef{:=}

\begin{document}

{\baselineskip=10pt \thispagestyle{empty} {Archived as
{\tt cond-mat/9804008} and {\tt mp\_arc 98-267} \hspace{\fill}
Preprint CPT-XXX-98}

\vspace{30pt}

\begin{center}
{\LARGE\bf Rigidity  of interfaces\\[10pt] 
in  the Falicov-Kimball model}\\[30pt] 
Nilanjana Datta\\
Institut de Physique Th\'eorique, EPFL, CH-1015 Lausanne, Switzerland.\\
E-mail: {\tt datta@dpmail.epfl.ch}\\
[15pt] Alain Messager\\
Centre de Physique Th\'eorique, CNRS-Luminy, Case 907, F-13288 Marseille,
France\\
E-mail: {\tt messager@cpt.univ-mrs.fr}\\[15pt]
Bruno Nachtergaele\\
Department of Mathematics, University of California, Davis, Davis,
CA 95616-8633, USA\\
E-mail: {\tt bxn@math.ucdavis.edu}\\[15pt]
(30 March 1998, revised 10 September 1999)\\[30pt]
{\it Dedicated to the memory of Roland Dobrushin}\\[15pt]
\end{center}

\noindent
{\bf Abstract}
\vspace{.2truecm}

We analyze the thermodynamic properties of interfaces in the three-dimensional
Falicov Kimball model, which can be viewed as a primitive quantum lattice
model of crystalline matter. In the strong coupling limit, the ionic subsystem
of this model is governed by the Hamiltonian of an effective classical spin
model whose leading part is the Ising Hamiltonian. We prove that the $100$
interface in this model, at half-filling, is rigid, as in the
three-dimensional Ising model. However, despite the above similarities with
the Ising model, the thermodynamic properties of its $111$ interface are very
different. We prove that even though this interface is expected to be unstable
for the Ising model, it is stable for the Falicov Kimball model  at
sufficiently low temperatures. This rigidity results from a phenomenon of 
``ground state selection'' and is a consequence of the Fermi statistics of the
electrons in the  model.

\vspace{20pt} \noindent {\bf Keywords:} Falicov-Kimball model, ground state
selection, rigidity  of interfaces, $100$-- and $111$ interfaces.

\vfill

\hrule width2truein \smallskip {\baselineskip=12pt \noindent Copyright
\copyright\ 1998, 1999 by the authors. Reproduction of this article 
in its entirety, by any means, is permitted for non-commercial purposes.\par }}

\newpage

\masection{Introduction}\label{sec:intro}

Domain boundaries can have important effects on the transport properties
of condensed matter materials. In some cases, transport is
believed to occur mainly or exclusively along domain boundaries
\cite{NW}. This is related to the drastic effect the presence of a domain
wall, or interface, can have on the low-lying excitations of the 
system. Therefore, it is important to study non-periodic equilibrium
states, in particular interface states, in statistical mechanics,
and to understand their stability, fluctuations, and low-lying
excitations.

Many phenomena in condensed matter physics, 
which are of current interest, are intrinsically quantum mechanical in origin.
Quantum effects also play a crucial role in the properties of interfaces. 
There are, however, very few rigorous results on interface
states in quantum statistical mechanics.  In \cite{BCF1,BCF2} a general
perturbation theory was developed, which, under certain assumptions, shows
that a small quantum perturbation does not destroy an existing interface
Gibbs state of a classical discrete spin model (the so-called
Dobrushin states).

In this paper we consider the opposite situation, namely, one in which
quantum fluctuations stabilize the interface against thermal fluctuations,
while the classical limit does {\it not} have a stable interface.
We demonstrate the occurrence of such an 
``order by disorder'' effect \cite{VBCC,Hen}, in a simple quantum lattice 
model  --- the three-dimensional Falicov Kimball (FK) model [See Section
\ref{sec:FK_model} for a description of the model]. 
We are motivated by recent work \cite{KN2} which demonstrates that, at 
zero temperature, 
in the two-dimensional ferromagnetic XXZ
Heisenberg model with Ising-like anisotropy the quantum fluctuations
stabilize the $11$--interface (i.e., an interface in the diagonal
direction).  In the present work we prove such an effect at
finite temperature for two interfaces in the FK model.

The FK model was chosen because its statistical mechanical properties 
have been studied
extensively (see, e.g., \cite{KL,gru,lebmac,ken94,GM,gruber,MMS,ku})
and, recently, convenient perturbation expansions have been developed for it
\cite{MMS,DFF4,ken97,Mes}. We expect that the XXZ model in three dimensions also
has a stable $111$ interface at sufficiently low temperatures. Its analysis
is, however, more involved due to the presence of gapless excitations in the 
interface \cite{KN2, Mat}.  

Our main result is that, in three dimensions, the FK model has a stable $111$ 
interface at sufficiently low temperatures. This should be compared with the
three-dimensional Ising model since, in the strong coupling limit, 
the FK model can be considered as a 
perturbation of the Ising model (see e.g. \cite{lebmac,MMS,Mes}). 
Dobrushin showed that the Ising model has
$100$--interface states, but its $111$--interfaces are expected to be {\it
unstable at any finite temperature}. It has been proved recently that in the
zero-temperature limit of the three-dimensional Ising model the $111$
interface fluctuates \cite{Ken}. This is related to the degeneracy of the
ground states with a $111$--interface which grows exponentially with the
volume (the rate of exponential growth may depend on the boundary conditions
however! See \cite{Pro}). In the FK model this degeneracy is lifted. In this
sense,  this is an example of the phenomenon  of ``ground state selection''
\cite{Hen} by quantum fluctuations. We refer to Section \ref{geom111} for a
detailed discussion of the $111$ interface configurations. 

In \cite{KL} Lieb and Kennedy showed how to study the FK model in terms of an
Ising-type model for the Ising configurations that is obtained by taking the
trace over the electron states for any  given ion configuration. The
Hamiltonian for the ions can be explicitly computed to any order in
perturbation theory together with a bound on the sum of the higher order terms
[see Appendix A and \cite{Mes}]. For the study of the $100$--interface one needs the explicit
form of this Hamiltonian up to second order. For the $111$--interface fourth
order perturbation theory is needed. In principle, our method could be used to
study interfaces with more general orientations, but higher order terms in the
perturbation series will be needed; e.g., the 112 interface is infinitely
degenerate at fourth order, but we expect it to be stabilized at sufficiently
low temperatures by the sixth order terms. Therefore, one should 
expect that the Falicov-Kimball model has an infinite number of interface
phase transitions.

We follow Dobrushin \cite{Dob} (see also \cite{BLPO,BLP})
in proving the existence of an  interface by
considering an effective two-dimensional model for the interface.  Although,
this two-dimensional model turns out to be quite complicated, and involves
many-body interactions of arbitrarily long range, we can analyze it with a
Peierls--type argument.  It is probably possible to develop a general
Pirogov-Sinai theory \cite{pirsin75, pirsin76, zah, BCF1} to treat this
situation, along the lines of \cite{DMS} and \cite{Par}, but  we found it more
convenient to make efficient use of simpler methods. The result is a more
transparent and relatively short proof.

Our main results are stated at the end of the next section. 
Our main technical tool is the convergence of certain cluster 
expansions proved in Appendix B. Appendix A contains the proof of 
a bound on the remainder term in the expansion of the effective 
Hamiltonians for the ions. In Section \ref{sec:rigidity100}
we discuss the rigidity  of a $100$--interface, which is much simpler than
the case of the $111$--interface treated in Section \ref{sec:rigidity}.

\masection{The FK model and effective Hamiltonians}
\label{sec:FK_model}

The Falicov Kimball (FK) model \cite{Falicov} is a lattice model of 
spin-polarized electrons and  classical particles (ions) \cite{Falicov}. The
electrons and ions interact via a purely on-site interaction. The electrons
can hop between nearest neighbour sites of the lattice, but the ions are
static. There is a hard-core repulsion between the ions, which prevents more
than one ion from occupying a single lattice site. The presence or absence of
an ion at a lattice site $x$ is described by a classical variable, $W(x)$,
which is equal to unity  if there is an ion at the site $x$ and is zero
otherwise. Let the number operator of an electron at the site $x$ be given by
$n = c^{\dagger}_{x} c_{x}$, where $c^{\dagger}_{x}$ and $c_{x}$ are the
creation and annihilation operators of the electron at the site $x$. Let
$\mu_i$ and $\mu_e$ denote the chemical potentials of the ions and electrons
respectively. Let ${\zed}^3$ denote an infinite cubic lattice, with unit
lattice spacing, such that the coordinates of the sites are given by
half--integers.  For notational simplicity in our description of the
interface, it is more convenient to consider this lattice instead of
$\szed^3$.  The Hamiltonian of the model defined on a finite lattice $\Lambda
\subset {\zed}^3$ is given by
\begin{equation}
\H_\Lambda (t, U) = \H_{0\Lambda}(U) + tV_\Lambda, \label{one}
\end{equation}
where $\H_{0\Lambda}(U)$ is a Hamiltonian which is given entirely in terms
of an on-site interaction as follows:
\begin{eqnarray}
\H_{0\Lambda}(U) &=&  2U \sum_{x \in \Lambda} W(x) n_x -
\mu_e \sum_{x \in \Lambda} n_x -
        \mu_i \sum_{x \in \Lambda} W(x), \nonumber\\
 &:=& \sum_{x \in \Lambda} \Phi_{0x} \label{hamfk}
\end{eqnarray}
The operator $V_\Lambda$ causes electrons to hop between nearest neighbour
sites of the lattice:
\begin{eqnarray}
V_\Lambda &=& - t \,\sum_{<xy> \subset \Lambda} \bigl(
c^{\dagger}_x c_y + c^{\dagger}_y c_x\bigr), \nonumber\\
  &=:& \sum_{X=<xy>\subset \Lambda} V_X, \label{hop}
\end{eqnarray}
Here $<xy>$ denotes a pair of nearest neighbour sites in the lattice.
The hopping amplitude of the electrons is denoted by $t\in\Rl$. 
The first rigorous study of the above model was
done by Kennedy and Lieb \cite{KL}. They considered the classical
particles to be nuclei and the on-site interaction to be the Coulomb
attraction between nuclei and electrons.  In accordance with this
interpretation, they chose the coupling constant $U$ to be
negative. They proved that the ground states of this model display
perfect crystalline ordering for the choice
\be
\mu_i= \mu_e = U 
\label{half-filling},
\ee
the ions being arranged in a checkerboard configuration.
The choice \reff{half-filling} corresponds to a neutral model: the
average number of electrons in the lattice is equal to the 
the average number of ions, both being equal to half the number of 
lattice sites (half-filling). 

The model described by the Hamiltonian $\H_\Lambda(t,U)$ for $U<0$, is 
mathematically equivalent to the model with $U>0$ (see \cite{KL}) and there is
a  simple relation between the properties of the attractive model, $U<0$, and
the repulsive one, $U>0$. In this paper we work with $U>0$. 

We study the FK model in the strong--coupling limit i.e., for $U>>|t|$,
and hence consider the hopping term $tV_\Lambda$ to be a perturbation to the
Hamiltonian $\H_{0\Lambda}(U)$. It is convenient to renormalize the
hopping amplitude $t$ to unity. This amounts to the following rescaling:
\bea
U/t & \longrightarrow& U \nonumber\\
\beta t &\longrightarrow& \beta,
\eea
where $\beta =1/k_B T$ ($k_B$ being the Boltzmann constant and $T$ the
absolute temperature). In our expansions $U^{-1}$ plays the role of a small
parameter.

For a fixed value of the coupling constant $U$, the zero temperature phase
diagram of the unperturbed Hamiltonian $\H_{0\Lambda}(U)$ in the plane of
chemical potentials can be easily obtained \cite{gruber}. To obtain the ground
states of $\H_{0\Lambda}(U)$, for any given set of values the chemical
potentials $\mu_i$ and $\mu_e$, we only need to find the single-site
configuration which minimizes $\Phi_{0x}$ \reff{hamfk}. When both the chemical
potentials are negative, the ground state corresponds to all sites of the
lattice being empty.  In the rest of the ($\mu_i$-$\mu_e$) plane, it is found
that for values of the chemical potentials such that $\mu_e < 2U$
{\em{and/or}} $\mu_{i} < 2U$, there is no doubly occupied site at zero
temperature. For these values of the chemical potentials, the ground state
corresponds to an all-ion configuration if $\mu_{i} > \mu_{e}$ with $\mu_{i} >
0$, and to an all-electron configuration if $\mu_{e} > \mu_{i}$ with $\mu_{e}
> 0$. For $0 < \mu_{i} = \mu_{e} < 2U$ all singly occupied configurations are
equally likely and hence the ground state is infinitely degenerate. The origin
and the point $\mu_{i} = \mu_{e}= 2U$ also correspond to infinitely many
ground states. At the origin, each site is either empty or singly occupied,
whereas at the point $(2U, 2U)$ each site is either singly- or doubly
occupied. At zero temperature, for $\mu_{i} > 2U$ and $\mu_{e} > 2U$, every
site is doubly occupied by an electron-ion pair. 

As in \cite{lebmac,MMS,Mes}, we will rely on the fact that for $U > c|t|$, 
where $c$ is a positive constant, the ionic subsystem of the FK model 
defined on a finite cubic lattice $\Lambda$ can be described by an effective 
classical Hamiltonian. We will study only the {\em neutral\/} model at
{\em half-filling\/},, i.e., $\mu_i= \mu_e=U$. Then, it follows
form the circuit representation of \cite{MMS} and the bounds proved in
Appendix A that the equilibrium states of the ions are described by
an effective classical Hamiltonian of the form
\be
\H^{\rm eff}_{\Lambda}(U) = 
\frac{1}{4U}\sum_{<xy> \in \Lambda} s'_x s'_y + \R_\Lambda(\beta,U),
\label{heff}
\ee
where 
\be
s'_x := 2W(x) -1 \label{spinvv}.
\ee 
The variable $s'_x$ can be interpreted as an on--site spin variable since
$s'_x = 1$ if there is an ion at $x$ and $s'_x = -1$ otherwise. Note that 
$\H^{\rm eff}_{\Lambda}(U)$ depends on $\beta$ through the remainder
term $\R_\Lambda(\beta,U)$. This is unavoidable if one wants an exact 
correspondence between the correlation functions of the ions in the 
Falicov-Kimball model and the same correlation functions of the effective 
classical spin model. In particular, if one writes $\R_\Lambda(\beta,U)$ 
as a sum of products of the $s'_x$ variables, one sees that it still contains
a nearest neighbour contribution with a coefficient that tends to zero
exponentially fast as $\beta\to\infty$. As all our results are for 
$\beta$ and $U$ sufficiently large, this temperature dependence will be of no 
consequence however. Similarly, as we will do later on [\reff{rel30}
and \reff{rel5}], one can extract from
$\R_\Lambda(\beta,U)$ the leading contributions for the next-nearest neighbour
and plaquette interactions, which are independent of $\beta$, and estimate
the temperature dependent corrections with the bounds proved in Appendix A.

All terms of higher orders in the perturbation
parameter $U^{-1}$, i.e., all terms of order $U^{-n}$, with $n \ge 3$,
are contained in the remainder $\R_\Lambda(\beta, U)$, which, just as the 
leading term, depends on $\beta$ in an inessential way. Hence we simply
write it as $\R_\Lambda(U)$. This remainder 
is expressible in terms of
local classical interactions $\{R_B(U)\}$:
\be
\R_\Lambda(U) = \sum_{B \cap \Lambda \ne \emptyset \atop {|B|\ge 2}} R_B(U)
\label{ru} \ee

Here $B$ denotes a {\it connected} set of lattice sites, i.e., if $x, \, y \,
\in B$ then there exists a sequence of sites $x_0=x, \, x_1, \ldots, x_n=y$
such that $x_i \in B$ and $|x_i - x_{i+1}|=1$ for all $i=0, \ldots, n-1$. The
number of sites in $B$ is denoted by $|B|$. We refer to such a set $B$ as a
{\em{bond}}. A bond can be represented by a graph, the vertices being the
sites and the lines of the graph representing  nearest neighbour bonds between
pairs adjacent sites. Let ${\cal{B}}$ denote the set of {\em{all}} bonds in
the lattice $\Lambda$. For each bond $B$ appearing in the above sum \reff{ru},
the interaction $R_B(U)$ can be expressed as a product of two or more on-site
spin variables $s'_x$ with $x \in B$.

Note that, for us, a {\em bond\/} is, by definition, a {\em connected set}.
This is convenient in combinatorial arguments and a natural choice in view of
the way perturbation theory produces long-range and multi-body interactions
as a composition of nearest neighbour hoppings. It does not exclude the
presence of terms of the form $s'_x s'_y$, with $\vert x-y\vert\geq 2$, in 
the effective Hamiltonian. Such terms {\em are included\/} in connected bonds
$B$ containing $x$ and $y$.

A configuration on ${\zed}^3$, denoted by $\omega$, is therefore given by a
set of assignments $\{s'_x\}_{x \in {\zed}^3}$ of $s'_x \in \{-1, 1\}$ to each
$x \in {\zed}^3$. For any finite subset $Y\subset {\zed}^3$, let $\omega_Y$
denote the restriction of the configuration $\omega$ to the subset $Y$. A
boundary condition (b.c.) will be specified by a configuration
${\bar{\omega}}$, meaning that for any finite volume $\Lambda\subset
{\zed}^3$, the system is considered with a fixed value for the spins on each
$x \in {\zed}^3 \setminus \Lambda$, determined by ${\bar{\omega}}$.

{From} Lemma \ref{expdec} of Appendix A (in particular 
\eq{used_bound}) it follows that for $\beta$ and $U$ large enough,
there exist positive constants, $c_1$ and $
{\tilde{c}}_2$ (with $c_1/U < 1$) such that:
\begin{itemize}
\item{for $|B| \ge 3$
\be
|R_B(U)| \le  {\tilde{c}}_2  \Bigl(\frac
{c_1}{U}\Bigr)^{g(B)} \label{bd2}
\ee
where
\begin{equation}
g(B) := n(B) - 1, \label{geebee}
\end{equation}
with $n(B)$ being defined as the minimum length (in units of lattice
spacing) of a closed path which passes through all sites of $B$.}
\item{while, for $|B|=2$,
\be
|R_B(U)| \le {\tilde{c}}_2 \Bigl(\frac{c_1}{U}\Bigr)^3,  \label{bd1}
\ee}
\end{itemize}
The latter bound [\reff{bd1}] results from the following fact: We have that
$g(B)= 1$ for $|B|=2$; however, the term in the effective 
Hamiltonian of order $U^{-1}$, has been extracted and is given by the
first term on the RHS of \reff{heff}. The contribution from electron hoppings
between nearest neighbour sites, to the remainder term, are therefore of order 
$U^{-3}$, as even powers of $U^{-1}$ do not occur in the expansion.

The above bounds imply that there exists a constant $r>0$ such that
\begin{equation}
\sum_{B \ni 0} |R_B(U)|e^{r g(B)} < \infty
\label{exporem}
\end{equation}
and hence the interaction $\{R_B(U)\}$ decays
exponentially.  Using the bound
\reff{bd1}, the effective Hamiltonian \reff{heff} can be written as
follows
\be \H^{\rm eff}_\Lambda(U)= \H_{0\Lambda}^{(2)}(U) + \R^{\ge 3}_\Lambda(U)
\label{heff1}\ee
with
\bea
\H_{0\Lambda}^{(2)}(U)  &=&\frac{1}{4U} 
\sum_{<xy>\subset\Lambda} s'_x s'_y +  \sum_{B \cap \Lambda \ne \emptyset
\atop {|B|=2}} R_B(\beta,U)\nonumber\\
&=&J(U) \sum_{<xy> \in
\Lambda} s'_x s'_y \label{h01}
\eea
where
\be
J(U):=J(\beta,U) \simeq \frac{1}{4U} + \ho > 0,
\label{ju}
\ee
the symbol ``$\ho$'' denoting terms of higher orders in $U^{-1}$ which are
bounded by $a U^{-3}$, for some constant $a>0$. Throughout the rest of the
paper, whenever the symbol $\ho$ appears in a sum involving powers of 
$U^{-1}$, it will denote the presence of a correction term which is bounded  by
a positive constant times the next odd power of $U^{-1}$. The correction  term
is an infinite series that can be computed order by order in perturbation 
theory, and that has a dependence on $\beta$. This temperature dependence is
inessential and we will routinely omit it from  the notations. We will only use
that, for $\beta U$ sufficiently large, a bound of the form \eq{bd2} holds. We
refer to Appendix A for the proofs of the bounds. A more general class of
models as well as a detailed discussion of the temperature dependence of the 
effective Hamiltonian is contained in \cite{Mes}.

We define
\be
\R^{\ge 3}_\Lambda(U) :=  \sum_{B \cap \Lambda \ne \emptyset 
\atop {|B|> 2}} R_B(U)
\ee
The leading part $\H_{0\Lambda}^{(2)}(U)$ of the effective Hamiltonian is
identical to the Hamiltonian of an antiferromagnetic Ising model with
nearest neighbour interactions of strength $J(U)$ in the presence of a
magnetic field of strength $h$.

The fact that the Hamiltonian $\H^{\rm eff}_\Lambda(U)$ \reff{heff1} 
is invariant under ``spin-flip'', is a consequence of half-filling.
The leading part, $\H^{(2)}_{0\Lambda}(U)$ [\reff{h01}],
of the effective Hamiltonian has two ground states in each of  which  the
$+$'s and $-$'s occupy alternate sites of the lattice, i.e., the
antiferromagnetic N\'eel states.  Thus, {\it{to order $U^{-1}$}}, the
effective Hamiltonian governing the ionic subsystem has a two-fold degenerate
ground state, while for the original unperturbed Hamiltonian
$\H_{0\Lambda}(U)$ (given by \reff{hamfk}) the ground state energy is
independent of the ion configuration. Hence, as far as the ionic subsystem is
concerned, the effect of the quantum perturbation $tV$ is to {\it{select}} two
ground state configurations from the infinitely many ground states of
$\H_{0\Lambda}(U)$.  This phenomenon in which the quantum perturbation lifts
the infinite degeneracy of the classical ground states is known as ``ground
state selection'' \cite{VBCC,Hen}. Moreover, for low enough temperatures,  the
characteristic long range order of the N\'eel states persists under the action
of the remainder ${\R}^{\ge 3}_\Lambda$ \cite{DFF4}.  The main purpose of this
paper is to prove that a similar phenomenon of ground state selection occurs
for the $111$ interface of the FK model and that the selected interfaces are
{\it{rigid}}, i.e., the interfaces persist in the thermodynamic limit at
finite but non-zero temperature.  See Section \ref{sec:rigidity}.

It will be more convenient for us to perform the transformation
\be
s'_x \longrightarrow  s_x = (-1)^{2\,|x|} s'_x \quad \hbox{for all} \quad 
x = (x_1, x_2, x_3) \in \Lambda,
\label{sxnew}
\ee
where $|x| = x_1 + x_2 + x_3$. The above transformation yields an 
equivalent Hamiltonian with the leading part
given by the {\it{ferromagnetic}} Ising model:
\be
\H^{\rm ferro}_\Lambda(U) = - J(U) \sum_{<xy> \in \Lambda} s_x s_y,
\ee
where $J(U)$ is given by \reff{ju}. Let us denote $s_x = 1$ by the 
symbol ``$+$'' and $s_x = - 1$ by the 
symbol ``$-$''.
The two ground states of $\H^{\rm ferro}_\Lambda(U)$ are
denoted by $s_+$ and $s_-$ and correspond to each site in the lattice
being occupied by a $+$ and a $-$ respectively.
We refer to these two ground state configurations and their
low-temperature analogues as {\it{homogeneous phases}}. 
For a finite volume $\Lambda
\subset {\zed}^3$, the boundary conditions defined 
\bea
{\hbox{either by}}\,\, s_x &=& +1 \quad \hbox{ for all } x\in{\zed}^3 
\setminus \Lambda,\nonumber\\
{\hbox {or by}}\,\, s_x &=& -1 \quad \hbox{ for all } x\in{\zed}^3 
\setminus \Lambda.\label{hombc}
\eea
are referred to as {\it{homogeneous b.c.}}

It is an interesting question what boundary configuration of the effective
spin system correspond to bona fide boundary conditions for the original
Falicov-Kimball model {\em without} introducing special boundary interaction
terms. It is not hard to see from the derivation of the effective 
Hamiltonian in \cite{Mes}, that the homogeneous b.c. discussed above, as well
as the boundary conditions employed later in this paper to construct
interface Gibbs states, can be achieved by imposing
Dirichlet boundary conditions for the electrons on a volume with includes 
the first boundary layer. On the effective Hamiltonian, this has the effect 
of truncating the interactions across the boundary to nearest neighbour
interactions only, which is inconsequential. This simple correspondence 
between boundary conditions, however, is only possible because we are 
considering spin-less fermions, at half-filling and with nearest neighbour 
hopping only. 

In more general situations natural boundary conditions in the 
original system will lead to modified boundary interactions in the effective 
spin system. The general strategy adopted in \cite{Mes} to address this problem
is to trace over all possible configurations of static fermions outside the
volume $\Lambda$ under consideration. Due to the Pauli principle, each such 
configuration of spinless fermions simply defines an excluded volume.
This means that the effective Hamiltonian describes a weighted average of
Gibbs states obtained with different Dirichlet-type boundary conditions
for the fermions. As the fermions can wander across the boundary of
$\Lambda$, the effective Hamiltonians thus obtained will have interaction terms
of arbitrary range across the boundary. As is clear from our analysis
here, this kind of ``averaged'' boundary conditions can equally well be used
to demonstrated to existence of Gibbs states with interfaces.

As mentioned before, the purpose of this paper is to study the thermodynamic
properties of the $100$ and $111$ interfaces of the three--dimensional FK
model. In finite volume, the presence of an interface can be enforced by a
suitable choice of {\it mixed} boundary conditions in the standard way
\cite{Dob} (see eqs.\reff{bc1} and \reff{bc2} below).

To construct the $100$ interface Gibbs state we consider $\Lambda \subset
{\zed}^3$ to be a  parallelepiped centered at the origin. The $100$ interface
is orthogonal to the vector ${\bf{n}}=(0,0,1)$.  The boundary condition
which leads to a $100$ interface (which we shall refer to as ``b.c. 1'') is
given as the configuration $\{s_x\}$ on the sites $x=(\x,\y,\z,)\in
\Lambda^c:= {{\zed}}^3 \setminus \Lambda$ defined by 
\be 
s_x= \cases{ +1 & if $\z \ge 1/2$\cr 
             -1 & if $\z \le -1/2$} \label{bc1} 
\ee

\begin{figure}[t]
\begin{center}
\resizebox{!}{7truecm}{\includegraphics{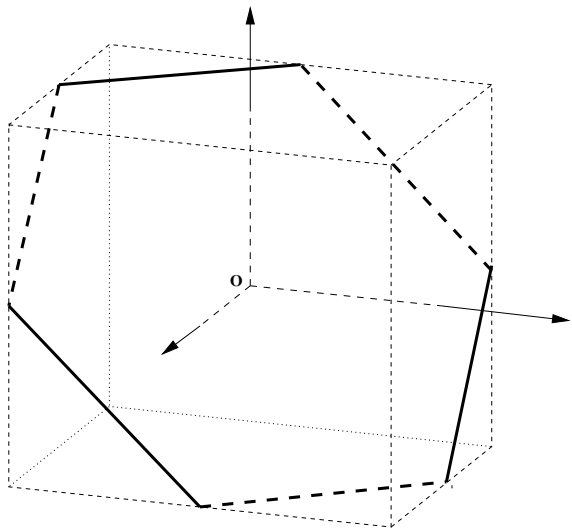}}
\parbox{14truecm}{\caption{\baselineskip=16 pt\small\label{fig:volume111}
The intersection of a finite cubic lattice $\Lambda \subset {\zed}^3$ centered
at the origin, with a plane passing through the origin and orthogonal to the
vector ${\bf{n}}=(1,1,1)$.}
}
\end{center}
\end{figure}

For an analysis of the $111$ interface consider the intersection of a a plane
passing through  the origin and orthogonal to the vector
${\bf{n}}=(1,1,1)$ with a cube $\Lambda \subset {\zed}^3$. This intersection
yields a plane bounded by a hexagon (as shown in Figure \ref{fig:volume111}) 
which divides $\Lambda$ into two equal volumes. To obtain a  $111$ interface
we  consider the spin variables $s_x$ [\reff{sxnew}] to have opposite values
on the two sides of this dividing plane.

More precisely, the boundary condition that
leads to a $111$ interface (which we shall refer to as ``b.c. 2'') is given as
the configuration $\{ s_x\}$ on the sites 
$x=(\x,\y,\z,)\in \Lambda^c \equiv {\zed}^3 \setminus \Lambda$ defined
by
\be
s_x= \cases{ +1 & if $\x+\y+\z \ge 1/2$\cr -1 & if $\x+\y+\z \le -1/2$}
\label{bc2}
\ee
Each of these boundary conditions divide the volume $\Lambda$ into two
subvolumes (the configuration in the latter being given by the two homogeneous
phases $s_+$ and $s_-$) and   enforce the occurrence of an interface (domain
wall) pinned to the boundary of the volume.  The residual free energy per unit
area of this interface is its surface tension and is denoted by the symbol
$\tau^{\rm{mixed\, b.c.}}$. It is defined by taking the difference of the free
energy of the system in the volume $\Lambda$ under mixed b.c. and the free
energy corresponding to the homogeneous b.c.
\be
\tau^{\rm{mixed\, b.c.}} =
\lim_{\atop{\Lambda \nearrow {\zed}^3}} - \frac{1}{\beta \vert I_\Lambda\vert}
\log \Bigl(\frac{\Xi^{\rm mixed}_\Lambda}{\Xi^{\rm hom}_\Lambda}\Bigr)
\label{sfce}
\ee 
where $\Xi^{\rm mixed}_\Lambda$ and ${\Xi^{\rm hom}_\Lambda}$ are the
partition functions of the system in the finite volume $\Lambda$ w.r.t mixed
b.c.  and homogeneous b.c. respectively (see Section
\ref{sec:partition_functions}).  The symbol $\vert I_\Lambda\vert$ denotes the
area of the portion of the ground state interface which is contained in the
volume $\Lambda$. Such a definition is justified \cite{mir} since the  volume
contributions proportional to the free energies  of the coexisting phases, as
well as the boundary effects, cancel and only the contributions  to the free
energy of the interface are left.

The limit $\lim_{\Lambda \nearrow {\zed}^3}$ is taken in a definite order :
e.g. if the ground state interface lies in a plane $P_{\bf{n}}$, passing
through the origin, which is orthogonal to a non-zero vector $\bf{n}$, then the
dimensions of the lattice perpendicular to the plane $P_{\bf{n}}$ are taken to
infinity before the dimensions parallel to it are taken to infinity. The
symbol $\vert I_\Lambda\vert$ denotes the area of the portion of the interface
which is contained in the volume $\Lambda$. 

It will be convenient to consider the {\it{relative Hamiltonian}} defined with
respect to the homogeneous phases $s_+$ and $s_-$. For any configuration $s=
\{s_x\}_{x \in {\zed}^3}$ the relative Hamiltonian is given by
\bea
H_\Lambda(s) &:=& - J(U) \sum_{<xy> \in \Lambda}
(s_x s_y - 1)
+ \sum_{B \cap \Lambda \ne \emptyset  \atop{|B| \ge 3}} \Phi_B(s)\nonumber\\
&:=& H^{(2)}_{0\Lambda}(s) + R_{\Lambda}^{\ge 3}(s) \label{rel1}
\eea
where
\be
\Phi_B(s) := R_B(s) - R_B(s_+) = R_B(s) - R_B(s_-), \label{rel2}
\ee
and $J(U)$ is given by \reff{ju}.
Here and henceforth the
explicit $U$-dependences of the relative Hamiltonian and its components
have been suppressed for notational simplicity.

Each configuration in a finite volume $\Lambda$, with respect to  fixed
boundary conditions (homogeneous or mixed) can be geometrically described by
specifying the Ising contours  which are defined as follows \cite{mir}: We
define a {\it{face}} to be a unit square which bisects a nearest neighbour
bond of the lattice perpendicularly. To each nearest neighbour bond we can
associate a face. A face $f$  belongs to $\Lambda$ if at least one site of the
corresponding nearest neighbour bond is in $\Lambda$. Given a configuration
$\omega_\Lambda$ on a finite lattice $\Lambda$, with b.c. $\bar{\omega}$, let
${\cal{S}}^{\bar{\omega}}(\omega_\Lambda)$ be the set of faces associated with
nearest neighbour bonds between opposite spins. Decompose
${\cal{S}}^{\bar{\omega}}(\omega_\Lambda)$ into maximally connected pairwise
disjoint components. Each such component is referred to as an {\em{Ising
contour}} (or simply {\em contour}, if confusion is not likely) and is denoted
by the symbol $\gamma$. For homogeneous b.c. the contours are closed surfaces
lying entirely within the volume $\Lambda$. However, for mixed b.c. there is
necessarily one (and only one) contour which is pinned to the boundary of the
volume $\Lambda$. This is the only infinite maximally connected component of
the set ${\cal{S}}^{\bar{\omega}}(\omega_\Lambda)$ and is referred to as the
interface. 

There is a one-to-one correspondence between spin configurations on the
lattice and non-intersecting families of contours $\Gamma = \{\gamma\}$:
$\omega_{\Lambda} = \omega_{\Lambda}(\Gamma)$. We shall refer to such families
as {\em{compatible}} families of contours. In the sequel we shall use the
symbol $\gamma$ to denote both the contour and its support. The number of
faces in a contour $\gamma$ is denoted by $|\gamma|$ and satisfies the bound
$|\gamma| \ge 6$, since the lattice is three-dimensional. 
The energy of a contour $\gamma$ is given by
\begin{equation}
E(\gamma) := H_{0\Lambda}^{(2)}(\Gamma' \cup \{\gamma\}) 
- H_{0\Lambda}^{(2)}(\Gamma').
\label{contenergy}
\end{equation}
for any set, $\Gamma'$, of non-intersecting contours, not containing $\gamma$,
such that $\Gamma' \bigcup \{\gamma\}$ is again a family of non-intersecting
contours.

It follows from \reff{rel1} that the relative energy of a configuration 
$\omega_\Lambda(\Gamma)$ is
given by 
\be
H_{\Lambda}(\Gamma) = H_{0\Lambda}^{(2)}(\Gamma) + R^{\ge 3}_\Lambda(\Gamma)
\label{rel}
\ee
with
\be
H_{0\Lambda}^{(2)}(\Gamma)
= \sum_{\gamma \in \Gamma} E(\gamma), \label{rel10}
\ee
where
\be
E(\gamma)
= 2 J(U) = {\Bigl(\frac{1}{2U} + \ho\Bigr)}  |\gamma| =: J_1(U) |\gamma| 
\label{j1}\ee
and
\be
R^{\ge 3}_\Lambda(\Gamma)
= \sum_{B \cap \Gamma \ne \emptyset \atop{|B| \ge 3}} \Phi_B(\Gamma).
\label{rgamma}
\ee
The condition $B \cap \Gamma \ne \emptyset$ denotes that the above sum runs
over all bonds $B$ such that $B \cap \gamma \ne \emptyset$ for some $\gamma
\in \Gamma$, i.e., the bond $B$ intersects at least one face of a contour
$\gamma \in \Gamma$. The quantity $E(\gamma)$ is the self-energy of the
contour $\gamma$ w.r.t. $H_{0\Lambda}^{(2)}$.  Its definition  \reff{j1}
implies that, for sufficiently large $U$, the Hamiltonian $H_{0\Lambda}^{(2)}$
satisfies the Peierls condition:
\begin{equation}
E(\gamma) \;\ge\; {\cal{J}} |\gamma|\;,
\label{peierls}
\end{equation}
with a Peierls constant
\begin{equation}
{\cal{J}} = c_0 U^{-1},
\label{s.5}
\end{equation}
where $c_0$ is a positive constant. Moreover, from the definition 
\reff{rel2} and the bound \reff{bd2} it
follows that for $|B| \ge 3$
\be
|\Phi_B| \le c_2 \Bigl(\frac{c_1}{U}\Bigr)^{g(B)}, 
\label{expo}
\ee
with $c_2 = 2{\tilde{c}}_2$. 
The effect of the term $R^{\ge 3}_\Lambda(U)$ is to modify the self-energy
of the contours and also to
introduce interactions between the contours. Hence the
spin model is reformulated as a model of interacting contours.

This contour Hamiltonian can be used to prove the rigidity  of the $100$
interface. However, we show in Section \ref{minimal} that, under the boundary
condition b.c.2 \reff{bc2}, the leading part  $H_{0\Lambda}^{(2)}$ of the
Hamiltonian yields infinitely many ground state interfaces in the thermodynamic
limit. These interfaces are characterized by the fact  that they all have
minimal area. We refer to such interfaces as {\it{minimal area interfaces}}.
Hence, to prove the rigidity of the $111$ interface we consider a more detailed
decomposition of the relative Hamiltonian in which all the terms up to order
$U^{-3}$ of the perturbation series are computed explicitly and retained in its
leading part $H_{0\Lambda}^{(4)}$:
\be
H_{\Lambda}= H_{0\Lambda}^{(4)} + R_{\Lambda}^{\ge 5} \label{rel3}
\ee
where
\bea
H^{(4)}_{0\Lambda} &=& - \Bigl(\frac{1}{4U}-\frac{11}{16U^3} + \ho\Bigr)
\sum_{<xy>\subset \vo}(s_x s_y - 1)
+\Bigl(\frac{3}{16 U^3} + \ho\Bigr) \sum_{x,y 
\in \vo \atop{\vert x-y\vert =\sqrt{2}}}
(s_x s_y - 1)\nonumber\\
&&+\Bigl(\frac{1}{8 U^3} 
+ \ho\Bigr)\!\sum_{x,y \in \vo \atop{\vert x-y\vert =2}}\!
(s_x s_y - 1)+ \Bigl(\frac{5}{16 U^3} + \ho\Bigr)\!\!\!\!
\sum_{x,y,z,t\subset P(\vo)}\!\!\!\!(s_x s_y s_z s_t -1) \label{rel30}
\eea
where $ P(\vo)$ is the set of {\it{plaquettes}}, each plaquette consisting
of four lattice sites forming a unit square.

The remainder $R_{\Lambda}^{\ge 5}$ is obtained from the series \reff{rgamma}
defining $R_{\Lambda}^{\ge 3}$ by subtracting all terms which depend on
$U^{-n}$ with $n \le 3$. It is given by
\be
R_{\Lambda}^{\ge 5}=\sum_{B \cap \Lambda \ne \emptyset
\atop{|B| > 3}}\tPhi_B,
\label{rel5}
\ee
with the potentials $\tPhi_B$ satisfying the 
bound
\be
|\tPhi_B| \le c_2 \left(\frac{c_1}{U}\right)^{m(B)},
\label{phibt}
\ee
where 
\be
m(B) := {\rm{max}} (5, g(B)).
\label{mb}
\ee 
In the expression for the relative Hamiltonian $H^{(2)}_{0 \Lambda}$
\reff{rel1} the terms up to order $U^{-1}$ are computed explicitly by second
order perturbation theory, while the terms of order $U^{-3}$ in 
 $H^{(4)}_{0 \Lambda}$ \reff{rel30} are obtained by fourth order 
perturbation theory. We shall refer to \reff{rel1} as the 
second order decomposition of the relative Hamiltonian, and \reff{rel3} 
as its fourth order decomposition.

In Section
\ref{sec:rigidity} we prove that from the
infinite set of minimal area interfaces, and up to translations, a unique
interface configuration is selected (i.e., attributed minimal energy) by
$H_0^{(4)}$. This interface and its translations are referred to as 
{\it{ground state interfaces}} of the three-dimensional FK model under the
boundary condition b.c.2 \reff{bc2}. 
Hence the $111$ interface exhibits ground state selection. Further, 
we prove that the selected interface 
is rigid in the sense that it persists under the action of the remainder
$R_{\Lambda}^{\ge 5}(U)$ at sufficiently low temperatures, in the 
thermodynamic limit. 

Our main results are that, for sufficiently large $U$, and at sufficiently low 
temperatures, the Gibbs states obtained in the thermodynamic  limit
$\Lambda\nearrow{\zed}^3$ with the boundary conditions b.c. 1 and b.c. 2,
describe rigid interfaces in the $100$-- and $111$ directions respectively. For
a precise statement of our main results we introduce the following notations:
Let  $< >_{\rm [b.c. 1]}$ and  $< >_{\rm [b.c. 2]}$ denote the expectation
values in 
the  (infinite--volume) Gibbs states with the mixed boundary condition b.c.1 
\reff{bc1} and b.c.2 \reff{bc2} respectively. Further, we recall that for a 
site $x=(x_1, x_2, x_3)$ in the lattice, $s_x$ denotes the  on--site spin
variable defined through  \reff{spinvv} and \reff{sxnew}. 
Using these notations and definitions, we state
our  main results through the following theorems:
\begin{Thm}
\label{theorem1A}
There exist positive constants $ U_0$ and $D_0$ such 
that for all $U > U_0$, and  $\beta/U > D_0$, the
following bounds are satisfied:
\be 
<s_x>_{\rm{[b.c. 1]}}
\geq \, 1- 2 {C_0} e^{-c' \beta/{U}} \quad \quad {\hbox{for 
$x_3 \ge 1/2$,}}
\label{s10}
\ee
and
\be 
<s_x>_{\rm{[b.c. 1]}}
\le \, -1 + 2 {C_0} e^{-c' \beta/{U}}  \quad \quad {\hbox{for 
$x_3 \le -1/2$,}},
\label{s20}
\ee 
where $C_0$ and $c'$ are positive constants given in terms of $U_0$ and $D_0$.
\end{Thm}

\begin{Thm}
\label{theorem1B}
There exist positive constants ${\tilde{U}}_0$, ${\tilde{D}}_0$ and 
${\tilde{D'}}_0$ such 
that for all $U >{\tilde{U}}_0 $, $\beta/U> {\tilde{D}}_0$
and $\beta/U^3 > {\tilde{D'}}_0$, the
following bounds are satisfied:
\be 
<s_x>_{\rm [b.c. 2]}
\geq \, 1- 2 \Bigl\{{\tilde{C_0}} e^{-c^{\prime\prime} \beta/{U^3}} 
+ {\tilde{C_1}} e^{-c_1^{''} \beta/{U}} \Bigr\}\quad \quad {\hbox{for 
$x_1+x_2+x_3 \ge 1/2$,}}
\label{s1}
\ee
and
\be 
<s_x>_{\rm [b.c. 2]}
\le \, -1 + 2 \Bigl\{{\tilde{C_0}} e^{-c^{\prime\prime} \beta/{U^3}} 
+ {\tilde{C_1}} e^{-c_1^{''} \beta/{U}} \Bigr\}
\quad \quad {\hbox{for 
$x_1+x_2+x_3 \le -1/2$}},
\label{s2}
\ee
where ${\tilde{C_0}}$, $c^{\prime\prime}$, ${\tilde{C_1}}$ and $c_1^{''}$  
are positive constants given in terms of ${\tilde{U}}_0$, ${\tilde{D}}_0$
and ${\tilde{D'}}_0$.
\end{Thm}

To prove these results we follow the method introduced by
Dobrushin \cite{Dob} and consider effective two--dimensional models of the
$100$-- and $111$ interfaces, obtained by projecting the interfaces 
on the planes defined by $x_3=0$ and $x_1+x_2+x_3=0$ respectively 
[see Section \ref{geom100} and \ref{geom111} 
for details]
The rigidity of the interfaces follows from an 
analysis of the low--temperature properties of these
effective two--dimensional models.

\masubsection{The geometry of $100$ interfaces.}
\label{geom100}
The geometry of the $100$ interfaces  of the FK model is the same as the
geometry of the $100$ interfaces described by Dobrushin for  the three
dimensional Ising model. Hence we refer to \cite{Dob} for the definitions of
geometrical objects which describe the interfaces and their significances as
configurations of a two dimensional (contour) model: the {\em{ceilings}} 
(which project on to 
ground states of the two dimensional model), the
{\em{walls}} (which project on to the contours), and the 
{\em{standard walls}} (which project on to the external contours).

\masubsection{The geometry of $111$ interfaces.}
\label{geom111}

The geometry of the $111$ interfaces  of the FK model is much  more involved. 
Let $\tcalI$ denote the family of interfaces under the b.c.2 \reff{bc2} and
$\tI$ denote its typical element. Such an interface is pinned at the boundary
of $\Lambda$ on the curve defined by
\be
\partial \Lambda \cap \{(x,y,z) \in {\szed}^3 \mid x+y+z = 0\} \label{curve}
\ee
In this section we describe the underlying geometrical structure necessary for
the definition and study of the effective two--dimensional model for such an
interface.

For each integer $n$ we define planes $P_n\subset{\szed}^3\subset\Rl^3$
orthogonal to  the vector $\hat{n}= (1,1,1)$, by
$P_n=\{(x,y,z)\in{\szed}^3\mid x+y+z=n\}$. Let $\P$ denote the plane in
$\Rl^3$ which contains $P_0$. Let $\Proj$ denote the orthogonal
projection onto $\P$ and let $\P_{\Lambda}$ denote the portion of the plane
$\P$ which is contained in the volume $\Lambda$.

An effective two-dimensional model for the $111$ interface is obtained by an
orthogonal projection of the interface $\tI$ onto $\P$. Its complete
description requires the following ingredients:
\begin{enumerate}
\item A set of {\it vertices\/} $\V=\V_0\cup\V_1\cup\V_2$, where
$\V_{\rm{n\, mod \, 3}}\equiv\Proj(P_n)$, $n\in{\szed}$, is a triangular
lattice in $\P$ with lattice constant $\sqrt{2}$. The set $\V$ also
forms a triangular lattice, but with lattice constant $\sqrt{2/3}$.
\item A set of {\it edges\/} $\E= \E_{01}\cup \E_{12}\cup \E_{20}$,
where $\E_{ij}$ is the set of nearest neighbour edges in the lattice
$\V_i\cup\V_j$. All edges have Euclidean length $\sqrt{2/3}$.
\item A set of {\it triangles\/} $\T$ consisting of all elementary
triangles in $\V$.
\item A set of {\it rhombi\/} $\R=\R_0\cup\R_1\cup\R_2$, where $\R_i$ is the
set of all rhombi formed by two triangles in $\T$ that share an edge
$e\in\E_{\{012\}\setminus\{i\}}$, $i=0,1,2$.
\end{enumerate}

The set of vertices $\V$ is the projection of the vertices in ${\szed}^3$,
and $\Proj((x,y,z))\in \V_i$ if and only if $i = (x+y+z)\, \bmod 3$.

The set of edges $\E$ is the projection of the set of nearest neighbour
bonds in ${\szed}^3$. They are the edges of a triangular lattice with
lattice constant $\sqrt{2/3}$. For each pair of distinct
$i,j\in\{0,1,2\}$, the set of vertices $\V_i\cup\V_j$ forms a triangular
lattice $\H_{ij}$ (also with lattice constant $\sqrt{2/3}$), and with
edges $\E_{ij}$. Together the three $\H_{ij}$ cover $\V$ twice.

As before, to each nearest neighbour bond in the lattice ${\zed}^3$ we
associate a unit square (face) which bisects it perpendicularly. Recall that
the vertices of such a face have integer coordinates. The rhombi in $\R$ are
the projections of these faces. Hence, an interface $\tI$ in the lattice
$\Lambda$ projects onto a covering of the plane $\P_\Lambda$ with rhombi in
$\R$. We refer to such a rhombus covering as a {\em{rhombus configuration}} or,
for brevity, as an $R$-configuration (to be distinguished from a configuration
of Ising contours). 

\begin{itemize}
\item A rhombus is said to be an {\em{overlapping}} one if it
contains the projection of more than one face of the interface.
Otherwise it is said to be non--overlapping. 
Each triangle $t$ in an $R$--configuration $\C$ necessarily belongs to
the projection of an odd number of faces of the interface. To each such
triangle $t$ we associate a number $o(t)$ which we 
refer to as its {\em{overlap 
number}}, and define as follows:
\be
{o(t) := \{{\hbox{the number of faces of the interface whose
projection contains}}\,  t}\}\,  - 1
\ee

A triangle with a non-zero overlap number is referred to as an overlapping
triangle. It is evident that each overlapping triangle has an even overlap
number.
\item Two non-overlapping rhombi which share an edge are said to form a 
{\it good pair} 
if the angle enclosed by them is $2 \pi/3$, i.e., if up to translations 
and rotations the pair is as shown in Figure \ref{fig:good_pos}.  
The edge shared by such a pair is referred to as a {\em{good edge}}.
We consider the good pairs as open complexes. This means that a good
pair is composed of an open edge together with the two adjacent open
rhombi. Two good pairs are connected if their intersection is an open
rhombus.

\begin{figure}[t]
\begin{center}
\resizebox{!}{3truecm}{\includegraphics{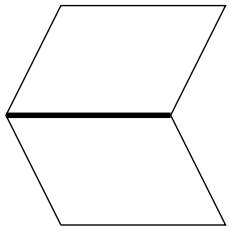}}
\parbox{14truecm}{\caption{\baselineskip=16 pt\small\label{fig:good_pos}
A pair of rhombi in {\em{good position}}. The edge shared by the two rhombi
is referred to as a {\em{good edge}}.
}}
\end{center}
\end{figure}

\item We define the {\it type} of a rhombus $r$, $\tau(r)$, to be $j$ if
$r\in\R_j$, with $j\in \{0,1,2\}$.
\end{itemize}

Let a {\it{tiling}} of the plane $\P$ be defined as a complete
covering of $\P$ with non-overlapping rhombi in $\R$. For each pair of
distinct $i,j\in\{0,1,2\}$ the set of edges $\E\setminus \E_{ij}$
drawn in the plane $\P$ yield a tiling of $\P$ with the rhombi in
$\R_{\{012\}\setminus\{ij\}}$. Minimal area interfaces and ground state
interfaces have simple geometric descriptions in terms of tilings
(see Section \ref{minimal}).

The mixed boundary condition b.c.2 translates into a boundary condition
for the $R$-configuration in $\P_{\Lambda}$. It is given by  a tiling 
of $\P \setminus \P_\Lambda$ with
rhombi of a single type, say $\R_0$.
We call this the {\it{standard b.c.}} for the rhombus model. 
\bigskip

\masubsection{\bf The minimal area interfaces}
\label{minimal}

An interface $\tI$ is of minimal area if and only if its projection 
$\Proj(\tI)$ is a tiling of $\P$. Equivalently each such tiling is in
one-to-one correspondence with a dimer covering of the hexagonal lattice dual
to the triangular lattice $\V$, i.e., of the lattice with set of sites given
by the centers of the triangles in $\T$.  It is less obvious, although also a
well-known fact in enumerative combinatorics \cite{Kup,Thu}, that all tilings
of $\P$ with rhombi in $\R$ correspond to a unique minimal area interface.
This is shown in the following proposition.  The proof is constructive, i.e.,
it provides an algorithm for obtaining the interface from the tiling and vice
versa.

\begin{Prop}\label{prop:1to1}
The tilings of the plane $\P$ with rhombi in $\R$ under standard
b.c. are in one-to-one correspondence with the minimal area interfaces
in the volume $\Lambda$ under the mixed boundary condition b.c.2
\reff{bc2}.
\end{Prop}
\begin{proof}
As noted above it is obvious that each minimal area interface under the
boundary condition b.c.2 \reff{bc2} projects onto a tiling of $\P_\Lambda$ with
rhombi in $\R$. Hence, we only need to show that to each tiling
there corresponds exactly one interface that has that tiling as its
projection. The interface will automatically be minimal.  This amounts
to associating a unique face of ${\szed}^3$ to each rhombus in the tiling
such that the resulting set of faces form a connected set which is
pinned at the boundary of $\Lambda$ along the curve defined by
\reff{curve}.  The projection of the set of faces constituting an
interface $\tI$ yields a set of rhombi in $\R$ that covers $\P_\Lambda$. For
each face $f$ we can number its vertices $a_1,a_2,a_3,a_4$ in such a
way that there is a unique integer $n(f)$ for which
$$
a_1\in P_{n(f)-1},\quad a_2\in P_{n(f)},\quad a_3\in P_{n(f)+1},\quad a_4\in
P_{n(f)}
$$
It is easy to see that if $\Proj(f)=r\in \R_i$, $i=0,1,2$,
then $i=n(f)\bmod 3$, and that $\Proj(f)$ and $n(f)$ uniquely determine
$f$.

For any tiling with standard boundary conditions we will construct a
unique  {\it height function\/} $h:\V\to {\szed}$ with the property that for
each rhombus $r$ in the tiling the heights of its vertices, when ordered
appropriately, and such that $\{v_i,v_{i+1}\}$ are edges of $r$, are
given by
$$
h(v_1)=n-1,\quad h(v_2)=n,\quad h(v_3)=n+1,\quad h(v_4)=n
$$
for some integer $n$ satisfying $\, i=n\bmod 3$ iff $r\in \R_i$. It follows
that $h$ satisfies $\,\vert h(v)-h(w)\vert =1\,$ for each edge $\{v,w\}$ of a
rhombus in the tiling. A minimal area interface in $\Lambda$ under the b.c.2
\reff{bc2} is an interface whose projection on the plane $\P_{\Lambda}$ is a
tiling.  It consists of all faces $\{f\}$ in $\Lambda$ for which:
\begin{enumerate}
\item $\Proj(f)$ is a rhombus in the tiling, and
\item the vertices $v_i$ of $\Proj(f)$ satisfy $\{h(v_i)\mid 1\leq i\leq
4\}=\{n(f)-1, n(f), n(f)+1\}$.
\end{enumerate}

It remains to construct the {\it height function\/} $h$ and to verify
that it is the unique function with the stated properties. Let us
denote by ${\vec{e}}_1,\ldots,{\vec{e}}_6$ the vectors of minimal
length $(=\sqrt{\frac{2}{3}})$, emanating from a single point of the
triangular lattice (spanned by the edges in $\E$) such that the tips
of these vectors are the vertices of a hexagon. See Figure
\ref{fig:vectors}.  For any two vertices $v, w \in \V$ for which
$\{v,w\}$ is an edge in a tiling of $\P$, 
${\vec{w}}- {\vec{v}}$ is one of ${\vec{e}}_i$.  Let us denote by
$\theta({\vec{e}}_i,{\vec{e}}_j)$ the angle between ${\vec{e}}_i$ to
${\vec{e}}_j$, which is a multiple of $\pi/3$. We claim that for each
tiling there is a unique height function $h$ satisfying
$$
h({{w}})-h({{v}})=\cases{+1
&if $\theta({\vec{e}}_1,{\vec{w}}-{\vec{v}} )$ is an even multiple of
$\pi/3$\cr -1
&if
$\theta({\vec{e}}_1,{\vec{w}} - {\vec{v}} )$ is an odd
multiple of $\pi/3$\cr}
$$
for each edge $\{v,w\}$ of a rhombus in the tiling. The height
function $h$ must then be obtained by summing up the above differences
along edges, starting from a convenient reference value on the
boundary of $\P_{\Lambda}$. Consistency of this definition follows
from the elementary observations that
\begin{enumerate}
\item  any two paths connecting the same pair of
vertices and consisting of edges in the tiling, together enclose a
bounded subset of the plane tiled with rhombi, and
\item for any rhombus the differences $h(w)-h(v)$ along the four edges
of a rhombi sum up to zero because the rhombus has two angles of
$\pi/3$ and two of $2\pi/3$.
\end{enumerate}
Then the  uniqueness of $h$ is also obvious.

\begin{figure}[t]
\begin{center}
\rotatebox{-90}{\resizebox{!}{3truecm}{\includegraphics{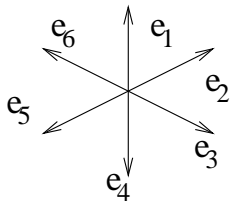}}}
\parbox{14truecm}{\caption{\baselineskip=16 pt\small\label{fig:vectors}
The labeling of the six unit vectors emanating from a site of the 
triangular lattice.
}}
\end{center}
\end{figure}

Note that the height function $h$ has been defined such that the height
of any vertex in the tiling is equal to the sum of the coordinates of the point
 in the interface of which it is the projection.
\end{proof}

\begin{Prop}\label{prop:degeneracy}
The number of minimal interfaces $N_\Lambda$
grows exponentially with the area of the interface and satisfies the bounds:
$$
2^{\vert I_\Lambda\vert\over 3}\leq N_\Lambda\leq 2^{2\vert I_\Lambda\vert}
$$
where $\vert I_\Lambda\vert$ is the area of the minimal area interfaces 
in the volume $\Lambda$.
\end{Prop}
\begin{proof}
Consider the interface defined by the $\R_0$ tiling, and pick one of
the three hexagonal sublattices of the vertices.  Note that,
independently of each other, each hexagon can be tiled in two ways
with three rhombi. This proves the lower bound.

The upper bound is obtained by considering the hexagonal lattice dual
to the triangular lattice.  Since every tiling is in one-to-one
correspondence with a dimer covering of this hexagonal lattice, it is
also in one-to-one correspondence with a path covering of the hexagonal
lattice, restricted by the condition that at each site the incidence of the
path is two. The upper bound is just the straightforward bound on the 
number of paths given by 
$$
2^l \quad {\mbox{where $l$ is the length of the path}}
$$
As the path is covering, the length of the path equals the number of bonds
in the hexagonal lattice which equals twice the number of rhombi needed
to tile the region, i.e., twice the area.
\end{proof}

Note that the exact rate of exponential growth of the degeneracy 
depends on the shape of the finite volumes \cite{Pro}.

{From} Proposition \ref{prop:degeneracy} it follows that in the limit 
$\Lambda \nearrow {\zed}^3$ there are infinitely many minimal area interfaces.
 
\masection{The relevant partition functions}
\label{sec:partition_functions}

We shall define the partition functions of the FK model in terms of the
second-- and fourth order decompositions of the relative Hamiltonian. These
are defined through  \reff{rel1} - \reff{rel2} and \reff{rel3} - \reff{rel5}
respectively.  Such definitions are justified because the quantity relevant
for the study of an interface is not a solitary partition function but rather
the quotient of two partition functions, namely, a partition function
corresponding to mixed b.c. and one corresponding to homogeneous b.c (see
\reff{sfce}). The use of relative Hamiltonians in the definition of the
partition functions corresponds to the simultaneous subtraction of the energy
of a homogeneous phase from the Hamiltonians appearing in the numerator and
denominator of the quotient and hence keeps the quotient unchanged.

Let $\Xi^+_{\Lambda}$ and $\Xi^-_{\Lambda}$ be the partition functions
in $\Lambda \subset {\zed}^3$ w.r.t the homogeneous boundary conditions
defined by \reff{hombc}.
The spin-flip symmetry of the
effective Hamiltonian $\H^{\rm eff}_\Lambda(U)$ \reff{heff1},
for the choice $h=0$, implies that
\be
\Xi^+_{\Lambda} = \Xi^-_{\Lambda} := \Xi^{\rm{hom}}_{\Lambda}.
\ee
Let $\Xi^{100}_{\Lambda}$ and $\Xi^{111}_{\Lambda}$ be the
corresponding partition functions under the mixed boundary conditions
b.c.1 \reff{bc1} and b.c.2 \reff{bc2} respectively.

A mixed b.c. (b.c.1 \reff{bc1} or b.c.2 \reff{bc2}) leads to the
appearance of an interface $I$ which divides the volume $\Lambda$ into
two subvolumes $\Lambda^a_{I}$ and $\Lambda^b_{I}$ which are,
respectively, the regions above and below $I$. The configuration in
these subvolumes are defined by finite sets of compatible
contours $\Gamma^a := \{\gamma^a_1, \ldots, \gamma^a_p\}$ and
$\Gamma^b = \{\gamma^b_1, \ldots, \gamma^b_q\}$ respectively. We
define $\Gamma:= \Gamma^a \cup \Gamma^b$ and denote a configuration on
the lattice by $(I \cup \Gamma)$.
Let $\I$ and $\tcalI$ denote the families of
interfaces resulting from b.c.1 and b.c.2 respectively. Let $I$ and
$\tI$ denote their typical elements.

In terms of the second order decomposition \reff{rel}
of the relative Hamiltonian $H_\Lambda$,
the partition function for homogeneous b.c. is given by
\be
\Xi^{\rm hom}_\vol=\sum_{\Gamma=\{\gamma_1,\dots,\gamma_n\} \subset\vol}
\,\prod_{i=1}^n e^{-\beta J_1(U) \vert\gamma_i\vert}
\,\prod_{\vert B\vert >2 \atop B\cap\Gamma\neq\emptyset} e^{- \beta
\Phi_B(\Gamma )},
\label{parthom}
\ee
where $ J_1(U) \simeq \frac{1}{2U} + \ho$
The partition function relevant for the mixed boundary
condition b.c.1 \reff{bc1} is also obtained by using eqns.\reff{rel} - 
\reff{rgamma}. It is given by
\begin{equation}
\Xi^{100}_\vol = \sum_{I\in\cal{I}} \, \sum_{\Gamma:=\Gamma^{a}\cup
\Gamma^{b}} e^{- \beta {\Ea}(I \cup \Gamma)},
\label{part100}
\end{equation}
where $\Ea(I \cup \Gamma)$ is the energy of the configuration
$(I \cup \Gamma)$. It is the value that the relative
Hamiltonian $H_\Lambda$ takes on the
configuration $(I \cup \Gamma)$:
\be
\Ea(I \cup \Gamma) = \sum_{\gamma^a \in \Gamma^a} J_1(U) |\gamma^a| +
 \sum_{\gamma^b \in \Gamma^b} J_1(U) |\gamma^b| + J_1(U)|I| +
\sum_{B:|B| \ge 3\atop{B\cap(I \cup \Gamma)
\ne \emptyset}} \Phi_B(I \cup \Gamma)
\label{Enn}
\ee
The first three terms on the r.h.s. of \reff{Enn} are the energies of
the contours in the configuration $(I \cup \Gamma)$ as defined through
\reff{contenergy}.  The third term is the energy of interaction among
these contours and arises from the long-range tail potential
$R_\Lambda^{\ge 3}$
\reff{rgamma} of the relative Hamiltonian $H_\Lambda$.

Let $(I \cup \emptyset)$ denote a configuration which consists only of
the interface $I$ and no other contours. The energy of such a
configuration can be interpreted as the ``bare'' energy of the 
interface, i.e., the energy of the interface in the absence of any
other contour. Let us denote this energy by $E^{\bare}_1(I)$. It is
given by
\be
E^{\bare}_1(I):= J_1(U) |I| +  \sum_{B: |B| \ge 3 \atop{B\cap I \ne \emptyset}}
\Phi_B( I \cup \emptyset).
\label{bare}
\ee
{From} \reff{j1} and \reff{expo} it follows that for sufficiently large $U$,
the energy $E^{\bare}_1(I)$ satisfies the bound
\be
|E^{\bare}_1(I)| \le {\rm{const.}} \frac{1}{U} |I|
\label{bare1bd}
\ee

It is convenient to isolate the ``bare'' energy of the interface from
the remaining terms in the expression \reff{Enn} for $\Ea(I \cup
\Gamma)$. From \reff{Enn} and \reff{bare} it follows that
\be
\Ea(I \cup \Gamma)
= \sum_{\gamma \in \Gamma} E(\gamma) + E^{\bare}_1(I) + \tE_1(I \cup \Gamma),
\ee
with
\bea
\tE_1(I \cup \Gamma)&=&\sum_{B:|B| \ge 3\atop{B\cap(I \cup \Gamma)
\ne \emptyset}} \Phi_B(I \cup \Gamma) - \sum_{B:|B| \ge 3 \atop{B\cap I \ne
\emptyset}} \Phi_B(I \cup \emptyset)  \nonumber\\
&=&  \sum_{B:|B| \ge 3\atop{B\cap(I \cup \Gamma)
\ne \emptyset}}\{\Phi_B(I \cup \Gamma) - \chi(B\cap I \ne \emptyset)
\Phi_B(I \cup \emptyset)\}\nonumber\\
&:=&  \sum_{B:|B| \ge 3\atop{B\cap (I \cup \Gamma)
\ne \emptyset}} \Phi_B'(I \cup \Gamma),
\label{tphi}
\eea
where $\chi(\cdot)$ denotes the characteristic function.
Note that
\be
\Phi_B'(I \cup \emptyset) = 0 \quad {\rm{if}} \quad B \cap \Gamma = \emptyset.
\label{pbd}
\ee
Hence, the functions $\Phi_B'$ for which $B$ intersects {\it{only}}
the interface, do not contribute to the energy $\tE_1(I \cup \Gamma)$.
The contributions of such bonds is included in the ``bare'' 
energy $E^{\rm bare}_1(I)$
of the interface. 

A non-zero $\Phi_B'(I \cup \emptyset)$ arises only from those bonds $B$
which intersect at least one contour in $\Gamma$. This observation allows
us to write  $\tE_1(I \cup \Gamma)$ as follows.
\be
\tE_1(I \cup \Gamma) =  \sum_{B:|B| \ge 3\atop{B\cap\Gamma \ne \emptyset}}
\Phi_B'(I \cup \Gamma)
\label{remE}
\ee
Further, it follows from \reff{expo} that $\Phi_B'(I \cup \Gamma)$
satisfies the bound
\bea
\vert \Phi_B'(I \cup \Gamma) \vert &\le& \vert \Phi_B(I \cup \Gamma) \vert +
\vert \Phi_B(I \cup \emptyset) \vert \nonumber\\
&<& 2 c_2 \Bigl(\frac{c_1}{U}\Bigr)^{g(B),}
\label{primebd}
\eea
for $|B| \ge 3$.
For bonds $B$ which do not intersect the interface $I$,
\be 
\Phi_B'(I \cup \Gamma) = \Phi_B(I \cup \Gamma)
\label{same}
\ee

The partition function for the boundary condition b.c.1, defined by
\reff{part100} is hence given by
\bea
\Xi^{100}_\vol&
=&\sum_{I\in\I}e^{-\beta E^{\bare}_1(I)}
\times\sum_{\Gamma^{a} = \{\gamma_1^{a}, \dots, \gamma_p^{a}\}
\subset\Lambda^{a}_{I}} \, \sum_{\Gamma^{b} =  \{\gamma_1^{b}, \dots,
\gamma_q^{b}\}\subset\Lambda^{b}_{I}}
e^{-\beta  \tE_1(I \cup \Gamma) }\nonumber\\
&&\prod_{j=1}^{p}  e^{-\beta J_1(U)\vert\gamma_j^{a}\vert}
\times\prod_{k=1}^{q} e^{-\beta J_1(U) \vert\gamma_k^{b}\vert},
\label{part100n}
\eea
with $E^{\bare}_1(I)$ and $\tE_1(I \cup \Gamma)$ being defined
through \reff{bare} and \reff{remE} respectively.
The partition function for the boundary condition b.c.2 \reff{bc2} can be
similarly written as
\begin{equation}
\Xi^{111}_\vol = \sum_{\tI\in{\tcalI}} \, \sum_{\Gamma:=\Gamma^{a}\cup
\Gamma^{b}} e^{- \beta \Eb(\tI \cup \Gamma)},
\label{part111}
\end{equation}
where $\Eb(\tI \cup \Gamma)$ is the energy of the configuration $(\tI
\cup \Gamma)$. In order to determine whether the $111$ interface is
rigid, it is necessary to use the fourth order decomposition 
[eqns. \reff{rel3} - \reff{rel5}]
of the relative Hamiltonian $H_\Lambda$, 
for computing the contribution of the interface $\tI$ to
the energy $\Eb(\tI \cup \Gamma)$.
However, it is sufficient
to consider the second order decomposition [eqns. \reff{rel} -
\reff{rgamma}] for evaluating the corresponding contribution of the
contours in $\Gamma$.
We first introduce some notations which are useful in evaluating 
$\Eb(\tI \cup \Gamma)$.

\begin{itemize}
\item{Let $P_{\tI} \subset P(\Lambda)$ denote the set of plaquettes in
${\zed}^3$ which
are intersected by the faces of $\tI$. Let $p$ denote a typical element in
this set.}
\item{Let $B^2_{\tI} \subset {\zed}^3$ denote the set consisting of 
pairs of next nearest neighbour sites,
\be
\{\{x,z\} \in {\zed}^3 | \,\vert x-z\vert = 2\},
\ee
such that the line joining each pair is intersected by a face in $\tI$.}
\end{itemize}

{From} \reff{rel3} - \reff{rel5}, it follows that the energy 
$\Eb(\tI \cup \Gamma)$ is given by
\bea
\Eb(\tI \cup \Gamma) &=&  {J}_2(U)|\tI| + (\frac{1}{8 U^3} + \ho)
\!\sum_{\{x,z\}\in B^2_{\tI}} h_{\{x,z\}}(\tI) +  
(\frac{1}{16 U^3} + \ho)\!\!\!\sum_{p=\{x,y,z,t\}\in P_{\tI}} h_p(\tI)
\nonumber\\
&+&  \sum_{B: |B|> 3\atop{B\cap \tI \ne \emptyset}}\tPhi_B( \tI \cup \Gamma)
+ \sum_{\gamma^a \in \Gamma^a} J_1(U) |\gamma^a| +
\sum_{\gamma^b \in \Gamma^b} J_1(U) |\gamma^b| 
+ \sum_{B:|B| > 2\atop{B \cap \Gamma \ne \emptyset\atop{B\cap \tI =\emptyset}}}
\Phi_B( \tI \cup \Gamma),\nonumber\\
\eea
where 
\be
h_p := 5 (s_x s_y s_z s_t - 1) +
3(s_x s_z + s_y s_t - 2)
\label{hp}
\ee
for each $p=\{x,y,z,t\}\in P_{\tI}$,
\be
h_{\{x,z\}} := s_x s_z - 1
\label{hxz}
\ee
for each $\{x,z\}\in B^2_\tI$, and
\be
{J}_2(U) \simeq  \frac{1}{2U} - \frac{11}{8U^3} + \ho 
\label{j22}
\ee
Let $(\tI \cup \emptyset)$ denote a configuration which has the
interface $\tI$ as its only contour. We denote the corresponding ``bare'' 
energy
of the interface by $E^{\bare}_2(\tI)$. It is defined as follows.
\bea
E^{\bare}_2(\tI)&:=& J_2(U)|\tI| +  (\frac{1}{8 U^3} + \ho)
\sum_{\{x,z\}(\tI)\in B^2_{\tI}}
h_{\{x,z\}}(\tI) \nonumber\\
&+& (\frac{1}{16 U^3} + \ho)\sum_{p=\{x,y,z,t\}\in P_{\tI}} h_p(\tI)
+ \sum_{B: |B| > 3 \atop{B\cap \tI \ne \emptyset}}
\tPhi_B( \tI \cup \emptyset).
\label{bare2}
\eea
The partition function for the boundary condition b.c.2 can be expressed as
follows:
\bea
\Xi^{111}_\vol&
=&\sum_{\tI\in\tcalI}e^{-\beta E^{\bare}_2(\tI)}
\times\sum_{\Gamma^{a} = \{\gamma_1^{a}, \dots, \gamma_p^{a}\}
\subset\Lambda^{a}_{\tI}} \, \sum_{\Gamma^{b} =  \{\gamma_1^{b}, \dots,
\gamma_q^{b}\}\subset\Lambda^{b}_{\tI}}
e^{-\beta  \tE_2(\tI \cup \Gamma) }\nonumber\\
&&\prod_{j=1}^{p}  e^{-\beta J_1(U)\vert\gamma_j^{a}\vert}
\times\prod_{k=1}^{q} e^{-\beta J_1(U) \vert\gamma_k^{b}\vert},
\label{part111f}
\eea
where
\bea
\tE_2(\tI \cup \Gamma) &:=& \sum_{B:|B| \ge 3\atop{B\cap \Gamma \ne
\emptyset\atop{B \cap \tI = \emptyset}}}\Phi_B(\tI \cup \Gamma) 
+ \sum_{B:|B| > 3\atop{B\cap \tI \ne
\emptyset}}\tPhi_B(\tI \cup \Gamma) -\sum_{B: |B| > 3
\atop{B\cap \tI \ne \emptyset}}\tPhi_B( \tI \cup \emptyset)
\nonumber\\
&=& \sum_{B:|B| \ge 3\atop{B\cap \Gamma \ne
\emptyset\atop{B \cap \tI = \emptyset}}}\Phi_B(\tI \cup \Gamma) 
+ \sum_{B:|B| > 3 \atop{B\cap \Gamma \ne
\emptyset}}\Bigl(\tPhi_B(\tI \cup \Gamma) - 
\tPhi_B( \tI \cup \emptyset)\Bigr)\nonumber\\
&=& \sum_{B:|B| \ge 3\atop{B\cap \Gamma \ne
\emptyset\atop{B \cap \tI = \emptyset}}}\Phi_B(\tI \cup \Gamma) 
+ \sum_{B:|B| > 3 \atop{B\cap \Gamma \ne
\emptyset}} \tPhi_B'(\tI \cup \Gamma),
\label{te2}
\eea
where we have defined the quantity
\be
\tPhi_B'(\tI \cup \Gamma) := \tPhi_B(\tI \cup \Gamma) - 
\tPhi_B( \tI \cup \emptyset).
\label{tphibp}
\ee
It satisfies the bound
\be
\vert \tPhi_B'(\tI \cup \Gamma)\vert \le 2 c_2 (\frac{c_1}{U})^{m(B)},
\label{phibp}
\ee
where $m(B)$ is defined through \reff{mb}.

We prove in Sections \ref{sec:rigidity100} and 
\ref{sec:rigidity} that the $100$ and $111$ interfaces are
rigid at low temperatures. 
The term dependent on the area of the interface, 
in the expression \reff{bare} for the ``bare'' 
energy $E^{\bare}_1(I)$ of the interface, is responsible for the 
rigidity of the  $100$ interface. 
However, for the $111$ interface, the corresponding area-dependent term 
in the expression for the ``bare'' energy $E^{\bare}_2(I)$ 
of the interface is not
sufficient for stabilizing it against thermal fluctuations.
The rigidity at low temperatures results instead 
from the geometry-dependent contribution of
the plaquette potential $h_p(\tI \cup \emptyset)$ 
to the energy of the interface (i.e., the third term on
the RHS of \reff{bare2}).

The proofs of these results involve an analysis of the convergence properties
of the partition functions $\Xi_\Lambda^{\hom}$, $\Xi_\Lambda^{100}$ 
and $\Xi_\Lambda^{111}$ in the limit $\Lambda \nearrow {\zed}^3$.  A
direct application of the method of  cluster expansion requires the contours
to be non-interacting. This means that the energy of a configuration is given
by a sum of terms, each depending on only one contour in the corresponding
compatible family. This is not true for the FK model, because the long range
interactions in its effective Hamiltonian induce  interactions among the
contours.  We overcome this technical difficulty by rewriting the partition
functions in terms of configurations of non-interacting but more complicated
contours called {\em{decorated contours}}. In the next section we define the
decorated contours and derive expressions for the partition functions in terms
of them. 

\masubsection{Decorated contours} 
\label{deccont}

Let us first explain how we express the partition function of a system
under homogeneous boundary conditions \reff{hombc} in terms of decorated
contours. The partition function for homogeneous b.c. is given by 
\reff{parthom}, which
we repeat here for convenience.
\be
\Xi^{\rm hom}_\vol=\sum_{\Gamma=\{\gamma_1,\dots,\gamma_n\} \subset\vol}
\,\prod_{i=1}^n e^{-\beta J_1(U) \vert\gamma_i\vert}
\,\prod_{\vert B\vert >2 \atop B\cap\Gamma\neq\emptyset} e^{- \beta
\Phi_B(\Gamma )}.
\label{parthomd}
\ee
 
By convention we treat empty products as unity. 
The idea is to analyze the effect of the dominant part 
$H_{0\Lambda}^{(2)}(U)$ of the relative 
Hamiltonian \reff{rel1} by a low temperature expansion in terms
of its contours $\gamma$, while treating the contribution of the long
range tail, $R_\Lambda^{\ge 3}$, by a high temperature expansion 
\cite{Par, DMS}. Hence, we write
\begin{eqnarray}
\Xi^{\hom}_{\Lambda} &=& \sum_{\Gamma \subset \Lambda } 
\prod_{\gamma \in \Gamma}\, e^{-\beta  J_1(U) \vert\gamma_i\vert}
\, \prod_{B \cap \Gamma \ne \emptyset \atop{|B| \ge 3}} 
\Bigl[ \bigl( e^{-\beta \Phi_B} - 1 \bigr) + 1 \Bigr] \nonumber \\
&= & \sum_{\Gamma \subset \Lambda}\prod_{\gamma \in \Gamma} e^
{-\beta  J_1(U) \vert\gamma_i\vert} \Bigl[ 1 + \sum_{n \ge 1}
\sum_{{B_1,\ldots , B_n} \atop {B_i \cap \Gamma \ne \emptyset \atop 
{|B_i| > 2 \,;\,(i=1 \ldots n)}}} \prod_{i=1}^{n} 
(e^{-\beta \Phi_{B_i}} - 1)\Bigr]
\label{expand}
\end{eqnarray}
To each term on the RHS of \reff{expand} we can associate a finite set of
compatible contours and a set of
bonds, each bond intersecting the support of at least one contour in the
set. More
precisely, a {\em{decorated contour}} is defined by the pair
\begin{equation}
D = (\Gamma_D, {\cal{B}}_D),
\end{equation}
where $\Gamma_D \subset \Gamma$ is a finite set of compatible contours and
${\cal{B}}_D := \{B_1, \ldots B_n\}$ is a finite set of bonds 
such that for each $B_i$, $1\le i \le n$, there is a $\gamma \in \Gamma_D$
with $B_i \cap \gamma \ne \emptyset$ and
\be
{\rm{supp}} D := \Bigl(\cup_{\gamma \in \Gamma_D}\gamma \Bigr) \cup 
\Bigl( \cup_{B \in {\cal{B}}_D} B \Bigr)
\label{suppd}
\ee
is a connected set. In \reff{suppd} ${\rm{supp}} D$ denotes the support of a
decorated contour $D$. In the sequel we shall use the symbol $D$ for both a
decorated contour and its support. Moreover,
\begin{equation}
|D| := \sum_{\gamma \in  D} |\gamma| +  \sum_{B \in {D}} g(B).
\end{equation}
Decorated contours have the following properties:
\begin{itemize}
\item The interiors of any two distinct closed Ising contours, which belong
to a decorated contour $D$, do not intersect. 
\item Each bond $B$ in a decorated contour intersects at least one contour
in $D$.
\item Any two contours in a decorated contour $D$ are connected through bonds
and other contours in $D$, i.e., for each pair of Ising
contours $\gamma$, $\gamma'$ in $D$, there are contours $\gamma_1, \ldots.
\gamma_k$ in $D$ such that $\gamma \sim \gamma_1$, $\gamma_1 \sim \gamma_2$, 
\ldots, $\gamma_k \sim \gamma'$, with the understanding that two contours  
$\gamma_1$ and $\gamma_2$ are connected, denoted by  $\gamma_1 \sim \gamma_2$, 
if one of the following holds:
\begin{itemize}
\item there is a bond $B$ in $D$ such that $B$ intersects both  $\gamma_1$
and  $\gamma_2$
\item there are two bonds $B_1$ and $B_2$ in $D$ such that $B_1 \cap B_2 \ne
\emptyset$, $B_1$ intersects $\gamma_1$ and  $B_2$ intersects $\gamma_2$.
\end{itemize}
\end{itemize}

To each decorated contour $D= (\Gamma_D, {\cal{B}}_D)$ we can associate a
weight  $W(D)$ as follows:
\begin{equation}
W(D):= \prod_{\gamma \in D} e^{-\beta J_1(U) \vert\gamma_i\vert}  
\prod_{B \in D \atop{|B| \ge 3}}
 (e^{-\beta \Phi_B} - 1).
\label{wd}
\end{equation}
Let ${\cal{D}}$ denote a finite family of compatible decorated contours,
i.e., a finite set of mutually non-intersecting contours. 
Then the partition
function can be expressed as follows.
\begin{equation}
\Xi^{\hom}_{\Lambda} = \sum_{{\cal{D}} \cap \Lambda \ne \emptyset}
\prod_{D \in {\cal{D}}} W(D).
\label{homdec} 
\end{equation}
where $ |W(D)| \le W_0(D)$ for all $D \in {\cal{D}}$, with
\begin{equation}
W_0(D) :=  \prod_{\gamma \in \Gamma_D} e^{-\beta 
c_0 U^{-1} |\gamma|} \prod_{B \in {\cal{B}}_{D}}\Bigl[\exp\Bigl({\beta c_2 
({\frac{c_1}{U}})^{g(B)}}\Bigr) - 1\Bigr].
\label{w0d}
\end{equation}
The above bound follows from the Peierls bound \reff{peierls} and 
the exponential decay \reff{expo}. In \reff{homdec} the 
partition function 
for the lattice model under homogeneous boundary conditions has been 
expressed in terms of a gas of {\em{non-interacting}}, pairwise disjoint, 
decorated contours. The methods of cluster expansion can now be applied
to analyze its convergence properties.

The partition functions corresponding to mixed boundary conditions \reff{bc1}--
\reff{bc2} can be expressed in terms of decorated contours in a similar manner.
However, under these boundary conditions there is a contour -- the interface --
which is pinned to the boundary of the volume $\Lambda$. In our definition of
decorated contours the interface is treated differently from the remaining
Ising contours in the volume $\Lambda$. It is not considered to be a part of a
decorated contour. Consequently, in the expressions for the partition functions
under mixed boundary conditions, there is an additional sum over all possible
interfaces. As described in Section \ref{sec:FK_model}, an interface $I$
divides the volume $\Lambda$ into two subvolumes $\Lambda^a_{I}$ and
$\Lambda^b_{I}$. For each interface, the set of non-interacting decorated
contours, corresponding to a mixed boundary condition, can be decomposed into
three subfamilies which consist, respectively, of contours which  intersect
$I$, which are above $I$ and which are below $I$. Hence, for a given interface
$I$, under the boundary condition b.c.1 \reff{bc1}, we define the following
compatible families of decorated contours.
\bea
\D_I &:=& \{ \{D_{l_1} \ldots D_{l_p} |D_{l_i} \cap I \ne \emptyset, 
1 \le i \le p, l_i \in {\szed}\} \} \nonumber \\
\D^a_I &:=& \{ \{D_{m_1} \ldots D_{m_q} | D_{m_i} \cap \Gamma^a \ne \emptyset,
D_{m_i} \cap I = \emptyset, 
1 \le i \le q, m_i \in {\szed}\} \} \nonumber \\
\D^b_I &:=& \{ \{D_{n_1} \ldots D_{n_r} | D_{n_i} \cap \Gamma^b \ne
\emptyset, D_{n_i} \cap I = \emptyset, 
1 \le i \le r, n_i \in {\szed}\} \},
\label{decfam} 
\eea
where $\Gamma^a$ and $\Gamma^b$ are the subfamilies of
compatible (Ising) contours which lie entirely in the subvolumes
$\Lambda^a_{I}$ and $\Lambda^b_{I}$ respectively. 
For a given interface $\tI$, under the boundary condition b.c.2
\reff{bc2}, the corresponding families of compatible decorated
contours are denoted by $\D_{\tI}$, $\D^a_{\tI}$ and $\D^b_{\tI}$.

{From} \reff{part100n}, \reff{tphi} and \reff{same} it follows that 
\bea
\Xi^{100}_\vol &=& \sum_{I\in\cal{I}} e^{-\beta E^{\bare}_1(I)}\,
\Bigl(\sum_{\D_I \cap \Lambda \ne \emptyset} \prod_{D \in \D_I} W_I(D)
\Bigr) \nonumber\\
&\times &  \Bigl(\sum_{\D^a_I \cap \Lambda \ne \emptyset} 
\prod_{D \in \D^a_I} W(D)
\Bigr) \Bigl(\sum_{\D^b_I \cap \Lambda \ne \emptyset} 
\prod_{D \in \D^b_I} W(D)
\Bigr)
\eea 
where $W(D)$ is defined through \reff{wd} and satisfies the bound 
$|W(D)| \le W_0(D)$, with $W_0(D)$ being defined through \reff{w0d}.
\smallskip

\noindent
For $D \in \D_I$ 
\be
W_I(D):= \prod_{\gamma \in D} e^{-\beta J_1(U) \vert\gamma_i\vert}  
\prod_{B \in D \atop{|B| \ge 3}}
 (e^{-\beta \Phi_B'} - 1).
\label{wId}
\ee
{From} the Peierls bound \reff{peierls} and the estimate \reff{primebd}
it follows that $|W_I(D)| \le W^0_I$ where
\begin{equation}
W^0_I(D) :=  \prod_{\gamma \in D} e^{-\beta 
c_0 U^{-1} |\gamma|} \prod_{B \in {D} \atop{|B| \ge 3}}
\Bigl[\exp\bigl({2 \beta c_2 
({\frac{c_1}{U}})^{g(B)}}\bigr) - 1\Bigr].
\label{w1}
\ee
Similarly the partition function $\Xi^{111}_\vol$, defined through
\reff{part111f} -- \reff{te2}, can be written as
\bea
\Xi^{111}_\vol &=& \sum_{\tI\in \tcalI} e^{-\beta E^{\bare}_2(\tI)}\,
\Bigl(\sum_{\D_{\tI} \cap \Lambda \ne \emptyset} \prod_{D \in
\D_{\tI}} W_{\tI}(D)
\Bigr) \nonumber\\
&\times &  \Bigl(\sum_{\D^a_{\tI} \cap \Lambda \ne \emptyset} 
\prod_{D \in \D^a_{\tI}} W(D)
\Bigr) \Bigl(\sum_{\D^b_{\tI} \cap \Lambda \ne \emptyset} 
\prod_{D \in \D^b_{\tI}} W(D),
\Bigr)
\eea
where for $D \in \D_{\tI}$ 
\be
W_{\tI}(D):= \prod_{\gamma \in D} e^{-\beta J_1(U) \vert\gamma_i\vert}  
\prod_{B \in D \atop{|B| \ge 3}}
 (e^{-\beta \tPhi_B'} - 1).
\label{wtId}
\ee
The bounds \reff{peierls} and \reff{phibp} 
imply that $|W_{\tI}(D)| \le 
W^0_{\tI}$ where
\begin{equation}
W^0_{\tI}(D) :=  \prod_{\gamma \in D} e^{-\beta 
c_0 U^{-1} |\gamma|} \prod_{B \in {D} \atop{|B| > 3}}
\Bigl[\exp\bigl({2 \beta c_2 
({\frac{c_1}{U}})^{m(B)}}\bigr) - 1\Bigr].
\label{w2}
\ee
where $m(B)$ is defined through \reff{mb}.

In order to proceed we need to analyze the convergence properties
of series of the form
\begin{equation}
S_\Lambda := \sum_{{\cal{D}}\cap \Lambda \ne \emptyset} 
\prod_{D \in {\cal{D}}} W(D) 
\label{series}
\end{equation}
where ${\cal{D}}$ is a finite set of compatible decorated contours
with weights given by
\begin{equation}
W(D):= \prod_{\gamma \in D} e^{-\beta E(\gamma)}  \prod_{B \in D}
 (e^{-\beta G_B} - 1),
\end{equation}
where the function $G_B$ is given by $\Phi_B$ \reff{rel2} for the homogeneous
boundary conditions \reff{hombc} and by the functions  $\Phi_B'$ \reff{tphi}
and $\tPhi_B'$ \reff{tphibp}  for the mixed boundary conditions \reff{bc1} and
\reff{bc2} respectively. The bounds satisfied by these functions are
respectively given by \reff{expo}, \reff{primebd} and \reff{phibp}. Since the
series in \reff{series} is expressed as a sum over compatible  families of
non-interacting decorated contours, its convergence properties can be studied
by the method of cluster expansions. The convergence of the above series (for
$G_B = \Phi_B$ and $G_B = \Phi_B'$) follows from Lemma  \ref{convergence} given
below. 

\begin{Lem}
\label{convergence}
Consider the series 
\begin{equation}
S_\Lambda := \sum_{{\cal{D}}\cap \Lambda \ne \emptyset} \prod_{D \in
{\cal{D}}} W(D), 
\label{series2}
\end{equation}
where 
\begin{equation}
W(D):= \prod_{\gamma \in D} e^{-\beta E(\gamma)}  \prod_{B \in D}
 (e^{-\beta G_B} - 1),
\label{deff}
\end{equation}
and assume that there exists positive constants $C_1$
and $C_2$ such that
\begin{equation}
E(\gamma) \ge C_1 \lambda |\gamma|
\end{equation}
and
\begin{equation}
|G_B| \le C_2 \lambda^{g(B)},
\label{defbd}
\end{equation}
where $g(B)$ is defined through \reff{geebee} and $\lambda < 1$. Then there
exists constants $b_0, \lambda_0 > 0$ such that for $\beta \lambda > b_0$ and
$\lambda < \lambda_0$, the series $S_\Lambda$ has a  convergent cluster
expansion, i.e.,
\begin{equation}
{\rm{log}} \, S_\Lambda
\;=\; \sum_{N\ge 1} {1\over N!}
\sum_{D_1 \cap \Lambda \ne \emptyset}\cdots 
\sum_{D_N \cap \Lambda \ne \emptyset} \Psi^\trunc(\{D_1,\ldots,D_N\})
\label{s.15}
\end{equation}
(the {\em cluster expansion}\/), where $\Psi^\trunc$ is a 
function on families of decorated contours with the property that
\begin{equation}
\Psi^\trunc(\{D_1, \cdots ,D_N\})\;=\; 0 \quad 
\hbox{if \, $\{D_1, \cdots ,D_N\}$ is not a cluster}\;,
\label{s.17}
\end{equation}
i.e., if $D_1\cup \cdots \cup D_N$ is not a connected set.
It satisfies the bound
\begin{equation}
\sum_{\{D_1,\ldots,D_N\}\ni 0}  {|\Psi^\trunc(D_1,\ldots,D_N)| \over
|D_1\cup\cdots\cup D_N|}\;\le\; 
s_N
\label{s.17.1}
\end{equation}
where $s_N$ is a constant of the order of 
$$\sup \prod_{i=1}^N |W(D_i)|,$$
the supremum being taken over all sets of $N$ decorated contours (not
necessarily pairwise disjoint).
\end{Lem}
The above lemma is proved in Appendix B. The proof of the convergence
of the series $S_\Lambda$ \reff{series} for $G_B = \tPhi_B$ 
is analogous to the proof of the above lemma and is hence
not included. The only difference is the replacement 
of the bound \reff{defbd} by the bound 
\be
|G_B| \le C_2 \lambda^{m(B)}.
\ee
where $m(B) := {\rm{max}}(5, g(B))$.

Let $P$ denote a cluster of decorated contours. It is a connected set of
intersecting decorated contours. A single decorated contour can occur 
several times in a cluster. 
Further we define
\be
|P| := \sum_{D \in P} |D| = \sum_{D \in P} \Bigl( \sum_{\gamma \in D} |\gamma|
+ \sum_{B \in D} g(B) \Bigr)
\label{pmod}
\ee
Let $\Psi^\trunc(P)$,  $\Psi_I^\trunc(P)$ and  $\Psi_\tI^\trunc(P)$ denote
truncated functions, defined on the cluster $P$, which satisfy the following
bounds. 
\be
\sum_{P\ni 0}  {|\Psi^\trunc(P)| \over
|P|}\;\le\; \sup \prod |W(D)|,
\label{psi1}
\ee
\be
\sum_{P\ni 0}  {|\Psi_I^\trunc(P)| \over
|P|}\;\le\; \sup \prod |W_I(D)|,
\label{psi2}
\ee
and
\be
\sum_{P\ni 0}  {|\Psi_\tI^\trunc(P)| \over
|P|}\;\le\; \sup \prod |W_{\tI}(D)|.
\label{psi3}
\ee
If the cluster $P$ consists of $N$ decorated contours then the 
product is over a set of $N$ decorated contours and the supremum 
in the above estimates is taken over all such of $N$ decorated contours.

To determine whether the $100$ and $111$ interfaces are rigid, we need to
analyze the following quotients.
\be
\Z_\Lambda^{100}=\frac{\Xi^{100}_{\vol}}{\Xi^+_{\vol}} \quad \mbox{and }
\quad \Z_\Lambda^{111}=\frac{\Xi^{111}_{\vol}}{\Xi^+_{\vol}}.
\ee
Using the results of cluster expansions (see, e.g., \cite{mir}) 
we can express these quantities in terms of the truncated functions 
defined above.

\begin{Prop}\label{prop:surfacetension}
There exist three constants $U_0$, $B_1$, and $B_2$, independent of
the volume $\vol$, such that for all $U>U_0$ and and $\beta/U > B_1$,
the quotient $\Z_\Lambda^{100}$ can be written as follows:
\be
\Z_\Lambda^{100} = \sum_{I\in {\cal I}} e^{-\beta E_1^{\bare}(I)} \times
\exp \left\{\sum_{P:\atop{P \cap I \ne \emptyset \atop{P \cap \Lambda \ne
\emptyset}}} \Psi_I^\trunc(P) -\sum_{P:\atop
{P \cap I \ne \emptyset \atop{P \cap \Lambda \ne
\emptyset}}} \Psi^\trunc(P)\right\}.
\label{zed100}
\ee
For all
$U>U_0$ and $\beta/{U^3} > B_2$, the quotient 
$\Z_\Lambda^{111}$ can be written as follows:
\be
\Z_\Lambda^{111} = \sum_{\tI\in {\tcalI}} e^{-\beta E_2^{\bare}(\tI)} \times
\exp \left\{\sum_{P:\atop{P \cap \tI \ne \emptyset \atop{P \cap \Lambda \ne
\emptyset}}} \Psi_\tI^\trunc(P) -\sum_{P:\atop
{P \cap \tI \ne \emptyset \atop{P \cap \Lambda \ne
\emptyset}}} \Psi^\trunc(P)\right\}.
\label{zed111}
\ee
\end{Prop}
\begin{proof}
The proof is standard. We expand the partition functions appearing 
in the numerator and in the denominator of $\Z_\Lambda^{100}$
and of $\Z_\Lambda^{111}$ by the cluster expansion performed in the
the Appendix B. The truncated functions which are left 
are precisely those
defined on the clusters which intersect the interface.
\end{proof}

Alternatively $\Z_\Lambda^{111}$ can be written as
\be
\Z_\Lambda^{111} = \sum_{\tI\in {\tcalI}} e^{-\beta E_2^{\dec}(\tI)}
\label{alt111}
\ee
where $ E_2^{\dec}(\tI)$ denotes the energy of the interface in the
presence of the decorated contours and is
defined as follows.
\be
E_2^{\dec}(\tI) := E_2^{\bare}(\tI) -\frac{1}{\beta} 
\Bigl[\sum_{P:\atop{P \cap \tI \ne \emptyset \atop{P \cap \Lambda \ne
\emptyset}}} \Psi_\tI^\trunc(P) -\sum_{P:\atop
{P \cap \tI \ne \emptyset \atop{P \cap \Lambda \ne
\emptyset}}} \Psi^\trunc(P)\Bigr]
\label{reldec}
\ee
We shall refer to  $E_2^{\dec}(\tI)$ as the energy of the decorated interface.

\masection{\bf Rigidity  of the $100$ interface}
\label{sec:rigidity100}

The study of the $100$ interface  requires  the second order decomposition of
the effective Hamiltonian of the FK model, in which  the second order
truncated  Hamiltonian $\H^{(2)}_0(U)$ is   the  Ising Hamiltonian, with
coupling constant $J_1(U)$. We point out   that  the Hamiltonian
$\H^{(2)}_0(U)$ is generated by the second order quantum fluctuations,   this
means that the rigidity  of the   $100$ interface of the FK model is of
quantum nature. The proof of  the rigidity  of the $100$ interface of the
three--dimensional FK model  is a generalization of Dobrushin's proof
of the rigidity of this interface in the three--dimensional Ising model 
at sufficiently low temperatures.
\cite{Dob}.

\noindent
{\bf Proof of Theorem \ref{theorem1A}:}\newline

Let ${\rm{Prob}}_{\Lambda}(I)$ denote the probability of occurrence of an
interface  $I$ in the FK model, defined on a finite cubic lattice $\Lambda$, 
under the boundary condition b.c.1 \reff{bc1}. It is defined as follows:
\be
{\rm{Prob}}_{\Lambda}(I)={1\over\Z_\Lambda^{100}}\,\,\exp\,
\{-\beta E_1^{\bare}(I)\} \times
\exp \left\{\sum_{P:\atop{P \cap I \ne \emptyset \atop{P \cap \Lambda \ne
\emptyset}}} \Psi_I^\trunc(P) -\sum_{P:\atop
{P \cap I \ne \emptyset \atop{P \cap \Lambda \ne
\emptyset}}} \Psi^\trunc(P)\right\}
\label{prob100}
\ee

To determine whether the $100$ interface is rigid, we need to analyze the
properties of this quantity in the thermodynamic limit.

We notice that the above expression \reff{prob100} is similar to the
corresponding probability of a $100$ interface  in the Ising model. This is 
because the leading term in the ``bare''  energy,  $E_1^{\bare}(I)$
\reff{bare}, of an interface $I$ is  exactly equal to the energy of an 
interface $I$ for an   Ising model with coupling constant $J_1(U)$.  We point
out that  this  similarity results from the fact that we have expressed the
probability ${\rm{Prob}}_{\Lambda}(I)$ directly in terms of an actual
interface $I$, separating the two coexisting phases, instead of expressing it
in  terms of a {\em decorated interface}. (The latter is given by a {\em
connected}  set $(I, \B_I, \Gamma_I)$, where $\B_I$ is a finite set of bonds
and $\Gamma_I$ is a finite set of Ising contours; [see \cite{mir} and
references therein].) This leads to a considerable simplification in the
calculations since an actual interface between two coexisting phases, under
the boundary condition b.c. 1, reduces to a flat interface orthogonal  to
${\bf{{n}}}= (0,0,1)$ at zero temperature. This property is,   however, not
satisfied by a decorated interface.

Let us describe the minor differences which arise in the description  of the
$100$ interface of the FK model with respect to that of the  Ising model. 
\begin{itemize} \item The first  difference lies in the contributions of the
truncated functions. For the FK model these functions are defined on sets of
decorated contours, whereas for the Ising model they are defined on sets of
Ising contours. However, there exists a positive constant $U_0$ for such that
for all $U>U_0$, the contribution of the truncated functions is exponentially
small for the FK model, as is the case for the Ising model. 
difference arises from the fact that the ``bare''  energy, $E_1^{\bare}(I)$
[\reff{bare}], of the interface consists of terms in  addition to the leading 
Ising--like term $J_1(U)|I|$. These terms are, however, small for  for $U >
U_0$, where $U_0$ is a positive constant. Hence they are treated  in the same
way as the truncated functions. \end{itemize}

The proof of  the rigidity  of the $100$ interface of the FK model is similar
to Dobrushin's proof of the rigidity  of the $100$ interface of the
three--dimensional Ising model \cite{Dob}.  The  proof of the rigidity can be
converted into the study of a  two--dimensional contour model, which resembles
an Ising model with long range interactions, in which the ground states are
the projections of the {\em{ceilings}} on the $100$ plane   and the contours
are the corresponding  projections of the {\em{walls}}. The rigidity of the
$100$ interface at low temperatures follows  from a Peierls argument on the
contours (walls) \cite{Dob}.  Taking into account the small modifications
described above, we deduce that there exists positive constants $U_0$ and
$D_0$ such that, for  all $U>U_0$ and  $\beta/U > D_0$, the assertions of
Theorem $1$ (see Section \ref{sec:FK_model}) are true.
\QED
We would like to remark that Theorem \ref{theorem1A}, stated for 
the FK model, is in general valid for a wide class of lattice Hamiltonians,
namely, those which can be expressed as a sum of two terms: 
a dominant nearest neighbour Ising Hamiltonian, and a 
remainder consisting of 
long--range many--body interactions satisfying exponential decay
\reff{exporem}.

\masection{\bf Rigidity  of the $111$ interface}
\label{sec:rigidity}

The case of the $111$ interface can be treated by using the ideas of Dobrushin 
but the situation is more involved. The main difference between the $111$
interface and the $100$ interface is that the second order decomposition of the
effective Hamiltonian \reff{rel10} does not lead to the existence of a unique
ground state interface in the $111$ direction. The energy of a $111$ interface
w.r.t the leading part,  $H^{(2)}_0(U)$, of the relative Hamiltonian, is
proportional to the area of the $111$ interface. Hence, the ground state
interface has minimal area. However, there are infinitely many such
interfaces in the infinite volume limit (Proposition
\ref{prop:degeneracy} of Section \ref{minimal}).  Hence, the {\em{ground
state}} of the $111$ interface has an {\em{infinite degeneracy}}. Thus, to
prove the rigidity  of the $111$ interface we require a more  refined
decomposition of the effective Hamiltonian, namely, the fourth--order
decomposition. In this section we prove that the degeneracy of the $111$ ground
state interface is lifted by  the fourth--order truncated effective 
Hamiltonian, $H^{(4)}_0(U)$, which takes into account the fourth--order quantum
fluctuations. The study of the properties of the $111$ interface can be reduced
to the analysis of a model defined on a two--dimensional triangular lattice
which is obtained by projecting the interface onto a fixed  plane (see Sections
\ref{geom111} and \ref{rhombus} for details). We refer to this model as the
{\bf{rhombus model}}. The rigidity of the $111$ interface at sufficiently low
temperatures can be deduced from the low temperature behaviour of the rhombus
model.

We first define the main quantity required in our proof of the rigidity of  the
$111$ interface:  the probability of occurrence of an interface $\tI$, in the
volume $\Lambda$, under the boundary condition b.c.2 \reff{bc2}:
\bea
{\rm{Prob}}_{\Lambda}(\tI) =\frac{1}{\Z_\Lambda^{111}} 
\exp\{-\beta E_2^{\bare}(\tI)\} \times
\exp \left\{\sum_{P:\atop{P \cap\tI \ne \emptyset \atop{P \cap \Lambda \ne
\emptyset}}} \Psi_\tI^\trunc(P) -\sum_{P:\atop
{P \cap \tI \ne \emptyset \atop{P \cap \Lambda \ne
\emptyset}}} \Psi^\trunc(P)\right\}.
\label{prob111}
\eea

\masubsection{Description of the Rhombus configurations}
\label{rhombus}

As we have seen, the projection of each face of the
interface $\tI$, onto the plane $\P$
yields a rhombus of one of the three types, i.e., belonging to
$\R_0$, $\R_1$,or $\R_2$.  An interface $\tI$ is projected onto a covering of
$\P$ with rhombi, which we refer to as an $R$-configuration (see Section
\ref{geom111}). In general, this rhombus covering is
not a tiling. Moreover, many interfaces are represented by the same $R$
configuration. Let $\C$ denote an $R$ configuration on $\P_\Lambda$.

The correspondence between the relative positions of the rhombi in an 
$R$--configuration and the local geometry of the corresponding interface
(as determined by the spin configurations on the plaquettes of $P_{\tI}$,
and the bonds of $B^2_{\tI}$) is given as follows:
\bea
\parbox{6cm}{\baselineskip=7pt{\bf 
local spin configuration on ${\zed}^3$}}
&\leftrightarrow& 
\parbox{7cm}{\baselineskip=7pt{\bf rhombus configuration on $\P$}}\nonumber\\ 
\arr{cc}{+&+\\+&-}&\leftrightarrow& 
\parbox{7cm}{\baselineskip=7pt two adjacent rhombi
of the same type i.e., a good pair of rhombi} \label{low}\\ 
\arr{cc}{+&-\\ +&-}&\leftrightarrow&
\parbox{7cm}{\baselineskip=7pt two adjacent rhombi of different types;
the common edge is then called a $\delta$--edge. See Figure 
\ref{fig:minimal}}\\ 
\arr{cc}{+&-\\-&+}&\leftrightarrow& 
\parbox{7cm}{\baselineskip=7pt 
four overlapping rhombi sharing an edge: the common edge is then 
called an $\omega$--edge.}\\
\arr{ccc}{+&-&+}&\leftrightarrow& 
\parbox{7cm}{\baselineskip=7pt two parallel rhombi sharing a vertex,
with one of them being necessarily overlapping: the two rhombi are
said to be connected by a $\lambda$--link (see below),}
\eea
and the same correspondences for spin configurations obtained by a spin flip
transformation or by rotations of the ones above. 
\smallskip

\begin{figure}[t]
\begin{center}
\rotatebox{-90}{\resizebox{!}{10truecm}{\includegraphics{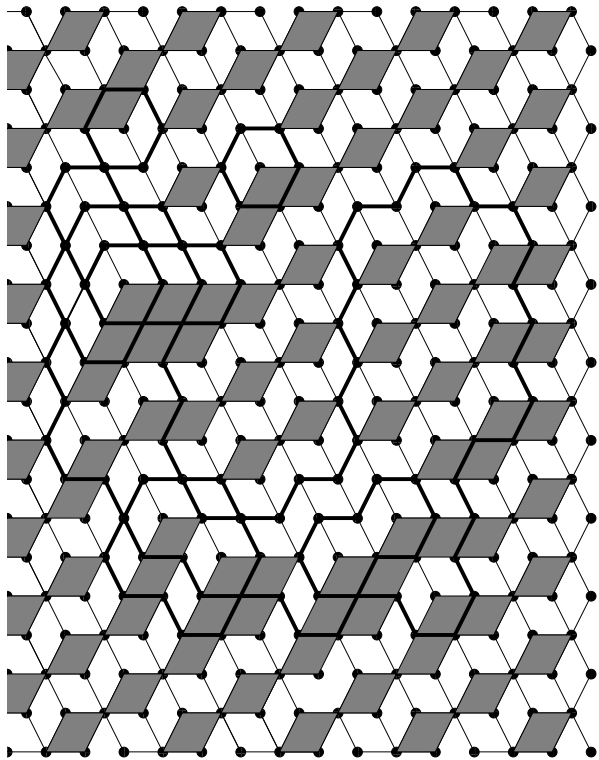}}}
\parbox{14truecm}{\caption{\baselineskip=16 pt\small\label{fig:minimal}
A minimal area interface with its $\delta$--edges (given by the dark lines).
The only purpose of the shading is to guide the eye.
}}
\end{center}
\end{figure}

\noindent
{\em{Definition of a $\lambda$--link}}: Consider a pair of parallel faces of an
interface, which are a unit distance apart. Consider a line segment which
connects the centers of these faces and is  perpendicular to them. The
projection of such a line segment on the plane $\P$ is referred to as a
$\lambda$--link. A single  $\lambda$--link can be the projection of several
such line segments. Hence, each  $\lambda$--link has a multiplicity, $m$, which
counts the number of line segments which  project onto it.

\masubsection{The energy of a rhombus configuration}
\label{select}

To order $U^{-3}$, the bare energy $ E^{\rm bare}_2(\tI)$ of an  interface
$\tI$, defined by \reff{bare2}, is a function of the local geometry of the
interface, and hence is a function of the $R$--configuration. The plaquette
potential $h_p$, (given by \reff{hp}),  as well as the next nearest neighbour
interaction $h_{\{x,z\}}$ with $|x-z|=2$,  (given by \reff{hxz}),  play 
crucial roles in determining this energy. Their contributions to the energy
for different spin configurations are given as follows:
\bea
h_p\left(\arr{cc}{+&+\\
+&-}\right)& =&-\, 16 \label{hp1}\\
h_p\left(\arr{cc}{+&-\\ +&-}\right)& =&- \, 12\\ h_p
\left(\arr{cc}{+&-\\
-&+}\right)&=&0
\\
 h_{\{x, z\}}\left(\arr{ccc}{+&-&+}\right)&=&0\\
h_{\{x, z\}}\left(\arr{ccc}{+&-&-}\right)=
h_{\{x, z\}}\left(\arr{ccc}{-&-&+}\right)&=& - \, 2\label{hp5}
\eea

The plaquette and next nearest neighbour configurations which correspond to 
the lowest energy for these potentials are those given in \reff{hp1} and 
\reff{hp5} respectively. Moreover, from \reff{low} it follows that  the
plaquette configuration with lowest energy corresponds to a good pair of
rhombi. Hence, a connected set of faces of the interface whose projection on
the plane $\P_\Lambda$ consists entirely of good pairs of rhombi defines a
{\em{local ground state configuration}} of the interface with respect to the
truncated Hamiltonian $H_0^{(4)}$ [\reff{rel30}]. Since two rhombi that form a
good pair are necessarily of the same type, it follows that there are three
such local ground state configurations corresponding to rhombi in the families
$\R_0$, $\R_1$ and $\R_2$ respectively.

We conclude from the above  that those minimal area interfaces whose
projections on the plane $\P$ are tilings with rhombi belonging to a single
family have minimum energy with respect to $H^{(4)}_0$ \reff{rel30} 
and are hence referred to as {\em{ground state interfaces}}.
Since we have chosen the standard b.c. for the $R$--configuration to be given
by a tiling of the plane $\P \setminus \P_\Lambda$ with rhombi in the family
$\R_0$ [see  Section \ref{geom111}], it follows that a ground state interface 
projects onto a tiling of $\P_\Lambda$ with rhombi in the family $\R_0$. We
hence denote a ground state interface as $\tI_{\R_0}$.  It is easy to see that
$\tI_{\R_0}$ has a {\em{perfect staircase structure}} (see Figure
\ref{fig:staircase}). Thus, under the boundary condition b.c.2  \reff{bc2}, the
geometry--dependent contribution of the plaquette potential $h_p$ leads to the
selection of a unique ground state interface (up to translations) from the
infinitely many minimal area interfaces.

\begin{figure}[t]
\begin{center}
\rotatebox{-90}{\resizebox{!}{7truecm}{\includegraphics{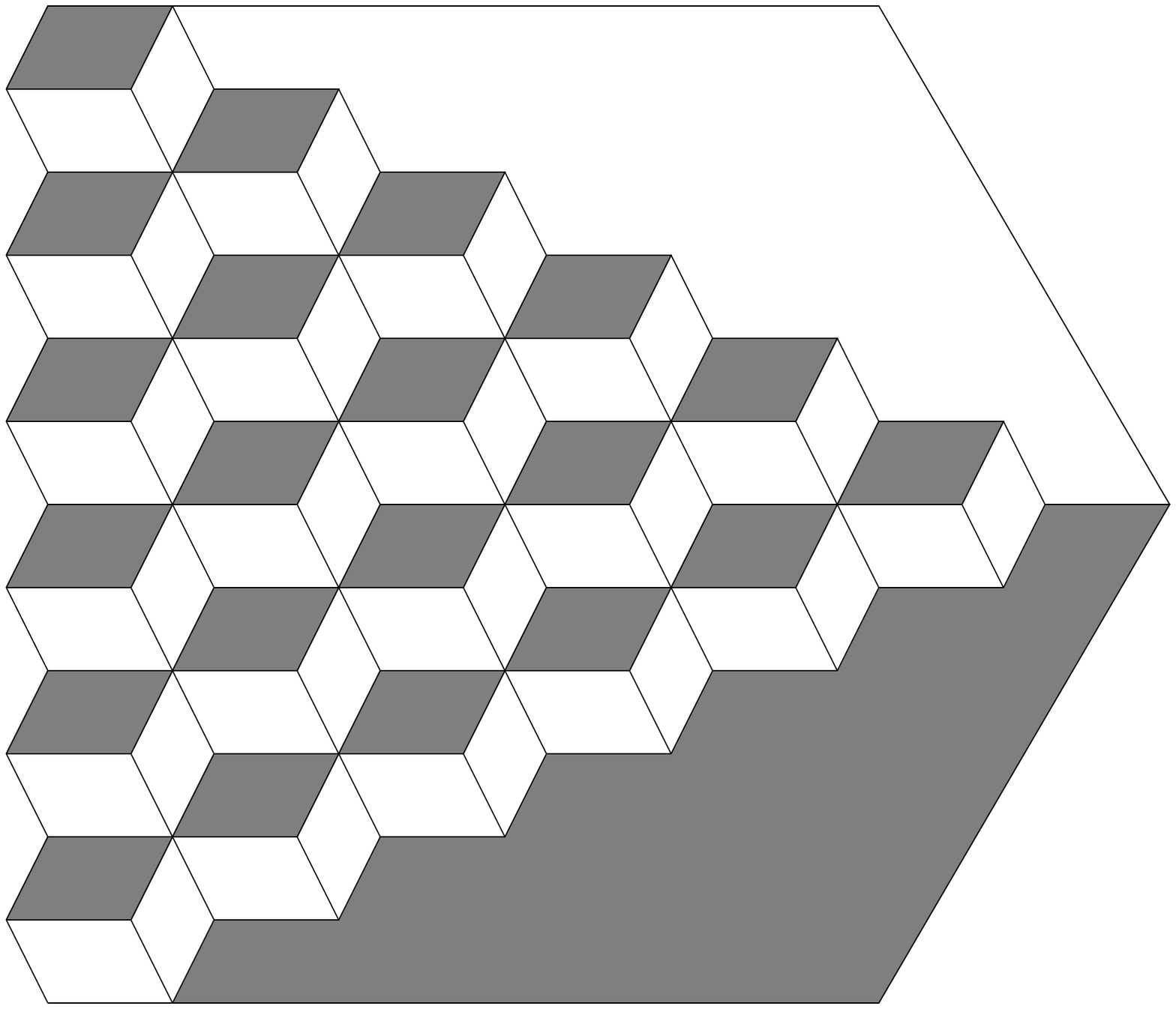}}}
\parbox{14truecm}{\caption{\baselineskip=16 pt\small\label{fig:staircase}
The ``perfect staircase'' structure of the interface $\tI_{\R_0}$.
}}
\end{center}
\end{figure}

For any interface $\tI$ the connected components of the 
set $\tI \setminus \tI_{\R_0}$ are referred to as {\em{pyramids}}. They
represent the local distortions of the interface from a perfect staircase 
structure. 

We would like to point out that  the above mentioned  phenomenon of ground
state selection is a  consequence of the Fermi statistics of the electrons in
the model. The anticommutation relations for the electron creation and
annihilation operators play a crucial  role in determining the exact
expressions for the potentials $h_p$ [\reff{hp}] and $h_{\{x,z\}}$ [\reff{hxz}]
which are responsible for lifting the infinite degeneracy.  If instead of
fermions we consider bosons, then the corresponding commutation rules for the
creation  and annihilation operators yield the following expression for the
plaquette potential:
\be
{\tilde{h_p}} =  1 - s_xs_ys_zs_w + 5 (s_xs_z + s_ys_w - 2),
\label{tp}
\ee
where $w,x,y$ and $z$ are four sites forming a plaquette. It is easy to see
that 
\bea
{\tilde{h_p}}\left(\arr{cc}{+&+\\
+&-}\right)& > & {\tilde{h_p}}\left(\arr{cc}{+&-\\
+&-}\right),
\eea
and hence, in the bosonic case, the perfect staircase configuration is not
favoured by the plaquette potential in fourth order of perturbation theory.
\bigskip

\masubsection 
{\bf{The energy of a ground state interface}}

The energy of the ground state interface $\tI_{\R_0}$ is
given by 
\be
E_2^{\dec}(\tI_{\R_0}):= E_2^{\bare}(\tI_{\R_0}) -\frac{1}{\beta} 
\Bigl[\sum_{P:\atop{P \cap \tI_{\R_0} \ne \emptyset \atop{P \cap \Lambda \ne
\emptyset}}} \Psi_{{\tI}_{\R_0}}^\trunc(P) -\sum_{P:\atop
{P \cap \tI_{\R_0}\ne \emptyset \atop{P \cap \Lambda \ne
\emptyset}}} \Psi^\trunc(P)\Bigr]
\ee
where,
\be
E_2^{\bare}(\tI_{\R_0}) 
= (J_2(U)- \frac{1}{4U^3})|\tI_{\R_0}| -(\frac{1}{ U^3}+\ho)
N_1(\tI_{\R_0}) + \sum_{\{ B\,:\, \vert B\vert > 3;
B \cap \tI_{\R_0}\neq \emptyset \}} \tPhi_B(\tI_{\R_0} \cup \emptyset).
\label{barei0}
\ee
Here $|\tI_{\R_0}|$ is the total number of faces in the interface $\tI_{\R_0}$,
which is equal to the number of rhombi in the corresponding tiling of
$\P_{\Lambda}$,
and $ N_1(\tI_{\R_0})$ is the total number of shared edges of rhombi in
the tiling. The above expression \reff{barei0} follows from \reff{bare2}.
It is obvious that $E_2^{\bare}(\tI_{\R_i})$, 
is the same for $i=0, 1$ and $2$.
Hence, we can define the relative energy of an interface $\tI$ as follows.
\be 
\epsilon_2^{\dec}(\tI) := E_2^{\dec}(\tI) - E_2^{\dec}(\tI_{\R_0}).
\label{vedec2}
\ee

\masubsection{The components of an $R$--configuration}

An $R$--configuration can be decomposed into {\it{bases}} -- which are local
ground state configurations of $H^{(4)}_0$ [\reff{rel30}], 
and $R$--{\em{contours}} -- which represent the
excitations. 
\smallskip

\noindent  
{\em{Bases}}: A { base} of a given $R$ configuration is a maximally connected
set of good pairs of rhombi. As each base is the union of open sets, it is an
open set, and by definition distinct bases do not intersect.  Denote by
$\{{\rm C}_1,..,{\rm C}_p\}$ the family of bases of a given $R$ configuration.
The type of a base ${\rm C}_i$ is defined as follows: $\tau({\rm C}_i)= j$, if
the rhombi it consists of belong  to $\R_j$.  Due to the mixed boundary
conditions one of the bases is connected to the boundary and hence becomes
infinite in the thermodynamic limit. Let $C_1$ denote this base. Since the
standard b.c. is chosen to be a tiling of  $\P \setminus \P_{\Lambda}$ with
rhombi in $\R_0$, we have $\tau(C_1) = 0$.
\smallskip

\noindent
{\em{$R$--contours}}: The maximally connected components of the complements of
the bases, i.e., of $\P_\Lambda\setminus\cup_{i=1}^{i=p}{\rm C}_i$, are called the
$R$--contours. Isolated vertices are not considered to be $R$--contours. Each
$R$--contour is a closed complex and is denoted by the symbol $\Upsilon$.

An $R$--contour $\Upsilon$ is defined by a pair
\be
\Upsilon = ({\rm{supp}}\Upsilon, \Theta_\Upsilon)
\ee
where 
\begin{itemize}
\item{${\rm{supp}}\Upsilon$ denotes the geometric support of the
$R$-contour and is a connected subset of the plane $\P$,}
\item{$\Theta_\Upsilon$ denotes the configuration on this support.
It is defined in terms of overlapping rhombi, $\delta$-- and $\omega$--
lines and $\lambda$--links that span ${\rm{supp}}\Upsilon$.}
\end{itemize}
For notational simplicity we shall often use the symbol $\Upsilon$
to denote both the $R$-contour and its geometric support 
${\rm{supp}}\Upsilon$. 

An $R$--configuration, under standard
boundary conditions, is given by a compatible family of $R$--contours,
$\C := \{\Upsilon_{1},...,\Upsilon_{n}\}$, i.e., a finite set of
non-intersecting, closed $R$--contours, along with a
specification of the types of the bases separating the $R$--contours. 
To avoid complicated notations, the types of the bases 
will be specified only when required. Those $R$--contours
whose supports are not in the interiors of any other $R$--contours in
$\C$ are referred to as the external contours of $\C$.

\masubsection{The structure of the $R$--contours}
\label{rcontours}

Each $R$--contour $\Upsilon$ has a detailed structure. It 
can be decomposed into two families of subcontours, one of which
can be empty.
\begin{itemize}
\item The {\it overlapping $R$--subcontours}
$\{\Upsilon^{ov}_1,...,\Upsilon^{ov}_p\}$ are
the  maximally connected sets  of overlapping rhombi contained in
$\Upsilon$. Each overlapping $R$--subcontour is a closed complex.
\item The complement  of the overlapping $R$--subcontours in
$\Upsilon$, i.e., $\Upsilon\setminus\cup_{i=1}^{i=p} 
\Upsilon^{ov}_i$ is considered as an open complex,
whose elements  $\{\Upsilon^{st}_1,...,\Upsilon^{st}_q\}$
are maximally connected open complexes. They  are called the {\em
{standard  $R$--subcontours}}. We refer to an $R$--contour as a standard 
$R$--contour if it has no overlapping components.
\end{itemize}

Hence, an $R$ contour $\Upsilon$ has the following decomposition:
\be
\Upsilon := \Upsilon(p,q) =  
\left\{\cup_{i=1}^{i= p} \Upsilon^{ov}_{i}\right\}
\bigcup \left\{\cup_{k =1}^{k=q} \Upsilon^{st}_k\right\}
\label{upsilon}
\ee

Let us describe the detailed  structure of the different $R$--subcontours.
A  standard  $R$--subcontour $\Upsilon^{st}_k$ is uniquely 
specified by a configuration $\Delta^{st}_k$  of  $\delta$--lines.
We refer to such $\delta$--lines as {\em{standard $\delta$--lines}}.
An  overlapping $R$--subcontour  $ \Upsilon^{ov}_i$ is
characterized  by  four families of configurations.

\begin{itemize}
\item  A  {\it configuration  of overlapping triangles} $\T^{ov}_i$.
It is easy to see that each overlapping triangle in the projection of an
interface has an even overlap number and each rhombus 
in $ \Upsilon^{ov}_i$ contains at least one overlapping triangle.
Let $|\T^{ov}_i|$ denote the sum of the overlap numbers of all the 
overlapping triangles contained in   $\Upsilon^{ov}_i$, i.e., 
\be
|\T^{ov}_i| := \sum_{ t \in \T^{ov}_i} o(t).
\ee
Let $a_i^{ov}$ denote the number of extra faces of an interface (in
comparison with the number of faces of a minimal area interface) which
project on to the support ${\rm{supp}} \Upsilon^{ov}_i$. It is given by
\be
a_i^{ov} = \frac{|\T^{ov}_i|}{2}.
\label{ai}
\ee
Since $|\T^{ov}_i|$ is even, $a_i^{ov}$ is an integer.
\item  A {\it  configuration  $\Delta^{ov}_i$  of  $\delta$ lines}.
The total length of the  $\delta$  lines in $\Upsilon^{ov}_i$ 
is denoted by $| \Delta^{ov}_i|$.
\item  A {\it  configuration $\Omega^{ov}_i$  of $\omega$ lines}.
The total length of the  $\omega$  lines in $\Upsilon^{ov}_i$
is denoted by $|\Omega^{ov}_i|$.
\item A {\it  configuration $\Lambda^{ov}_i$  of  $\lambda$ links}.
The total length of the  $\lambda$  links of   $\Upsilon^{ov}_i$
is denoted by $|\Lambda^{ov}_i|$.
\end{itemize}
The number of rhombi spanning ${\rm{supp}}\,\Upsilon^{ov}_i$ is 
bounded above by $a_i^{ov}$, since each rhombus in 
${\rm{supp}}\,\Upsilon^{ov}_i$ is an overlapping one. Each of these
rhombi contribute at most three distinct sites to 
${\rm{supp}}\,\Upsilon^{ov}_i$ since the latter is a connected set.
The number of sites in the $R$ contour $\Upsilon(p,q)$ 
(defined through \reff{upsilon}) hence satisfies the bound
\be
|\Upsilon| \le \sum_{i=1}^{p}\{3|a^{ov}_i|  
              + (|\Delta^{ov}_i| + 1)
              + (|\Lambda^{ov}_i| + 1)
              + (|\Omega^{ov}_i| + 1)\}
              + \sum_{k=1}^{q}(| \Delta^{st}_k| + 1).
\label{numsites}
\ee
The relative energy of the interface $\tI$
can be expressed in terms of the energy of the corresponding
$R$ configuration $\C = \{\Upsilon_{1},...,\Upsilon_{n}\}$.
It is given by
\be 
\epsilon_2^{\dec}(\tI)
\equiv \epsilon_2^{\dec}(\{\Upsilon_{1},...,\Upsilon_{n}\})
:= \sum_{i=1}^{i=n} F(\Upsilon_i)
+W(\{\Upsilon_{1},...,\Upsilon_{n}\}),
\ee
where  $F(\Upsilon_i)$ is the energy of the $R$--contour
$ \Upsilon_i$  computed with the fourth  order truncated relative Hamiltonian
${H}^{(4)}_{0 \Lambda}$ defined in \reff{rel30}, and
$ W(\{\Upsilon_{1},...,\Upsilon_{n}\})$ is the contribution to the
energy of the $R$ configuration arising from the long--range tail
$R^{\ge 5}_\Lambda$ defined in \reff{rel5}. The latter consists of higher
order corrections to the energy of each individual contour, as well
as interaction energies between the different $R$ contours in the
compatible family. From \reff{vedec2}, \reff{bare2} and 
\reff{reldec} it follows that
\bea
\sum_{i=1}^{n} F(\Upsilon_i) &=& J_2(U) \bigl(\vert \tI \vert -
\vert {\tI}_{\R_0} \vert \bigr) + \bigl( \frac{1}{8U^3} + \ho \bigr)
\!\Bigl[\sum_{\{x,z\}\in B^2_{\tI}} h_{\{x,z\}}(\tI) - 
\sum_{\{x,z\}\in B^2_{{\tI}_{\R_0}}} h_{\{x,z\}}({\tI}_{\R_0})\Bigr]
\nonumber\\
&+&  
\bigl(\frac{1}{16 U^3} + \ho \bigr) \quad \Bigl[
\sum_{p=\{x,y,z,t\}\in P_{\tI}} h_p(\tI) - \sum_{p=\{x,y,z,t\}\in
P_{{\tI}_{\R_0}}} h_p({\tI}_{\R_0})\Bigr],
\label{fup}
\eea
and
\bea
W(\{\Upsilon_{1},...,\Upsilon_{n}\}) = 
&=&\Bigr\{ \sum_{B : \vert B\vert > 3
\atop {B \cap \tI\neq \emptyset}} {\tPhi_B}(\tI\cup \emptyset) 
- \sum_{B : \vert B\vert > 3;
\atop{B \cap \tI_{\R_0}\neq \emptyset}} \tPhi_B(\tI_{\R_0} \cup \emptyset)
\Bigl\} 
\nonumber\\
&-& \frac{1}{\beta} \Bigl\{ 
\sum_{P:\atop{P \cap \tI \ne \emptyset \atop{P \cap \Lambda \ne
\emptyset}}} \Psi_\tI^\trunc(P) -
\sum_{P:\atop
{P \cap \tI_{\R_0} \ne \emptyset \atop{P \cap \Lambda \ne
\emptyset}}} \Psi_{\tI_{\R_0}}^\trunc(P) \Bigr\} \nonumber\\
&+& 
 \frac{1}{\beta} \Bigl\{\sum_{P:\atop{P \cap \tI 
 \ne \emptyset \atop{P \cap \Lambda \ne
\emptyset}}} \Psi^\trunc(P) -
\sum_{P:\atop
{P \cap \tI_{\R_0} \ne \emptyset \atop{P \cap \Lambda \ne
\emptyset}}} \Psi^\trunc(P)\Bigr\}, \nonumber\\
\label{wup}
\eea
\begin{Lem}
Let $\Upsilon$ be an $R$--contour defined through \reff{upsilon}
and let $\C$ denote an $R$--con\-figu\-ra\-tion.
The energies $F(\Upsilon)$ and $W(\C)$, defined through \reff{fup}
and \reff{wup} respectively, are given by the following expressions:
\bea
F(\Upsilon)&=& J_2(U) \sum_{i=1}^{i=p}a^{ov}_i
+ K_2(U)\Bigl(\sum_{k=1}^{k=q} |\Delta^{st}_k|
 + \sum_{i=1}^{i=p} | \Delta^{ov}_i| \Bigr) \nonumber\\
&+& \Bigl( \frac{1}{U^3}+\ho\Bigr)\sum_{i=1}^{i=p}|\Omega^{ov}_i|
+\Bigl(\frac{1}{4U^3}+\ho\Bigr)\sum_{i=1}^{i=p}| \Lambda^{ov}_i|,
\label{fup2}
\eea
where
\bea
J_2(U) &=& \frac{1}{2U}-\frac{11}{8U^3}+\ho,\label{j2} \\
K_2(U) &=& \frac{1}{4U^3} + \ho \label{k2}
\eea
and
\bea 
W(\C) &=&
\sum_{B: |B| > 3 \atop{ B_{\P} \cap \C
\ne \emptyset}}
\widehat{\Phi_B}(\C) \nonumber\\
&-& \frac{1}{\beta} \Bigl\{ 
\sum_{P:\atop{P_{\P} \cap \C \ne \emptyset
}} \Psi_\tI^\trunc(P) -\sum_{P:\atop
{P_{\P} \cap \C \ne 
\emptyset }} \Psi^\trunc(P)\Bigr\},
\label{wup2}
\eea
where
\be
\widehat{\Phi_B}(\C):= {\tPhi_B}(\tI \cup
\emptyset) - \tPhi_B(\tI_{\R_0} \cup \emptyset)
\label{hatphi}
\ee
and $B_{\P}$ is the projection of the bond $B$ on the plane $\P$. The notation
$B_{\P} \cap \C\ne \emptyset$ is used to denote the condition  that $B_{\P}$
intersects either of the following in the $R$ configuration: an overlapping
triangle, a  $\delta$-line, an  $\omega$-line, or a $\lambda$-link. Similarly,
$P_{\P}$ denotes the projection of the cluster $P$ of decorated contours, on
the plane $\P$. 
\end{Lem}
\begin{proof}
We use \reff{hp1}--\reff{hp5} and the definition of the different components
of an $R$--contour to obtain the expression for the energy $F(\Upsilon)$ from
\reff{fup}.
It is a sum of the
the energies of the overlapping-- and standard  $R$--subcontours in
$\Upsilon$. The energy of an overlapping $R$--subcontour is given in terms 
of the number of overlapping
triangles, $ \delta$-lines, $\omega$-lines, and $\lambda$-links 
which constitute it. The energy of a standard 
$R$--subcontour is  proportional to 
the number of $\delta$-lines in it.

The expression \reff{wup2} follows from the definition
\reff{wup}, since the only terms which survive in each of the three 
paranthesis 
on the RHS of \reff{wup}
are those in which the projections of
the bonds $B$ or the clusters $P$ on the plane $\P$, 
intersect at least one $R$--contour in the
compatible family $\C$. 
This
concludes the proof. 

In the definition \reff{hatphi} we have made use of the fact that while the
interface $\tI$ corresponds to the $R$--configuration $\C$, containing the
$R$--contour $\Upsilon$, the projection of the ground state interface
$\tI_{\R_0}$ contains no $R$--contours.
\end{proof}

More generally, the energy of an $R$--configuration $\C$ can be expressed as
follows:
\be 
\epsilon_2^{\dec}(\C) = \sum_{\Upsilon \in \C} \epsilon(\Upsilon \mid \C),
\label{engup}
\ee
where $\epsilon(\Upsilon \mid \C)$ denotes the total energy of an
$R$--contour $\Upsilon$ belonging to the $R$--configuration $\C$. It
is defined as follows:
\be
\epsilon(\Upsilon \mid \C) := F(\Upsilon)
+W(\Upsilon \mid \C),
\label{egc}
\ee
where  $F(\Upsilon)$ is given by \reff{fup2} and $W(\Upsilon\mid \C)$
is given by
\bea 
W(\Upsilon\mid\C) &=&
\sum_{B: |B| > 3    \atop{ B_{\P} \cap \Upsilon
\ne \emptyset}}
\widehat{\Phi_B}(\C) \nonumber\\
&-& \frac{1}{\beta} \Bigl\{ 
\sum_{P:\atop{P_{\P} \cap \Upsilon \ne \emptyset
}} \Psi_\tI^\trunc(P) -\sum_{P:\atop
{P_{\P} \cap \Upsilon \ne 
\emptyset }} \Psi^\trunc(P)\Bigr\}.
\label{wup3}
\eea

\masubsection {\bf The relevant probabilities}

The probability of occurrence of an interface $\tI$ in the volume $\Lambda$ 
can be identified with the probability  of the corresponding compatible family
of $R$--contours $\C \equiv \{\Upsilon_{1},...,\Upsilon_{n}\}$ in the rhombus
model.
\be
\prob_{\Lambda}(\tI) = \prob_{\Lambda}(\{\Upsilon_{1},...,\Upsilon_{n}\})
=\frac{e^{-\beta  \epsilon^{\dec}_{2}(\{\Upsilon_{1},...,\Upsilon_{n}\})}}
{\sum_{\{\Upsilon_{1},...,\Upsilon_{n}\}\subset {\P}_\Lambda}
 e^{-\beta  \epsilon^{\dec}_{2}(\{\Upsilon_{1},...,\Upsilon_{n}\})}}
\ee
Further, let $\prob_{\Lambda}\Upsilon$ denote the probability of
occurrence of an $R$--contour $\Upsilon$.
It is given by
\be 
\prob_{\Lambda}\Upsilon = \frac{\sum^\prime e^{-\epsilon_2^{\dec}(\C)}}
{\sum e^{-\epsilon_2^{\dec}(\C)}},
\label{probl}
\ee
where the sum in the numerator is over all $R$--configurations
which contain the given $R$--contour $\Upsilon$, while the
denominator has an unrestricted sum over all $R$--configurations. 

To prove the rigidity of the $111$ interface, we need to find an
upper bound to the probability $\prob_{\Lambda}\Upsilon$ defined 
in \reff{probl}. This is given in the 
following proposition .

\begin{Prop}\label{prop:estimate}
There exists positive constants $U_0$ and $b_0$, such that for all $U> U_0$
and $\beta/U > b_0$, the
probability, $\prob_{\Lambda}(\Upsilon(p,q))$, of occurrence of an 
$R$--contour $\Upsilon(p,q)$ (defined through \reff{upsilon})
satisfies the following bound:
\bea
{\rm{Prob}}_{\Lambda}(\Upsilon(p,q))
\leq  \prod_{i=1}^{i=p} \left\{
e^{-\beta {J}_3(U) a^{ov}_i}
\right\}
\times   \prod_{k=1}^{k=q}\left\{ e^{-\beta {K}_3(U)
|\Delta^{st}_k|}\right\}
\eea
where 
\bea
{J}_3(U) &=& J_2(U) - A_1 U^{-5}-
 \frac{2}{\beta} e^{-6 \beta A_2 U^{-1}}\nonumber\\
{K}_3(U) &=& {K}_2(U)  - A_1 U^{-5}-
\frac{2}{\beta} e^{-6 \beta A_2 U^{-1}}
\eea
and $A_1$ and $A_2$ are positive constants depending on $c_1$, $c_2$ and $U_0$. 
The constants $J_2(U)$ and $K_2(U)$ are defined through \reff{j2} and 
\reff{k2} respectively.
\end{Prop}

The proof of this proposition requires two steps.  The first is to
obtain a lower bound on the total energy, $\epsilon(\Upsilon \mid \C)$,
of an $R$--contour $\Upsilon$ which belongs to an $R$ configuration
$\C$. This energy is defined through \reff{egc}.

The second step is a generalization of the Peierls argument analogous to
Dobrushin's treatment of the antiferromagnetic Ising model \cite{dobafising}.
A unique specification of an $R$ configuration requires the specification of
not only a compatible family of $R$--contours, but also the type of the bases
adjacent to the inner and outer boundaries of each $R$--contour. In other
words contours in the $R$ configuration are not only required to be pairwise
disjoint, but there is the additional requirement of matching of boundary
conditions. One way of analyzing such contour expansions would be to use the 
Pirogov Sinai theory extended to interacting contours \cite{DMS, Par}.
However, instead of doing this we resort to a much simpler method. We use a
recipe for removing a contour from a compatible family which is a
generalization of the one introduced by Dobrushin in the study of the
antiferromagnetic Ising model. The idea is to map a configuration of
$R$--contours $\{\Upsilon_1,...,\Upsilon_p\}$ to a new one
$\{\Upsilon'_2,...,\Upsilon'_p\}$ with one contour less, where $\Upsilon_r$
and $\Upsilon'_r$ are either the same, or related to each other by a simple
geometric transformation. The transformation preserves the energy of the
$R$--contour, at least to order $U^{-3}$. By using this generalization of 
Dobrushin's construction we avoid using the Pirogov-Sinai theory.
\bigskip

\noindent
{\bf Step 1: A lower bound to the total energy of an $R$--contour}
The following lemma is necessary to determine a lower bound to
the total energy $\epsilon(\Upsilon \mid \C)$, [defined through \reff{egc}].

\begin{Lem}
\label{wbound}
There exist positive constants $U_0$, $b_0$, such that for 
all $U>U_0$ and $\beta/U > b_0$ the following bound is satisfied:  
\be
\vert W(\Upsilon \mid \C)\vert \le W_0(\Upsilon \mid \C),
\ee
where
\be
W_0(\Upsilon \mid \C) := \vert\Upsilon\vert
\left\{a_1 U^{-5} + \frac{c}{\beta} e^{-6 \beta a_2 U^{-1}}\right\},
\label{woo}
\ee
with $c$, $a_1$ and $a_2$  being positive constants depending 
on $c_1$, $c_2$, $U_0$ and $b_0$.
\end{Lem}
\begin{proof}
{From} the definition \reff{wup3} of $W(\Upsilon \mid \C)$ it follows that 
\bea
\vert W(\Upsilon \mid \C)\vert
&\leq& 
\sum_{B: |B| > 3 \atop{ B_{\P} \cap \Upsilon
\ne \emptyset}}
\vert \widehat{\Phi_B}(\C)\vert \nonumber\\
&+& \frac{1}{\beta} \Bigl\{ 
\sum_{P:\atop{P_{\P} \cap \Upsilon \ne \emptyset
}} \vert \Psi_\tI^\trunc(P)\vert + \sum_{P:\atop
{P_{\P} \cap \Upsilon \ne 
\emptyset }} \vert \Psi^\trunc(P)\vert \Bigr\} \nonumber\\
&\leq& 
\sum_{x^* \in \Upsilon} \sum_{B: |B| > 3\atop{ B \ni x}}
\frac{\vert \widehat{\Phi_B}(\C)\vert}{\vert B \vert} \nonumber\\
&+& \sum_{x^* \in \Upsilon} \frac{1}{\beta} \Bigl\{ 
\sum_{P:\atop{P\ni x}} \frac{\vert \Psi_\tI^\trunc(P)\vert}{\vert P \vert}
 + \sum_{P:\atop
{P \ni x}} \frac{\vert \Psi^\trunc(P)\vert}{\vert P \vert} \Bigr\},
\label{wup4}
\eea
where $x^*$ is the projection of the site $x$ of the lattice on the plane $\P$.
{From} the definition \reff{hatphi} of $\widehat{\Phi_B}$ and the bound
\reff{phibt} it follows that for all $U > c_d \, c_1$,
\be
\sum_{x^* \in \Upsilon} \Bigl\{\sum_{B: |B| > 3\atop{ B \ni x}}
\frac{\vert \widehat{\Phi_B}(\C)\vert}{\vert B \vert }
\Bigr\} \leq a_1\, \vert \Upsilon \vert U^{-5},
\ee
where $a_1$ is a positive constant depending on $U_0$, $c_1$ and $c_2$.
Further, using the bounds \reff{psi1} and \reff{psi3}, and the
definitions \reff{w0d} and \reff{w2},  we obtain the following 
bound:
\be
\frac{1}{\beta} \sum_{x^* \in \Upsilon} \Bigl\{
\sum_{P:\atop{P\ni x}} \frac{\vert \Psi_\tI^\trunc(P)\vert}{\vert P \vert} + 
\sum_{P:\atop
{P \ni x}} \frac{\vert \Psi^\trunc(P)\vert}{\vert P \vert} 
\Bigr\} \leq \frac{c \vert \Upsilon
\vert }{\beta} e^{-6 \beta a_2 {U^{-1}}},
\label{wup5}
\ee
where $c$ and $a_2$ denote positive constants depending on 
$c_1$, $c_2$, $U_0$ and $b_0$.
The factor of six in the exponent arises from the fact that the smallest 
Ising contour has six faces.
\end{proof}

The lower bound to $\epsilon(\Upsilon \mid \C)$ is given by the following
corollary.

\begin{Cor}
There exist positive constants $U_0$ and $b_0$, such that for all $U> U_0$
and $\beta/U > b_0$, the total energy $\epsilon(\Upsilon \mid \C)$ 
[\reff{egc}],
of an $R$--contour $\Upsilon$ (defined through \reff{upsilon}) 
which belongs to an $R$--configuration
$\C$, satisfies the following bound:
\be
\vert \epsilon(\Upsilon \mid \C)\vert \geq J_3(U) \sum_{i=1}^{p} a_i^{ov}
+ K_3(U) \sum_{k=1}^{q} |\Delta_k^{st}|.
\label{coro}
\ee
where 
\bea
{J}_3(U) &=& J_2(U) - 6 a_1 U^{-5}-
 \frac{6c}{\beta} e^{-6 \beta a_2 U^{-1}}\label{j3}\\
{K}_3(U) &=& {K}_2(U)  - 2a_1 U^{-5}-
\frac{2c}{\beta} e^{-6 \beta a_2 U^{-1}} \label{k3},
\eea
and $c$, $a_1$ and $a_2$ are the positive constants of
Lemma \ref{wbound}. The constants
$J_2(U)$ and $K_2(U)$ are defined through \reff{j2} and \reff{k2} respectively.
\end{Cor}
\begin{proof}
{From} the definition \reff{egc} of $\epsilon(\Upsilon \mid \C)$ it follows
that  $$\vert \epsilon(\Upsilon \mid \C)\vert \geq F_0(\Upsilon) -
W_0(\Upsilon|\C),$$ where $F_0(\Upsilon)$ is a lower bound to $F(\Upsilon)$,
and $W_0(\Upsilon|\C)$ is an upper bound to $W(\Upsilon | \C)$. Obtaining
$F_0(\Upsilon)$ from \reff{fup2}, using Lemma \ref{wbound} and the
bound \reff{numsites} yields the bound
\reff{coro}.  The leading contributions to the energy  of the overlapping $R$
subcontours $\Upsilon^{ov}_i$  of the $R$--contour $\Upsilon$ stem from the
{\it overlapping triangles} $\T^{ov}_i$ of $\Upsilon^{ov}_i$. An edge of an
overlapping triangle can in general coincide with either one of the following
-- a $\delta$--edge, an $\omega$--edge, or a $\lambda$--link. A uniform lower
bound to the energy $\epsilon(\Upsilon \mid \C)$ is obtained by  omitting the
positive energies of these additional edges. 
\end{proof}
\bigskip 

\noindent
{\bf Step 2: A generalized Dobrushin's transformation.}

Consider an $R$-configuration $\C$ 
defined by the  $R$--contours
$\Upsilon_1,...,\Upsilon_l $. Let  $\Upsilon_1$ be the
contour we want to remove;  
$\Upsilon^c_1 = \P \setminus \Upsilon_1$
is an open set, which has maximally connected components
$\{O_0,...,O_r\}$, where $O_0$ is its exterior
and  $\{O_1,...,O_r\}$ are the components of its interior.
As $\Upsilon_1$ is a maximally connected component of
the complement of the bases, any $R$--contour in $O_i$
is not connected to  $\Upsilon_1$.
Therefore, we can uniquely define a type $\tau(O_i)$
of each interior, according  to the type of base that
separates  $\Upsilon_1$ and the contours in the interior of $O_i$.
For simplicity assume that $\tau(O_0) = 0$. We would
like to lift  $\Upsilon_1$ out of the $R$-configuration 
and fill the gap thus created with
bases of type $0$. This can be done only if
$\tau(O_i) = 0$ for every $i$, which is not the case in general.
The extension of Dobrushin's trick \cite{dobafising} is first to apply
a translation $S_i$ to $O_i$ such that
\be 
\tau(S_i[O_i]) =0, i=1,\ldots,r.
\ee
It is easy to see that such translations $S_i$
over one lattice spacing exists; 
e.g., we can use vertical translations over one lattice unit in
the upward downward direction. Let us denote these translations
by $S$ and $S^{-1}$, respectively. Then the following
relation holds for a base ${\rm C}_0$ of type $0$:
\be 
\tau(S^n[C_0]) = -n \bmod 3.
\ee
Now we will apply  $S$ to each $O_i$ of type $1$, and $S^{-1}$
to each $O_i$ of type $2$. Let $n_i =0,\pm 1$ be the exponent such that:
\be  
\tau(S^{n_i}[O_i]) = 0, \qquad i=1,\ldots, r.
\ee
In general the translations of the different $O_i$ can now overlap, i.e.,
\be 
S^{n_i}[O_i] \cap S^{n_j}[O_j] \neq  \emptyset.
\ee
As long as all intersections are bases of type $0$, there is
no problem, and a new configuration with the contour $\Upsilon_1$
removed can be defined. This corresponds to the situation in which 
the $R$-contours in $ S^{n_i}[O_i]$ and $S^{n_j}[O_j]$ do not
intersect each other. We now prove that this is indeed the case. Denote
the inner $R$--contours in $O_i$ by $\Upsilon_{i,1},\ldots,\Upsilon_{i,q_i}$.
Then we have the following property of the interiors. 

\begin{Lem} 
For every $R$--contour $\Upsilon_{i,j}$ in the interior $O_i$  we have:
\be 
S^{n_i}[\Upsilon_{i,j}] \subset \overline{O_i}; \qquad j=1,\ldots,q_i,
\ee
where  $\overline{O_i}$ denotes the closure of $O_i$, and where
$n_i$ is an integer such that $\tau(S^{n_i}[O_i]) = 0$.
\end{Lem}
\begin{proof} 
As $O_i$ is a simply--connected set, and
$\Upsilon_{i,j}$ is closed, it is enough to show that, for every edge
$e\subset \Upsilon_{i,j}$, we have $S^{n_i}(e) \subset \overline{O_i}$.
Let  $P_e$ be the union of all closed triangles intersecting $e$ either
with an edge or a vertex. Then we have
\be P_e\subset \overline{O_i},\qquad \mbox{for all }e\subset \Upsilon_{i,j},
\ee
because otherwise $ \Upsilon_{i,j}$ would be connected to $O^c_i$.
It is then obvious that
\be  S^{n_i}[e]\subset P_e \subset \overline{O_i}.
\ee
\end{proof}

\begin{Lem}\label{lem:dob}
For any pair of $R$--contours $\Upsilon_{i_1,j_1}$ and
$\Upsilon_{i_2,j_2}$, the following is true:
\be S^{n_{i_1}}[\Upsilon_{i_1,j_1}] \cap S^{n_{i_2}}[\Upsilon_{i_2,j_2}]
=  \emptyset, \qquad \mbox{for all }
i_1 \neq i_2;\quad  1\leq j_1 \leq q_{i_1};\quad  1\leq j_2\leq q_{i_2}.
\ee
\end{Lem}
\begin{proof} As the $R$--contours $\Upsilon_{i_1,j_1}$ and
$\Upsilon_{i_2,j_2}$ are closed complexes in two distinct interiors
$O_{i_1}$ and $O_{i_2}$, we have
\be d(\Upsilon_{i_1,j_1}, \Upsilon_{1}) \geq 1. \ee
Hence we have
\be d(\Upsilon_{i_1,j_1}, \Upsilon_{i_2,j_2})\geq 2.
\ee

As $S^{n_{i_1}}$  and  $S^{n_{i_2}}$ are both translations over one
lattice unit, it is clear that
\be S^{n_{i_1}}[\Upsilon_{i_1,j_1}] \cap S^{n_{i_2}}[\Upsilon_{i_2,j_2}]
= \emptyset,
\ee
whenever  $d(\Upsilon_{i_1,j_1}, \Upsilon_{i_2,j_2})\geq 3$.
Hence the only case that we need to investigate is
$$
d(\Upsilon_{i_1,j_1}, \Upsilon_{i_2,j_2})=2\quad.
$$
Let us suppose that
\be  S^{n_{i_1}}[\Upsilon_{i_1,j_1}] \cap S^{n_{i_2}}[\Upsilon_{i_2,j_2}]
\neq \emptyset.\ee
Then there should exist a vertex $v$ such that
\be 
v\in \Upsilon_1,\quad  v\in  S^{n_{i_1}}[\Upsilon_{i_1,j_1}],
\quad v\in S^{n_{i_2}}[\Upsilon_{i_2,j_2}].
\ee
\begin{itemize}
\item 
If $n_{i_1} = n_{i_2}$, then we conclude that $\Upsilon_{i_1,j_1}$ and
$\Upsilon_{i_2,j_2}$ have to intersect, which is excluded by hypothesis. \item
Hence, we must have $n_{i_1} = 1$ and $n_{i_2} = -1$ (or $n_{i_1} =-1$ and
$n_{i_2} = 1$). This means that there is $v\in \Upsilon_1$, $v_1 \in
\Upsilon_{i_1,j_1}$ and $v_2 \in \Upsilon_{i_2,j_2}$ as  Figure \ref{fig:dob1}.

\begin{figure}[t]
\begin{center}
\vbox{
\rotatebox{-90}{\resizebox{4truecm}{!}{\includegraphics{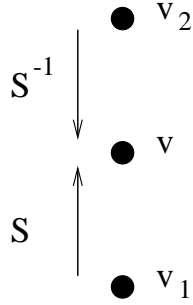}}}}
\parbox{14truecm}{\caption{\baselineskip=16 pt\small\label{fig:dob1}
The relative postions of $v$, $v_1$ and $v_2$ used in the proof
of Lemma \ref{lem:dob}.
}}
\end{center}
\end{figure}

The points $v$, $v_1$ and $v_2$ are related: $S(v_1) = v = S^{-1}(v_2)$. As
the three vertices belong to non--intersecting $R$--contours, there are only
the three possibilities shown in Figure \ref{fig:dob2}. Using the fact that
the contour boundaries cannot subtend angles of ${\pi\over 3}$ when the
$R$--contours do not intersect, we complete these diagrams as shown in Figure
\ref{fig:dob3}. In each case we can show that this leads to contradicting
assignment of types to the rhombi containing the vertex $v$. For example, in
the case (a) we are forced to assign the types of the rhombi as shown in
Figure \ref{fig:dob4}. This contradicts the fact that $S^{-1}[2] = 0$ and
$S(1) = 0$, and hence is not allowed. We exclude the case (b) in the same way.
For case (c), consider the dotted hexagon. It is easy to see that all rhombi
with diagonals that are edges of the same hexagon must be of the same type.
This implies  that: $ \tau(O_{i_1}) = \tau(O_{i_2})$ contradicting the
condition $n_{i_1} = - n_{i_2}$. 
\end{itemize}
\end{proof}

\begin{figure}[t]
\begin{center}
\vbox{
\rotatebox{-90}{\resizebox{4truecm}{!}{\includegraphics{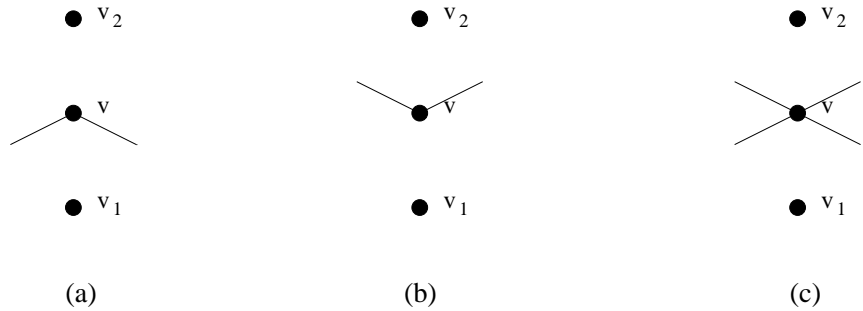}}}
}
\parbox{14truecm}{\caption{\baselineskip=16 pt\small\label{fig:dob2}
The three cases used in the proof of Lemma \ref{lem:dob}.
}}
\end{center}
\end{figure}

\begin{figure}
\begin{center}
\vbox{
\rotatebox{-90}{\resizebox{5truecm}{!}{\includegraphics{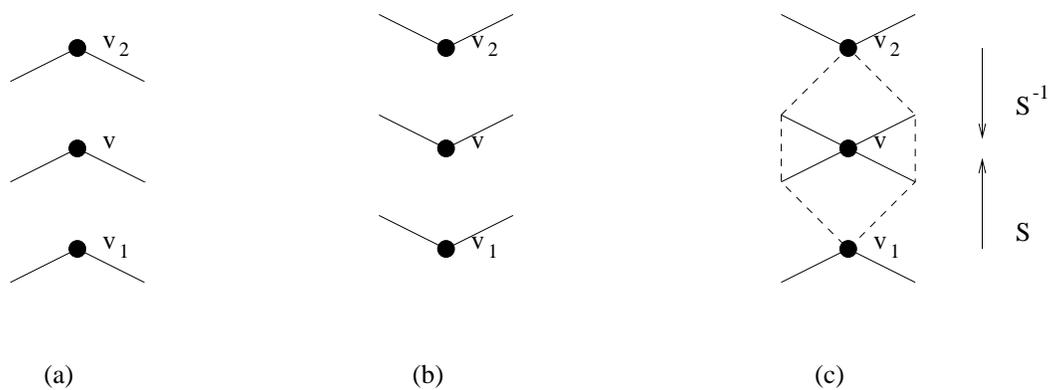}}}
}
\parbox{14truecm}{\caption{\baselineskip=16 pt\small\label{fig:dob3}
The assignment of edges corresponding to the three cases shown in
Figure \ref{fig:dob2}.
}}
\end{center}
\end{figure}

\begin{figure}
\begin{center}
\vbox{
\rotatebox{-90}{\resizebox{4truecm}{!}{\includegraphics{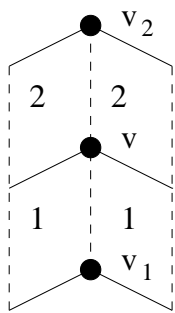}}}
}
\parbox{14truecm}{\caption{\baselineskip=16 pt\small\label{fig:dob4}
The assignment of types to the rhombi of Figure \ref{fig:dob3} (a).
}}
\end{center}
\end{figure}

Now we can complete the Dobrushin argument and the proof of Proposition
\ref{prop:estimate}. We have shown that any $R$--contour $\Upsilon$ can be
removed, meaning that the following operations were performed to obtain a new
configuration of non--overlapping $R$--contours.

\begin{itemize}
\item Erase $\Upsilon$
\item Translate  the interior $O_i$ by $S^{n_i}$ such that
\be 
\tau(S^{n_i}[O_i])  = \tau(O_0) = 0,
\ee
where $O_0$ is the exterior of $\Upsilon$. After the translations
have been performed some parts of the bases,
which are now all of the type $0$, will overlap.
\item Fill up the gaps that were left with the base of type $0$
\end{itemize}

The essential point of this construction is that the $R$ contours which lie in
the interiors of $\Upsilon_1$ have been translated without intersecting each
other.  The effect of the translation  is to modify only the energies
corresponding to the potentials of range greater than two, which are
corrections to the leading terms. This concludes the proof of Proposition
\ref{prop:estimate}.

\masubsection{The Peierls Condition for the  geometric $R$--contours.}

There may be many $R$--contours which have the same support. Different
$R$--contours whose supports coincide differ from each other in the
configuration of their constituent overlapping $R$--subcontours, e.g. in the
overlap numbers of the overlapping triangles, the multiplicity of the
$\lambda$--links etc.

It is convenient to group the $R$--contours into equivalence classes
depending on their support. This allows us to obtain bounds on
relative probabilities entirely in terms of the supports of
the $R$--contours. We define {\em{equivalent contours}} as follows.
\smallskip

\noindent
Two  $R$--contours 
$$
\Upsilon:=(\cup_{i=1}^{i=p}\Upsilon^{ov}_i)\cup 
(\cup_{k=1}^{k=q}\Upsilon^{st}_k) \quad{\rm{and}}\quad
\Upsilon^{\prime}:=(\cup_{i=1}^{i=p}\Upsilon^{\prime ov}_i)\cup
(\cup_{k=1}^{k=q}\Upsilon^{\prime st}_k)
$$
are said to be {\it{equivalent}} iff they fulfill the following conditions:
\begin{itemize}
\item  The  standard subcontour  $\Upsilon_k^{st} \in \Upsilon$
is identical to the standard  subcontour $\Upsilon^{\prime st}_k
\in \Upsilon^{\prime}$ for $k=1, \ldots, q$.
Each standard $R$--subcontour is given by a unique configuration of 
$\delta$-lines. 
\item The overlapping $R$--subcontour  $\Upsilon^{ov}_i \in \Upsilon$ 
and  the  overlapping $R$--subcontour $\Upsilon^{\prime ov}_i \in
\Upsilon^{\prime}$
have  the same support, denoted by $\supp(\overline{\Upsilon}^{ov}_i)$,
for $i=1, \ldots, p$.
\end{itemize}

{\sl A geometric contour} $ \overline{\Upsilon} \equiv
\overline{\Upsilon}(p,q)$ is an equivalence  class of $R$--contours, i.e., 
\be  
\overline{\Upsilon}(p,q) \equiv \overline{\Upsilon} 
:=\left\{\Upsilon_{\alpha}=\Bigl(\cup_{i=1}^{i=p} 
{\Upsilon}^{ov}_{\alpha i}\Bigr) \bigcup \Bigl(\cup_{k=1}^{k=q} 
{\Upsilon}^{st}_k
\Bigr) \mid \supp\Upsilon_{\alpha i}^{ov} =
\supp\overline{\Upsilon}_i^{ov} \,\, {\rm{for}} \,\, 1\le i \le p \right\},
\label{geomcont}
\ee
where the subscript $\alpha$ is used to label the different $R$--contours
which have overlapping subcontours of identical support. Like an $R$--contour,
a geometric contour can also be decomposed into  overlapping and standard
subcontours:
\bea
\overline{\Upsilon}(p,q) &=&
\Bigl(\cup_{i=1}^{i=p}\overline{\Upsilon}^{ov}_i\Bigr) \cup
\Bigl(\cup_{k=1}^{k=q}{\Upsilon}^{st}_k\Bigr), \nonumber\\
&=:& \overline{\Upsilon}^{ov} \cup \overline{\Upsilon}^{st},
\eea
where the symbol $\overline{\Upsilon}^{ov}_i$ denotes the $i$-th 
overlapping geometric subcontour. It follows from \reff{geomcont} that
all $R$-contours constituting a geometric contour have the same support.

We can associate a unique number $r_i^{ov}$ to the support  $\supp(
\overline{\Upsilon}^{ov}_i)$ which is defined as follows:
\be
r_i^{ov}:= \hbox{minimum number of (distinct) rhombi 
in which} \, \supp\overline{\Upsilon}^{ov}_i \, \hbox{can be decomposed.}
\label{unique}
\ee
For simplicity we shall often use the symbol $\overline{\Upsilon}$ to denote
both the geometric contour and its support.

Each overlapping $R$--subcontour $\Upsilon^{ov}_{\alpha i}$ is characterized
by a configuration of overlapping triangles, each of which is labeled by an
even overlap number. The  $R$--subcontours belonging to a given geometric
subcontour $\overline{\Upsilon}^{ov}_i$ differ from each other in the
distribution and overlap numbers of their overlapping triangles.  Since each
of these $r_i^{ov}$ rhombi is an overlapping rhombus, the following bound is
satisfied:
\be
a_{\alpha\,i}^{ov} \ge r_i^{ov},
\label{totov}
\ee
where $a_{\alpha\,i}^{ov}$ is defined through \reff{ai}, the subscript
$\alpha$ labelling the different $R$--subcontours belonging to 
$\overline{\Upsilon}^{ov}_i$. Let $\prob_{\Lambda}\overline{\Upsilon}$ denote
the probability of occurrence of an $R$-contour with support
$\supp\overline{\Upsilon}$, 
with support 
through \reff{geomcont}) in $\P_\Lambda$. To compute this probability we need
to sum over all possible $R$--contours which belong to $\overline{\Upsilon}$.
\begin{Prop}
\label{prop10}
There  exist positive constants $U_0$, $d_0$ and $d_1$  such that,
for all $U> U_0$, $\beta/{U} > d_0 $ and $\beta/{U^3} > d_1 $, 
the probability
$\prob_{\Lambda}(\overline{\Upsilon}(p,q))$ satisfies the bound
\be
\prob_{\Lambda}(\overline{\Upsilon}(p,q))
\leq \prod_{i=1}^{i=p}e^{-\beta U^{-1} D_1\, r^{ov}_i}
\times\prod_{k=1}^{k=q}e^{-\beta U^{-3} D_2 \,|\Delta^{st}_k|},
\label{qed}
\ee
where $D_1$ and $D_2$ are positive constants depending on $U_0$,
$d_0$ and $d_1$. 
\end{Prop}
\begin{proof}
The probability $\prob_{\Lambda}(\overline{\Upsilon}(p,q))$ can be expressed
in terms of the probability $ \prob_{\Lambda}(\Upsilon(p,q))$ that an
$R$--contour $\Upsilon(p,q)$ occurs in a given $R$--configuration. Using
Proposition  \ref{prop:estimate} we obtain the following bound.
\bea
\prob_{\Lambda}(\overline{\Upsilon}(p,q)) & =&\sum_{ \{\Upsilon \in
\overline{\Upsilon}\}} \prob_\Lambda (\Upsilon(p,q))\nonumber\\ 
&\leq& 
\prod_{i=1}^{i=p}\Bigl(\sum_{\alpha \ge 1 :\atop{\Upsilon^{ov}_{\alpha i} 
\in \overline{\Upsilon}^{ov}_i}} e^{(-\beta {J}_3(U)
a^{ov}_{\alpha i})}\Bigr) \times \prod_{k=1}^{k=q}
e^{(-\beta {K}_3(U) |\Delta^{st}_k|)}
\label{bound} 
\eea

Next we need to estimate the sum over $\alpha$ in the last line of
\reff{bound}. Each overlapping subcontour with a given support is obtained by
the projection of a set of faces (not necessarily connected) of an interface.
The number of faces of an interface which projects onto the overlapping
$R$--subcontour $\Upsilon^{ov}_i$ is equal to $(a_{\alpha i}^{ov} +
r_i^{ov})$. Since all the overlapping $R$--subcontours in
$\overline\Upsilon^{ov}_i$ have the same support, it is possible to find an
edge, which belongs either to a base or to a standard $R$--subcontour, such
that it intersects all these  overlapping $R$--subcontours at a fixed vertex.
(The choice of such an edge is, however, not unique). From these considerations
it follows that we can replace the sum over $\alpha$ by a sum over the
variables $a_i$, which take integer values and correspond to the  distinct
sets of faces (each set belonging to an interface $\tI$) which satisfy the
following properties:

(1) The projection of the (disjoint) union of the faces in each such 
set is connected to a fixed end of a fixed edge in the triangular
lattice spanning the plane $\P$.

(2) Each set has $a_i + r^{ov}_i$ faces with $a_i \geq r^{ov}_i$.

Since the projection of each such set of faces is a connected set
of rhombi, at least one vertex of each rhombus is shared by another
rhombus. Hence each rhombus has at most three vertices to which
the fixed edge can be attached. The number of ways of placing $a_i + r^{ov}_i$
rhombi on the $\supp\overline\Upsilon^{ov}_i$, which is
connected to a fixed end of a fixed edge in the triangular
lattice spanning $\P$, is bounded by the number 
of ways of constructing a 
connected set which consists of $a_i + r^{ov}_i$ rhombi and is 
connected to this fixed vertex. This latter number, which we denote by 
$N^{ov}(a_i)$, is easily seen to satisfy the bound
\be
N^{ov}(a_i) \le m^{a_i + r^{ov}_i},
\label{num}
\ee
with $m=36$, by the K\"onigsberg Bridge Lemma \cite{simon}. 
Hence,
\bea
\prob_\Lambda(\overline{\Upsilon}(p,q))
& \leq&  \sum_{a_1\geq r^{ov}_1, \dots,  a_p\geq r^{ov}_p}\!\!\!\!
m^{\ a_1+  r^{ov}_1 + \dots + a_p+ r^{ov}_p}\times e^{-\beta(a_i + \ldots +
a_p)
{J}_3(U)}
\times\prod_{k=1}^{k=q}e^{-\beta {K}_3(U)\times|\Delta^{st}_k|}
\nonumber \\
&\leq&  \prod_{i=1}^{i=p} m^{2 r^{ov}_i} e^{-\beta r^{ov}_i J_3(U)}
\Bigl( \sum_{a_i \ge 0} m^{a_i} e^{-\beta a_i J_3(U)}\Bigr)
\times\prod_{k=1}^{k=q}e^{-\beta {K}_3(U)\times|\Delta^{st}_k|}
\label{serie}
\eea
It is easy to see that there exists positive constants $U_0$,  $d_0$, such
that  for all $U>U_0$ and $\beta/{U} > d_0$, the geometric series in
paranthesis (on RHS of \reff{serie}) converges. Hence, using the definitions
\reff{j3} and \reff{k3} of  $J_3(U)$ and $K_3(U)$ we find that for all
$U>U_0$, $\beta/U > d_0$ and $\beta/{U^3} > d_1$ (where $d_1$ is a positive
constant) the following bound holds.
\be
\prob_\Lambda (\overline{\Upsilon}(p,q))
 \leq \prod_{i=1}^{i=p}e^{-\beta U^{-1} D_1\,r^{ov}_i}
\times\prod_{k=1}^{k=q}e^{-\beta U^{-3} D_2 \,|\Delta^{st}_k|},
\ee
where $D_1$ and $D_2$ are positive constants depending on $U_0$, $d_0$ and 
$d_1$. 
\end{proof}

\masubsection{Proof of the rigidity  of the $111$ interface}
\label{proof_rigidity}

The rigidity of the $111$ interface can be expressed in terms of the
probability, $\prob_\Lambda(s_{x_0} =-1| {\rm{b.c.2}})$,  that a  lattice site
$x_0=(x_1, x_2, x_3)$, such that $x_1 + x_2 + x_3 \ge 1/2$, is occupied by a
``$-$'' spin under the boundary condition b.c.2 \reff{bc2}. To have $s_{x_0}=
-1$, the site $x_0$ must be enclosed either by a pyramid of the interface or
by at least one Ising contour.  As a result we have that
\be
\prob_\Lambda \Bigl(s_{x_0} =-1| {\rm{b.c.2}}\Bigr) \le \sum_{\ovup \ni \xx} 
\prob_\Lambda{\ovup} + \sum_{\gamma \ni x_0} \prob_\Lambda{\gamma},
\label{probres}
\ee
where $x_0^*$ denotes the projection of the site $x_0$ on the plane 
$\P_\Lambda$. The first term on the RHS of \reff{probres} arises when the
interface has at  least one  pyramid whose projection on $\P_\Lambda$ encloses
the point $x_0^*$. The second term arises when the ``$-$'' spin at $x_0$ is
enclosed by one or more Ising contours, whose presence does not lead to a
distortion of the  interface from its perfect staircase structure around the
site $x_0$.  The above bound \reff{probres} is also satisfied by the
probability  $\prob_\Lambda \Bigl(s_{{\tilde{x}}_0} = 1| {\rm{b.c.2}}\Bigr)$,
for ${{\tilde{x}}_0}= (x_1, x_2, x_3)$ with $x_1 + x_2 + x_3 \le -\,1/2$.

Hence, to prove the rigidity of the $111$ interface we need to estimate the
terms on the RHS of \reff{probres}. An estimate of the first term is given by
the following proposition.

\begin{Prop}
\label{prop:summable}
There exist positive constants  $U_0$ and  $b_1$ and $b_2$,  such that for all
$U >U_0$ $\beta/U > b_1$ and $\beta/U^3 > b_2$, the  following estimate is
true:
\be
\sum_{\overline{\Upsilon} \ni \xx}
{\rm{Prob}}_\Lambda \overline{\Upsilon} 
\leq   {\tilde{C_0}}\,e^{-\,c^{\prime\prime}\,\beta\,U^{-3}}, 
\label{sumrigid}
\ee
where ${\tilde{C_0}}$ and $c^{\prime\prime}$ are positive constants depending 
on $U_0$, $b_1$ and $b_2$. 
\end{Prop}

\begin{proof}
{From} Proposition \ref{prop10} it follows that 
\be
\sum_{\ovup \atop{\overline{\Upsilon}\ni \xx}}
{\rm{Prob}}_\Lambda \overline{\Upsilon} 
\le
\sum_{\ovup \ni \xx \atop {\ovup:= \ovup^{ov} \cup \ovup^{st}}}
\prod_{\Upsilon_i^{ov} \in \ovup^{ov}_i}
e^{-\beta U^{-1} D_1  r^{ov}_i}
\times\prod_{\Upsilon^{st}_k \in \ovup^{st}}
e^{-\beta U^{-3} D_2 |\Delta^{st}_k|},
\label{eq1}
\ee
where we have used the following notations: $\ovup^{ov}_i$ 
denotes an overlapping geometric subcontour (see \reff{geomcont}),
while $|\Delta^{st}_k|$ denotes the total number of
$\delta$--lines in the standard geometric subcontour 
$\ovup^{st}_k \equiv \Upsilon^{st}_k$. Further, an empty product is equal to
unity.
\bea
{\hbox{RHS of \reff{eq1}}} &=&  
\sum_{\ovup \ni \xx \atop {\ovup:= \ovup^{ov} \cup \ovup^{st}
\atop{ \ovup^{ov} \ne \emptyset}}}
\prod_{\Upsilon_i^{ov} \in \ovup^{ov}}
e^{-\beta U^{-1} D_1  r^{ov}_i}
\times \prod_{\Upsilon^{st}_k \in \ovup^{st}}
e^{-\beta U^{-3} D_2 |\Delta^{st}_k|}\nonumber\\
& & \quad + 
\sum_{\ovup \ni \xx \atop {\ovup:= \ovup^{st}\atop{ \ovup^{ov} = \emptyset}}}
e^{-\beta U^{-3} D_2 |\Delta^{st}|}
\nonumber\\
&=&
\sum_{\ovup \ni \xx \atop {\ovup:= \ovup^{ov} 
\cup \ovup^{st}\atop{ \ovup^{ov} \ne \emptyset}}}
\prod_{\Upsilon_i^{ov} \in \ovup^{ov}}
e^{-\beta U^{-1} D_1  r^{ov}_i}
\Bigl[\prod_{\Upsilon^{st}_k \in \ovup^{st}}
\bigl\{\bigl(e^{-\beta U^{-3} D_2 |\Delta^{st}_k|} - 1 \bigr) + 1 \bigr\}
\Bigr] \nonumber\\
& & \quad + 
\sum_{\ovup \ni \xx \atop {\ovup:= \ovup^{st}\atop{ \ovup^{ov} = \emptyset}}}
e^{-\beta U^{-3} D_2 |\Delta^{st}|}
\nonumber\\
&=&
\sum_{\ovup \ni \xx 
\atop {\ovup:= \ovup^{ov} \cup \ovup^{st}
\atop{ \ovup^{ov} \ne \emptyset}}}
\prod_{\Upsilon_i^{ov} \in \ovup^{ov}}
e^{-\beta U^{-1} D_1  r^{ov}_i}
\Bigl\{ 1 + \sum_{n \ge 1}
\sum_{\Upsilon_1^{st}, \ldots, \Upsilon_n^{st}
\atop{
\Upsilon_i^{st} \cap \ovup^{ov} \ne \emptyset 
\atop{(i=1,..,n)}}}
\prod_{k=1}^{n}(e^{-\beta U^{-3} D_2 |\Delta^{st}_k|} - 1) 
\Bigr\}\nonumber\\
& & \label{spi1}\\
& & \quad + 
\sum_{\ovup \ni \xx \atop {\ovup:= \ovup^{st}\atop{ \ovup^{ov} = \emptyset}}}
e^{-\beta U^{-3} D_2 |\Delta^{st}|},
\label{spi2}
\eea
where  
$|\Delta^{st}|$ denotes the number of $\delta$--lines
in the standard $R$--contour $\Upsilon^{st}$.

The term on the RHS of \reff{spi1} is similar to \reff{expand} of  Section
\ref{deccont}. This allows us to prove the bound \reff{sumrigid} by using a 
method analogous to the one used in that section.  Each geometric contour
$\ovup$ appearing in the sum on the RHS of \reff{spi1} consists of a finite
set of non-intersecting overlapping geometric subcontours  and a finite set of
standard geometric subcontours such that each standard geometric subcontour
intersects at least one overlapping geometric subcontour in the set. We can
alternatively express each such $\ovup$ as a connected set of auxiliary
polymers, each polymer being a connected set consisting of only a
{\em{single}} overlapping geometric subcontour and a finite set of standard
geometric subcontours which intersect it. To each term on the  RHS of
\reff{spi2} we can associate a standard geometric contour which is given by a
standard $R$-contour. These considerations allow us to express the RHS of 
\reff{eq1} in terms of the elements of a more general polymer system, the
polymers being referred to as {\it{spider contours}} (or $S$ contours, for
brevity) and defined as follows: 
\smallskip

\noindent
An $S$ contour, $\spi$ is a finite connected set of geometric
$R$--subcontours, containing at most one overlapping geometric
$R$--subcontour. Hence, in general
\be
\spi := \overline{\Upsilon}^{ov}\bigcup \{\cup_{i=1}^{i=s}
\Upsilon^{st}_i\},
\label{spider}
\ee
where $\overline{\Upsilon}^{ov}$ denotes an overlapping geometric subcontour.
 
In particular, an $S$ contour can reduce to a single overlapping $R$--contour
(with no standard part) or to one standard $R$--contour. We refer to an $S$
contour which has no overlapping part as a {\em{standard}} $S$ contour and
denote it by the symbol $\spi^{st}$. The other $S$ contours are said to be
overlapping and are denoted by the symbol $\spi^{ov}$. Every standard
$R$--subcontour which belongs to an overlapping $S$ contour, (defined through 
\reff{spider}), necessarily intersects the overlapping geometric  subcontour
$\overline{\Upsilon}^{ov}$.

A given geometric $R$--contour $\ovup$, with support $\supp{\ovup}$,
can be built from a finite family of intersecting $S$ contours,
For an $S$ contour $\zeta$ given by \reff{spider}, we define
\be 
|\spi| := r^{ov} + \sum_{i=1}^s |\Delta^{st}_i|,
\label{spisize}
\ee
where $r^{ov}$ denotes the minimal number of rhombi needed to cover the 
support of $\ovup^{ov}$ (see \reff{unique}). For a standard $S$-contour 
$\spi^{st}$ we define
\be
|\spi^{st}| := |\Delta^{st}|,
\ee
which is the number of $\delta$-lines in it.

We define the weight $\eta(\cdot)$ of an {\it{overlapping}} $S$ contour
$\spi^{ov}$ (defined through \reff{spider}) to be
\be
\eta(\spi^{ov}):= e^{-\beta U^{-1} D_1 r^{ov}}\times \prod_{\Upsilon_k^{st}
\cap \ovup^{ov} \ne 
\emptyset}
e^{-\beta U^{-3} D_2 |\Delta^{st}_k|}.
\label{wt1}
\ee
In \reff{wt1} we use the convention that an empty product is unity. Hence the
case in which the $S$ contour does not contain any standard part, is included
in \reff{wt1}. The corresponding weight for a {\it{standard}} $S$ contour
$\spi^{st} \equiv \Upsilon^{st}$ is defined as
\be
\eta(\spi^{st}) := e^{-\beta U^{-3} D_2 |\Delta^{st}|}
\label{wt2}
\ee
{From} \reff{spi1}, \reff{spi2}, \reff{wt1} and \reff{wt2} it follows that  
\bea
\sum_{\ovup \atop{\overline{\Upsilon}\ni \xx}}
\prob(\overline{\Upsilon})
&\le& 
 \sum_{n \ge 1} \frac{1}{n!} \sum_{{\spi_1, \ldots , \spi_n} \atop
{\spi_1 \cup \ldots \cup \spi_n = \ovup \atop {\ovup \ni \xx, 
{\rm{connected}}}}} \prod_{j=1}^n \eta(\spi_j)\nonumber\\
&\le& \sum_{n\ge 1} \frac{ C^n}{n},
\label{same2}
\eea
where
\begin{equation}
C:=  \sum_{\spi \ni \xx} \eta(\spi) e^{|\spi|},
\label{pfist}
\ee
In obtaining the bound \reff{same2}, we have made use of Lemma 3.5 of 
\cite{pfister} (as in \reff{pfist0} of the Appendix B). The
proof of the bound \reff{sumrigid} reduces to the proof of the following 
lemma.
\end{proof}
\begin{Lem}
\label{spi}
For each $D'>0$, there exist positive constants $b_1$ and $b_2$ such that for
all $\beta/U > b_1$ and $\beta/U^3 > b_2 $ the following bound holds:
\begin{equation}
\sum_{\spi \ni \xx} \eta(\spi) e^{|\spi|} \leq D'
\label{wtsp} 
\end{equation}
\end{Lem}
\begin{proof}
\be
{\hbox{LHS of \reff{wtsp}}} =  \sum_{\spi^{ov} \ni \xx} \eta(\spi^{ov})
e^{|\spi^{ov}|}
+  \sum_{\spi^{st} \ni \xx} \eta(\spi^{st}) e^{|\spi^{st}|}
\label{wtsp2}
\ee
Let us first evaluate the second term on the RHS of \reff{wtsp2}.
To do this we use the fact that
the smallest standard $R$--contour consists of six $\delta$--lines. Hence,
\be
 \sum_{\spi^{st} \ni \xx} \eta(\spi^{st}) e^{|\spi^{st}|} \le
\sum_{k \ge 6} \sum_{\spi^{st} \ni \xx \atop{|\spi^{st}| = k}}
e^{-\beta U^{-3} D_2 k} e^k.
\label{wtsp3}
\ee
Consider each standard $S$ contour, appearing in the above sum, as a connected
graph which contains a fixed vertex $\xx$. The maximum coordination number of
each vertex in the graph is five, because a $\delta$--line which intersects a
given $\delta$--line at a fixed end, can emerge in any one of five directions
(in the triangular lattice spanning $\P$). Then by the  K\"onigsberg Bridge
Lemma \cite{simon}
\be
{\rm{RHS}}\,\, {\rm{of}}\,\, \reff{wtsp3} \le 6\,\sum_{k \ge 6} 
5^{2k} e^{-\beta U^{-3} D_2 k} e^k.
\label{wtsp4}
\ee
The factor of six arises from the fact that there can be at most six different
$\delta$--lines in the rhombus model which contains the fixed site $\xx$.  
Let $b>0$ be a constant, such that for all $\beta/U^3 > b$, the geometric
series in \reff{wtsp4} converges. Then for all  $\beta/U^3 > b$,
\be
\sum_{\spi^{st} \ni \xx} \eta(\spi^{st}) e^{|\spi^{st}|} \le 
e^{-6D_4 \beta U^{-3}}
\label{wtsp5} 
\ee
where $D_4$ is a positive constant depending on $b$.  Next we need to
evaluate the first term  on the RHS of \reff{wtsp2}. This term can be further
decomposed into two sums, depending on whether the fixed site $\xx$ belongs to
the overlapping subcontour of an $S$ contour or not. Using the definition
\reff{spider} of an overlapping $S$ contour, we can write
\be
\sum_{\spi^{ov} \ni \xx} \eta(\spi^{ov})
e^{|\spi^{ov}|} =  \sum_{\spi^{ov} \atop{ \ovup^{ov}\ni \xx}} \eta(\spi^{ov})
e^{|\spi^{ov}|} + \sum_{\spi^{ov} \ni \xx \atop{ \ovup^{ov}\not\ni \xx}} 
\eta(\spi^{ov}) e^{|\spi^{ov}|}.
\label{wtsp6}
\ee
The second term on the RHS of \reff{wtsp6} corresponds to the situation in
which the fixed vertex is necessarily contained in a $R$--standard subcontour
of the overlapping $S$ contour. 
\medskip

\noindent
{\bf{Evaluation of the first term on the RHS of \reff{wtsp6}}}:

Let us construct  $\ovup^{ov}$ starting from $\xx$. Once a rhombus which has
$\xx$ as one of its vertices is chosen, the next rhombus which  intersects it
can be placed in twenty four different ways; it can intersect the  first
rhombus either at  any one of its four vertices or along any one of its four
edges. Further, for intersection either along an edge, or at a vertex, there
are three  possible orientations of the pair of rhombi. This allows us to
consider each rhombus as a vertex of a connected graph which contains a fixed
vertex. The fixed vertex of the graph  corresponds to the rhombus which
contains the site $\xx$. There are twelve different rhombi in the triangular
lattice spanning $\P$  which contains a given site. Then, by the K\"onigsberg
Bridge Lemma \cite{simon} the number, $N(\ovup^{ov}|n;\xx)$, of different ways
of constructing an overlapping geometric contour $\ovup^{ov}$, such that it
contains a fixed site $\xx$, and has $r^{ov} = n$, satisfies the bound
\be
N(\ovup^{ov}|n;\xx) \le 12\,d_1^{\,2n}  \quad {\rm{with}} \,\, d_1= 24.
\label{wtsp7}
\ee
The overlapping geometric contour $\ovup^{ov}$ has at most $(3n +1)$ vertices
at which a $\delta$--line can intersect it.   Moreover, from each vertex of
$\ovup^{ov}$ there can emerge at most four standard  $\delta$--lines. Further,
each such standard $\delta$--line can  correspond to either one of two
different pairs of rhombi.  These considerations yield the bound
\bea
\sum_{\spi^{ov} \atop{ \ovup^{ov}\ni \xx}} \eta(\spi^{ov})
e^{|\spi^{ov}|} &\le& \sum_{n\ge 1} 12\, d_1^{\,2n} e^{-\beta U^{-1} D_1 \,n} 
\times
\Bigl\{8 \times (3n +1) \sum_{k\ge 1} 5^{2k} e^{-(\beta U^{-3} D_2\, k)}\, 
e^k \Bigr\}\label{wtsp8} 
\eea
Let $b'_1$ and $b'$ be positive constants such that for all $\beta/U > b'_1$
and $\beta/{U^3}> b'$ the series in $n$ and $k$ converges. Then for such
values of $\beta$ and $U$ the RHS of \reff{wtsp8} satisfies the bound
\be
{\hbox{RHS of \reff{wtsp8}}} \le 
e^{- \beta U^{-3} D_6} \, e^{- \beta U^{-1} D_5},
\label{star1}
\ee
where $D_5$ and $D_6$ are positive constants depending on $b'_1$ and $b'$
respectively.
\medskip

\noindent
{\bf{Evaluation of the second term on the RHS of \reff{wtsp6}}}: 

To evaluate
this sum we make use of the fact that each standard $R$--subcontour in
$\spi^{ov}$ intersects $\ovup^{ov}$. We construct  $\zeta^{ov}$ starting from
the site $\xx$ which now belongs to a standard $R$--subcontour. If the
standard $R$--subcontour which contains $\xx$ has  $m$ $\delta$--lines, then
there are at most $m+1$ vertices at which $\ovup^{ov}$ can intersect it.
Moreover, there can be at most four standard $\delta$--lines emerging from
each of the $(3r^{ov} + 1)$ vertices of $\ovup^{ov}$ and there are two
possible orientations of each such pair of $\delta$ lines. From these
considerations we obtain
\bea
&&\sum_{\spi^{ov}\ni \xx \atop{ \ovup^{ov}\not\ni \xx}} \eta(\spi^{ov})
e^{|\spi^{ov}|}\nonumber\\
&&\quad \le  \sum_{m\ge 1} 5^{2m} e^m e^{-\beta U^{-3} D_2\,m}
\Bigl[ (m + 1)\sum_{n\ge 1} d_1^{2n} e^{-\beta U^{-1}\, n }
\Bigl\{8 \times (3n +1) \sum_{k\ge 1} 5^{2k} e^k e^{-\beta U^{-3} D_2\,k}
\Bigr\}\Bigr].\label{wtsp9}\nonumber\\
\eea
Let $b^{''}_1$ and $b^{''}$ be positive constants such that for all 
$\beta/U >b^{''}_1$ and $\beta/U^3 > b^{''} $ the series on the RHS of 
\reff{wtsp9} converges. Then, for such values of $\beta$ and $U$
the following bound is satisfied:
\be
{\hbox{ RHS of \reff{wtsp9}}} \le e^{-\beta U^{-3} D_8} \, 
e^{- \beta U^{-1} D_7},
\label{star}
\ee
where $D_7$ and $D_8$ are positive constants depending on $b^{''}_1$
and $b^{''}$ respectively. Let 
\be
b_1:= {\rm{max}}(b'_1, b^{''}_1),
\ee
and
\be
b_2:= {\rm{max}}(b, b', b^{''})
\ee
Then from \reff{wtsp5}, \reff{star1} and \reff{star} it follows that 
for all $\beta/U > b_1$ and $\beta/U^3 > b_2$ 
\be
\sum_{\spi \ni \xx} \eta(\spi) e^{|\spi|} \le C_0 \, 
e^{- c^{''}\beta / U^3} =:D',
\ee
where $C_0$ and $c^{''}$ are positive constants depending on
$b_1$ and $b_2$ respectively. Hence 
$D'$ is a positive constant which can be made arbitrarily small by 
making $b_1$ and $b_2$ large enough. This concludes the proof.
\end{proof}

To estimate the second term on the RHS of \reff{probres} we make use of the 
result of the cluster expansion given in Lemma \ref{convergence}. This yields
the following proposition.
\begin{Prop}\label{prop:second}
There exist constants $U_1, b_3 >0$ such that for all $U>U_1$ and 
$\beta/U > b_3$ the  following estimate is true:
\be
{\sum_{\gamma \ni x_0}}
{\rm{Prob}}_\Lambda(\gamma)
\leq   {\tilde{C_1}}\,e^{-\,c_1^{''}\,\beta\,U^{-1}}, 
\label{sumgamma}
\ee
where ${\tilde{C_1}}, c_1^{''}$ are positive constants depending on $U_1$ and 
$b_3$.
\end{Prop}
\begin{proof}
We have that 
\be
\sum_{\gamma \ni x_0} \prob_\Lambda \gamma \le 
\sum_{D:\atop{\D \ni \gamma \atop{ \gamma \ni x_0}}} \prob_\Lambda D
\label{ft1}
\ee
where $\prob_\Lambda D$ denotes the probability of occurrence of a decorated 
contour $D$ in $\Lambda$.
\be
{\hbox{RHS of}} \, \reff{ft1} \le \frac{\Bigl[\sum_{D:\atop{\D \ni \gamma 
\atop{ \gamma \ni x_0}}} W(D) \Bigl\{ \sum_{\D}^{\prime}
\prod_{D' \in \D} W(D')]\Bigr\}\Bigr]}{\Bigl[\sum_{\D} \prod_{D' \in \D}W(D')
\Bigr]}.
\label{ft2}
\ee
The symbol $\sum_{\D}^{\prime}$ is used to denote a sum  over all finite 
compatible sets
$\D = \{D'\}$ of decorated contours which are  {\em{compatible}} with $D$, the
latter being a decorated contour which encloses the site $x_0$. In the
denominator we have an unrestricted sum. The result of the cluster expansion
[Lemma \ref{convergence}] can be applied to both these sums to yield
\be{
\hbox{RHS of}} \, \reff{ft2} \le\sum_{D:\atop{\D \ni \gamma 
\atop{ \gamma \ni x_0}}} W_0(D) \exp \Bigl\{ - \sum_{N \ge 1} \frac{1}{N!}
\sum_{D_1,\ldots,D_N\atop{(i) \& (ii)}} \Psi^T (\{D_1,\ldots,D_N\})\Bigr\},
\label{ft3}
\ee
where $(i) \,\& \,(ii)$ refer to the following conditions:
\smallskip

\noindent
(i) $D_1 \cup \ldots \cup D_N \equiv P$ is a connected set (a cluster),
\smallskip

\noindent
(ii)$P \cap D \ne \emptyset$.
\smallskip

\noindent
We have made use of the fact that $|W(D)| \le W_0(D)$, where $W_0(D)$ is
given  by \reff{w0d}. 
\be 
{\hbox{RHS of} \, \reff{ft3}}\, \le\sum_{D:\atop{\D \ni \gamma 
\atop{ \gamma \ni x_0}}} W_0(D) \exp \Bigl\{- |D| \sum_{N \ge 1} 
\frac{1}{N!} \sum^*_P \frac{|\Psi^T(P)|}{|P|}\Bigr\}
\label{ft4}
\ee
where $$|D| = \sum_{\gamma \in D} \{|\gamma| + \sum_{B \cap \gamma \ne 
\emptyset} g(B)\}$$ and  the symbol $\sum^*_P$ denotes a sum over all 
clusters $P$ which contain a fixed site. The bound  \reff{sumgamma} is then
obtained by making use of \reff{w0d} and the bound  \reff{psi1} on the RHS of
\reff{ft4}.  
\end{proof} 
{From} the Propositions \ref{prop:summable} and \ref{prop:second} it follows
that for $x_0=(x_1, x_2, x_3)$ such that $x_1 + x_2 +x_3 \ge 1/2$, we have
the following upper bound on the probability $\prob_\Lambda(s_{x_0} = -1)$:

\begin{Lem}
\label{lemfin}
There exist positive constants ${\tilde{U_0}}, {\tilde{D_0}}$ and 
${\tilde{D'_0}}$ such that for all $U > {\tilde{U_0}}$, 
$\beta/U > {\tilde{D_0}}$ and $\beta/U^3 > {\tilde{D'_0}}$ the following 
bound is satisfied:
\be 
\prob_\Lambda(s_{x_0} = -1) \le 
\Bigl\{{\tilde{C_0}} e^{-c^{\prime\prime} \beta/{U^3}} +
 {\tilde{C_1}} e^{-c^{\prime\prime}_1 \beta/{U}}\Bigr\}
\ {\rm{for}}\, 
x_0 = (x_1,x_2,x_3) \, {\rm{and}}\, x_1+x_2+x_3 \ge 1/2,
\label{bdfin}
\ee
\end{Lem}

\masubsection{Proof of Theorem \ref{theorem1B}}

We now have all the estimates necessary to prove our main result on the
rigidity  of the $111$ interface, i.e., Theorem \ref{theorem1B}. 

\begin{proof} 
For $x=(x_1, x_2, x_3)$ with $x_1+x_2+x_3 \ge 1/2$, we have that 
\be
<s_x>_{\rm{[b.c. 2]}} = 1 - 2 \times \lim_{\atop{\Lambda \nearrow {\zed}^3}}
\prob_\Lambda (s_{x} = -1 | {\rm{b.c.2}}),
\label{prr}
\ee
where $x^*$ denotes the projection of the site $x$ on the plane $\P$.

Lemma \ref{lemfin} provides an upper bound to the probability 
$\prob_\Lambda (s_{x} = -1 | {\rm{b.c.2}})$ which is uniform in the volume
$\Lambda$. Introducing this bound, \reff{bdfin}, in \reff{prr} yields
\reff{s1}.

Similarly for $x=(x_1, x_2, x_3)$ with $x_1+x_2+x_3  \le -1/2$, we have that
\be
<s_x>_{\rm [b.c. 2]} = - 1 + 2 \times \lim_{\atop{\Lambda \nearrow {\zed}^3}}
\prob_\Lambda \Bigl(s_{x} = 1 | {\rm{b.c.2}}\Bigr) 
\ee
which reduces to \reff{s2} by the analogue of Lemma \ref{lemfin} for
$\prob_\Lambda \Bigl(s_{x} = 1 | {\rm{b.c.2}}\Bigr)$ with 
$x_1+x_2+x_3  \le -1/2$ . 
\end{proof}

\section*{Appendix A: Bound on the remainder term of the effective Hamiltonian}
\label{appendixA}
\renewcommand{\thetheorem}{\mbox{A.\arabic{Thm}}}
\renewcommand{\theequation}{\mbox{A.\arabic{equation}}}
\setcounter{equation}{0}

In this appendix we present a complete proof of a bound on the remainder
term for the effective Hamiltonians derived from the circuit representation
of \cite{MMS}, which is missing in this reference. A proof of a similar
bound for a more general class of Hamiltonians will be given in \cite{Mes}.
The bound is essential to control the temperature dependence of the
effective Hamiltonians. Therefore, we present the proof here in considerable
detail, although we certainly do not claim that the proof is new:
the proof of Lemma \ref{lem:factorization} (given below) closely follows 
ideas of \cite{datferfro},
and, in general, our discussion follows the lines of \cite{MMS} and
\cite{UGM}.

\subsection*{A.1 Definitions}

The Hamiltonian defined on a finite lattice 
$\Lambda \subset {\szed}^d$, $d \geq 2$ is given by
\be
H_\Lambda = H_0 + tV -\mu_e N_e -\mu_i N_i,
\ee
where
\be 
H_0 = 2U\sum_{x\in \Lambda} W(x)  c^\dagger_x c_x
\label{hamt}
\ee
and
\be
V = - \sum_{\neighbours x y} c^\dagger_x c_y + h.c.,
\ee
where ${\neighbours x y}$ denotes a pair of nearest neighbour sites 
on the lattice. 
Let us restrict our attention to the {\em neutral case} at {\em{half-filling}},
i.e., $\mu_e=\mu_i=U$. 
In order to make 
the origin of various terms in the series expansions more transparent,
we do not set $t=1$ in this appendix, unlike in the main text.

The effective hamiltonian (for a given configuration $S=\{s_x\}$ of ions)
is defined through the relation
\be
\exp [-\beta H_{\rm{eff}}(\beta,S)]= {\rm{Tr}}_{{\cal{F}}_e} 
\exp [-\beta H_\Lambda]
\ee
where the trace is over the electronic Fock space (denoted by 
${\cal{F}}_e$).\\
Iterating Duhamel's formula
\begin{equation}
e^{-\beta ({H_0 + tV})}\;=\;
e^{-\beta H_0} \,+\, \int_0^\beta d\tau\,
e^{-(\beta-\tau) H_0} \,{(-tV)}\,
e^{-\tau (H_0 + tV)} \;,
\end{equation}
we obtain the Dyson series
\bea
e^{-\beta H_\Lambda} &:=& e^{\beta\,U(N_e + N_i)} \sum_{n=0}^\infty
\int_{0}^\beta d\tau_n\,\int_0^{\tau_n} d\tau_{n-1} \cdots \int_0^{\tau_2}
d\tau_1 \, e^{(\beta- \tau_n)H_0}\,(-tV)\nonumber\\
& & e^{-(\tau_n - \tau_{n-1})H_0} (-tV) \cdots e^{-(\tau_2 - \tau_1)H_0}
(-tV) e^{-\tau_1 H_0}
\eea
Hence
\bea
e^{-\beta H_\Lambda} &:=& e^{\beta\,U(N_e + N_i)} \sum_{n=0}^\infty t^n
\sum_{\neighbours{x_1}{y_1}, \dots, \neighbours{x_n}{y_n}} 
\int_{0}^\beta d\tau_n\,\int_0^{\tau_n}d\tau_{n-1} \cdots \int_0^{\tau_2}
d\tau_1 \, \nonumber\\
& &\qquad e^{(\beta- \tau_n)H_0}\,c^\dagger_{x_n}c_{y_n}
\cdots e^{-(\tau_2 - \tau_1)H_0} c^\dagger_{x_1}c_{y_1}e^{-\tau_1 H_0}
\eea
As in \cite{MMS} we introduce the set ${\cal{C}}(\Lambda)$ of (classical)
configurations associated to the electron subsystem. An element 
$X\in{\cal{C}}(\Lambda)$ is a finite sequence $X=(x_1,\ldots,x_m)$
of distinct sites in $\Lambda$. The state $|X\rangle \in {\cal{F}}_e$, associated
to $X \in {\cal{C}}(\Lambda)$ is defined as follows:
\be
|X> := \bigl[c^\dagger(x_1) \ldots c^\dagger(x_1)\bigr]_\preceq \, |0>,
\label{order}
\ee
where $|0> \in  {\cal{C}}(\Lambda)$ denotes the vacuum.

The symbol$\preceq$ denotes a total ordering of the sites in $\Lambda$, 
chosen to avoid ambiguities in the definition of the phase 
in \reff{order}. For convenience we choose the spiral order \cite{datferfro} 
for $d=2$ and an analogous ordering for $d\ge 3$. This ordering is chosen
to have the property that, for any finite set $X\subset \Lambda$, the
set $\{X'\} =\{x\in {\szed}^d; x\preceq X\}$ of lattice sites which
are smaller than $X$, or belong to $X$, is finite.

For a given sequence of pairs of nearest neighbour sites,
$(\neighbours{x_1}{y_1}, \dots, \neighbours{x_n}{y_n})$, 
let 
\be
\epsilon_j|X_j> = c^\dagger_{x_j}\,c_{y_j} \, |X_{j-1}> \quad {\hbox{for}} 
\quad 1\le j \le n,
\label{conda}
\ee
where each $\epsilon_j = \pm 1$. 
[Note that $|X_j| = |X_{j-1}|$ for $j=1,\ldots,n$.]\\
Setting $\tau_0=0$, $\tau_{n+1} =\beta$ and $X_0=X_n=X$ we obtain
\bea
{\rm{Tr}}_{{\cal{F}}_e}e^{-\beta H_\Lambda} &:=& e^{\beta\,U(N_e + N_i)} \sum_{n=0}^\infty t^n
\{\prod_{j=0}^n\langle X_j|e^{-(\tau_{j+1} - \tau_j) H_0}|X_j \rangle\}\,
\nonumber\\
&\times &\langle X|\bigl[\prod_{j=0}^n c^\dagger_{x_j}c_{y_j}\bigr]_\preceq |X\rangle
\label{trace}
\eea

Note that 
\be
H_0|X_j> = e(X_j) |X_j>,
\ee
where 
\be
e(X_j):= 2U \times \bigl[ {\hbox{number of sites in $X_j$ for which 
$W(x) = 1$,
(i.e., $s_x = 1$)}}\bigr]
\ee
As in \cite{MMS}, we introduce the notion of {\em{trajectories}}. 
Let $\tau_i$
be a positive integer variable (with $\tau_0=0$ and $\tau_{n+1}=\beta$) 
which we refer to as the ``time''. A trajectory $\zeta=\zeta(\tau_i)$ is a 
sequence $x(\tau_0),x(\tau_1),\ldots,x(\tau_{n+1})$ of sites in $\Lambda$ such 
that either
$$\tau_i \ne \tau_{i+1} \quad {\hbox{with}} \quad x(\tau_i)= x(\tau_{i+1})$$
or
$$\tau_i = \tau_{i+1} \quad {\hbox{with $x(\tau_i)$ and $x(\tau_{i+1})$
being nearest neighbour sites on the lattice.}}$$
This last case we describe as a {\em{jump}}. Let
\be
{\cal{J}}(\zeta):= \{ \langle x(\tau_i)\,x(\tau_{i+1})\rangle\,|\tau_i = \tau_{i+1};
x(\tau_i),\,x(\tau_{i+1})\in \zeta\}
\label{jump}
\ee
denote the set of jumps in the trajectory $\zeta$. Let ${\cal{T}}=\{\zeta\}$
denote a set of non--intersecting trajectories. If $g$ is a function on the trajectories, 
then we define a ``sum'' over sets of trajectories as follows:
\begin{eqnarray}
\intsum_{\zeta} g(\zeta) &\bydef& 1 \,+\, \sum_{n\ge 1} 
\, \sum_{(\llB_1,\ldots,\llB_n)} \, 
\sum_{(X_0,\ldots,X_n)\atop{X_i \in {\cal{C}}(\Lambda)}}
\indic[{\rm (i), (ii)}]\nonumber\\
&& {}\times
\int_{0}^\beta d\tau_n\, \cdots \int_0^{\tau_2}
d\tau_1 \,
g(\zeta)\;.
\label{intsum}
\end{eqnarray}
where $\llB_i =\{\langle x_i\,y_i\rangle\}$ denotes a set of nearest neighbour sites on the
lattice.  By $\indic [E]$ we mean the indicator function of the
event $E$; in particular, $\indic[{\rm (i), (ii)}]$ in \reff{intsum} vanishes
unless 
\begin{enumerate}
\item{the relation \reff{conda} holds}
\item{$X_0 = X_n$}.
\end{enumerate}
Hence the RHS of \reff{trace} can be expressed as a ``sum''over sets of non--intersecting
trajectories:
\be
\exp [{-\beta H_{\rm{eff}}(\beta,S)}]= e^{\beta U N_i} \intsum_{\cal{T}} e^{\beta\,U\,
|{\cal{T}}|} \, \epsilon({\cal{T}})\, \prod_{\zeta\in {\cal{T}}} e^{-2U|\zeta|}\,
\bigl( \prod_{\langle xy\rangle\in \zeta} t\bigr),
\label{eqtraj}
\ee
where $|{\cal{T}}|$ is the number of trajectories in ${\cal{T}}$, and $|\zeta|$ is 
defined as follows:
\be
|\zeta|:= \bigl({\hbox{vertical length of $\zeta$}}\bigr) \cap \bigl(\{x : s_x=+1\} \times
[0, \beta] \bigr),
\ee
and 
\be
\epsilon({\cal{T}}):= \langle X| {\bf{T}} \prod_{\zeta \in {\cal{T}}} 
\Bigl(\prod_{\langle x(\tau_i)\,x(\tau_{i+1})\rangle \in {\cal{J}}(\zeta)}
c^\dagger_{x(\tau_{i+1}} \, c_{x(\tau_i)}\Bigr)|X\rangle = \pm 1.
\ee
The symbol ${\bf{T}}$ denotes that the product is ``time-ordered'', and
$\epsilon({\cal{T}})$ denotes the sign of the permutation of the electrons
under the action of ${\cal{T}}$.

\subsection*{A.2 Circuit Representation}
To a given ion configuration $S$ and a given set of trajectories ${\cal{T}}$ we associate 
a set,  $\Omega$,  of oriented circuits as follows:

Vertical segments of trajectories located on sites $x \in \Lambda$ with $s_x =
+1$, will be considered as up-oriented components of circuits; vertical
segments of the complement  of the set of trajectories in $\Lambda \times
[0,\beta]$, located on sites with  $s_x = -1$, are considered as down-oriented
components of circuits.  On each horizontal bond at which a jump takes place,
we draw a segment with an arrow in the direction of the jump. The vertical
segments together with the horizontal  jumps form  oriented closed circuits.
More precisely, an {\em{oriented closed circuit}}, $\omega$, is a maximally
connected component of the oriented segments of the  trajectories. Let $\Omega
=\{\omega_1 \ldots \omega_n\}$ denote a finite set of such circuits. The space
of all circuits compatible with an ion configuration $S$ 
is denoted by $\mathcal W(S)$ [compatibility means that segments of circuits
are oriented upwards if they are located on sites occupied by ions
and are oriented downwards otherwise]. 

Let $|\omega|$ be the total length of the vertical segments of a circuit 
$\omega$, and let $p(\omega) \in \szed$ be the winding number of $\omega$. The 
latter can take positive or negative integer values, with negative values 
indicating that $\omega$ winds around the ``time''-axis $[0,\beta]$ with a 
downward orientation.

The following two relations hold:
\bea
\sum_{\tau \in \mathcal T} |\zeta| &=& \frac12 \sum_{\omega \in \Omega} 
\bigl( |\omega| + \beta p(\omega) \bigr) \label{one_a}\\
|\mathcal T| &=& |\Lambda| - N_i + \sum_{\omega \in \Omega} p(\omega)
\label{two}
\eea
As we are concerned with the half-filled case, we always have that
$\sum_{\omega\in\Omega} p(\omega) = 0$.

{From} \reff{eqtraj}, \reff{one_a} and \reff{two} we have that 
\be
e^{-\beta H_{\rm{eff}}(\beta,S)} = e^{\beta\,U\,|\Lambda|} 
\intsum_\Omega \epsilon(\Omega)
\,\prod_{\omega \in \Omega} e^{-U|\omega|}\bigl
(\prod_{\neighbours x y \in \omega} t\bigr),
\ee
where $\intsum_\Omega$ denotes a ``sum'' over all sets, $\Omega$, of 
non--intersecting closed circuits (defined analogous to \reff{intsum}) 
and $\epsilon(\Omega)$ is the sign of permutation of the electrons under 
the action of the circuits. The following lemma is crucial for the next
step. Its proof closely follows \cite{datferfro}, and ideas from
\cite{UGM}.

\begin{Lem}\label{lem:factorization}
There exists a function $\epsilon(\omega)$ such that $\forall \Omega$:
$$
\epsilon(\Omega) = \prod_{\omega \in \Omega} \epsilon(\omega).
$$
\end{Lem}

So we can write 
\be
e^{-\beta\heff(\beta,S)} = e^{\beta U |\Lambda|} \intsum_\Omega 
\prod_{\omega\in \Omega} z(\omega)
\ee
with the {\em{weight}}, $z(\omega)$, of a circuit, $\omega$, defined as follows:
\be
z(\omega) = \epsilon(\omega) e^{-U|\omega|} \, t^{j(\omega)},
\ee
where $j(\omega)$ denotes the number of jumps in $\omega$.

\noindent
{\bf Proof of Lemma \ref{lem:factorization}:}
Let $n_x(\tau_n)$ denote the number of electrons at the site $x$ at the
time $\tau_n$. Then the set of sites $x \in \Lambda$ for which
\be
W(x) + n_x(\tau_n) \ne 1
\label{number}
\ee
is said to define the {\em{defect set}}, $D_n(S)$, for a given ion
configuration $S$, at the time $\tau_n$.

The {\em{section}}, $\Gamma_n(\omega)$, of a circuit $\omega$, at a time 
$\tau_n$, is defined as follows:
\be
\Gamma_n(\omega) := \{x\in \bigl[\omega \cap ({\hbox{the plane}} \,\,
\tau=\tau_n)\bigr]\}.
\ee
In particular, $\Gamma_0(\omega)$ is referred to as the initial (``time-zero'')
section of the circuit $\omega$, since $\tau_0=0$.
It is clear that each $x \in \Gamma_n(\omega)$ belongs to the 
defect set $D_n(S)$ (for $n\ge 0$).
	
Let $|\Gamma_n(\omega)>$ denote the state in the electron Fock space, 
${\cal{F}}_e$, defined as follows:
\be
|\Gamma_n(\omega)> := \bigl(\prod_{x\in \Gamma_n(\omega)\atop{s.t. \, 
s_x=1}}c^\dagger_x\bigr)_{\preceq} |0>,
\label{gamman}
\ee
where $|0>$ denotes the vacuum. 
More generally, for a set, $\Omega$, of circuits, we define the section
at time $\tau_n$ to be 
\be
\Gamma_{n}(\Omega) := \{x\in \bigl[\Omega \cap ({\hbox{the plane}} \,\,
\tau=\tau_n)\bigr]\}
\ee
and the corresponding state to be
\be
|\Gamma_n(\Omega)> := \bigl(\prod_{x\in \Gamma_n(\Omega)\atop{s.t. \, s_x=1}}
c^\dagger_x\bigr)_{\preceq} |0>,
\ee
The following relations hold:
\smallskip

\noindent
(a)
\be
|\Gamma_i(\omega)> = \uphi{i}\,|\Gamma_{i-1}(\omega)> \quad {\hbox{for}} \quad 1 \le i\le n,
\label{sec1}
\ee
where $\llB_i =\{<x_i\,y_i>\}$ denotes a finite set of pairs of nearest
neighbour sites, and $\uphi{i}$ denotes the corresponding set of
operators
\be
\uphi{i} := \{c^\dagger_{x_i}\,c_{y_i}\}.
\ee
(b)
\be
\Gamma_0(\omega) = \Gamma_n(\omega)
\label{pdc}
\ee
The above relation is due to the periodicity in the ``time'' direction.

Each circuit $\omega$ is uniquely determined by the following
\begin{itemize}
\item{its initial section $\Gamma_0(\omega)$ (and hence on the state 
$|\Gamma_n(\omega)>$);}
\item{a sequence of operators $\uphi{i}, \ldots,\uphi{n}$;}
\item{a sequence of ``times'' $\tau_1, \ldots, \tau_n$ at which these operators
act.}
\end{itemize}
The sign $\epsilon(\omega)$ of a circuit $\omega$ is given by
\bea
\epsilon(\omega)&=& \prod_{i=1}^n <\Gamma_i(\omega)|\uphi{i}|\Gamma_{i-1}
(\omega)>\nonumber\\
&=& <\Gamma_0(\omega)|\uphi{n}\cdots \uphi{1}|\Gamma_0(\omega)>
\label{sign1}
\eea
The second line follows from (a) and (b) above.

To prove the
factorization property it is enough to consider a set, $\Omega$, of
circuits (compatible with an ion configuration $S$) that can be
divided into two mutually non-interesecting (``time''-periodical) 
subsets, denoted by $\omega_B$ and
$\omega_C$. Each subset can be made up of several circuits.  
To each of $\Omega$, $\omega_B$ and
$\omega_C$ is associated a sign. 
 
The sign of $\omega_B$ is given by
\be
\epsilon(\omega_B) = <\Gamma_0(\omega_B)|\uphi{n}\cdots \uphi{1}|
\Gamma_0(\omega_B)>
\label{sign2}
\ee

The sign, $\epsilon(\omega_C)$, of the component $\omega_C$ is defined in an
analogous fashion, with the sequence of operators $\uphi{n}\cdots \uphi{1}$
replaced by $\cphi{m}\cdots \cphi{1}$. The set $\Omega$ is defined by an 
initial section 
$\Gamma_{0D} (= \Gamma_0(\omega_D))$ and a sequence of operators
$\Phi_{{\llD}_{1}},\ldots,\Phi_{{\llD}_{n+m}}$, which is a
uniquely determined permutation of the sequence
$\Phi_{{\llB}_{1}},\ldots,\Phi_{{\llB}_{n}},
\Phi_{{\llC}_{1}},\ldots,\Phi_{{\llC}_{m}}$.
Its sign $\epsilon(\Omega)$ is also defined by \reff{sign2} with the
obvious changes.
We need to
show that the sign for $\Omega$ factorizes, i.e.,
\begin{equation}
\epsilon(\Omega) \;=\; \epsilon(\omega_B) \, \epsilon(\omega_C)\;.
\label{vg.1}
\end{equation}
By iteration of the argument we obtain the
factorization into the signs of the individual circuits.  

\medskip

Let us first discuss the computation of the sign of a single component
$\omega_B$.  
In terms of the vacuum $|0\rangle$, the expression of the
sign $\epsilon(\omega_B)$
takes a very simple form.  To obtain it, we
observe that
\begin{equation}
\langle \Gamma_{0B} 
|\sphiq{n}\cdots\sphiq{1}| \Gamma_{0B} \rangle \;=\;
\langle 0| \,{\bf C}_{\Gamma_{0B}}\, \sphiq{n}\cdots\sphiq{1} 
\,{\bf C}^*_{\Gamma_{0B}} \,|0\rangle\;,
\label{v.20}
\end{equation}
where ${\bf C}^*_{\Gamma_{0B}}$ is a product of creation
operators that creates the section $\Gamma_{0B} 
\bigl(= \Gamma_0(\omega_B)\bigr) $, i.e.,
\be
{\bf C}^*_{\Gamma_{0B}}:=
\bigl(\prod_{x\in \Gamma_{0B}\atop{s.t. \, s_x=1}}c^\dagger_x\bigr)_{\preceq}.
\ee

The product 
\begin{equation}
{\bf A}_B \;\bydef\; {\bf C}_{\Gamma_{0B}}\, \uphi{n}\cdots\uphi{1} 
\,{\bf C}^*_{\Gamma_{0B}}
\label{zetab}
\end{equation}
is a monomial in creation and destruction operators that 
can be combined into {\em number}\/ operators, because of
the required periodicity,
\begin{equation}
{\bf A}_B\,|0\rangle \;=\; \epsilon(\omega_B) \,|0\rangle \;,
\label{aab}
\end{equation}
of $\omega_B$.  The factorization of
signs is a consequence of the fact that this combination can be made
in a well-defined fashion which is not affected by the presence of
other (compatible) circuits.  

We choose the following procedure
which we refer to as {\em circuit collapsing}\/. Consider the string 
of operators appearing in the product ${\bf A}_B$. For brevity we shall
refer to each appearance of a destruction or creation operator supported
on a site $x$ as an {\em occurrence}\/ of $x$.
We denote by ${\cal S}({\bf A}_B)$ (``shadow'' of
${\bf A}_B$) the set of sites $x$ occurring in ${\bf A}_B$.
We start with the leftmost operator in the product ${\bf A}_B$ 
\reff{zetab}. In order to yield a non-zero contribution this has
to be a destruction operator with support on some site $x$, i.e.,
$c_{x}$. We now move this operator through the operators present 
to its right (i.e., downwards in time), using the anticommutation relations, until we encounter the 
next occurrence of the site 
$x$ in the product \reff{zetab}. This is necessarily a 
creation operator, $c^\dagger_{x}$, since otherwise the successive
actions of these two operators would yield zero. Hence we obtain a factor
\begin{equation}
{\bf c}_x\,{\bf c}^\dagger_x
\;=\;{\bf 1} - {\bf n}_x
\label{cd*}
\end{equation}
times a phase $\alpha_x^{(1)}(\omega_B)$ (${\bf n}_x$ being the number operator
for electrons at the site $x$). This phase arises 
due to the anticommutation of the initial ${\bf c}_x$ with 
intermediate operators. All the operators which appear to the left of 
this factor are {\em not}\/ supported in $x$ and, hence, we can
move this factor to the leftmost end of the string to obtain
\begin{equation}
{\bf A}_B \;=\; \alpha_x^{(1)}(\omega_B)\, ({\bf 1} - {\bf
n}_x)\, \widehat{\bf A}_B\;,
\label{ahatb}
\end{equation}
where $\widehat{\bf A}_B$ satisfies
\begin{equation}
\widehat{\bf A}_B\,|0\rangle \;=\; \alpha_x^{(1)}(\omega_B)\,
\,\epsilon (\omega_B) \,|0\rangle \;,
\label{ahab}
\end{equation}
and has two fewer occurrences of $x$.

Next, we repeat the above procedure for the string of operators
defining $\widehat{\bf A}_B$ and continue
pulling out, in the same way,
successive phases and factors ${\bf 1} - {\bf n}_x$.   
Once all occurrences of $x$ have been dealt with, 
we obtain a product of factors 
$({\bf 1} - {\bf n}_x)^k={\bf 1} - {\bf n}_x$
and an overall phase $\alpha_x(\omega_B)$, so that
\begin{equation}
{\bf A}_B \;=\; \alpha_x(\omega_B)\, 
({\bf 1} - {\bf n}_x)\, 
\widetilde{\bf A}_B\;,
\label{ahatbx}
\end{equation}
where $\widetilde{\bf A}_B$ satisfies
\begin{equation}
\widetilde{\bf A}_B\,|0\rangle \;=\; \alpha_x(\omega_B)\,
\epsilon (\omega_B)\,|0\rangle \;,
\label{aahab}
\end{equation}
and has {\em no}\/ occurrence of the site $x$:
${\cal S}(\widetilde{\bf A}_B) = {\cal S}({\bf A}_B)\setminus
\{x\}$.

We then repeat the whole procedure for $\widetilde{\bf A}_B$.  At the
end, once all the sites $x$ in ${\cal S}({\bf A}_B)$ have been
exhausted, one 
obtains that the whole of ${\bf A}_B$ has ``collapsed'' into factors 
${\bf 1} - {\bf n}_x$ times a numerical factor.  A simple
replacement in \reff{aab} shows that this factor must equal 
$\epsilon (\omega_B)$.  That is,
\begin{equation}
{\bf A}_B\;=\;
\epsilon (\omega_B)\, \prod_{x\in {\cal S}(\omega_B)} 
({\bf 1} - {\bf n}_x)\;.
\label{v.25}
\end{equation}

\medskip

We are now ready to prove the factorization property that we need:  
Let $\Omega$ be a family of circuits and let it be
decomposed into two (``time-periodical'') subfamilies of circuits,
$\omega_B$ and $\omega_C$. Then 
\be
\epsilon (\Omega) \;=\; \epsilon (\omega_B)\, \epsilon (\omega_C)\;.
\label{v.40}
\ee

We can write $\omega_C$ via an operator
\begin{equation}
{\bf A}_C \;\bydef\; {\bf C}_{\Gamma_{0C}}\, \Phi_{\llC_m}
\cdots\Phi_{\llC_1} \,{\bf C}^*_{\Gamma_{0C}}\;.
\label{zetac}
\end{equation}
[with $\Gamma_{0C} = \Gamma_0(\omega_C)$],
such that
\begin{equation}
{\bf A}_C\,|0\rangle \;=\; \epsilon (\omega_C) \,|0\rangle \;.
\label{aac}
\end{equation}

We shall use the
following two consequences of the compatibility (i.e., mutual non-intersection)
of the two components
$\omega_B$ and $\omega_C$:
\begin{itemize}
\item[{\bf (C1)}]  The monomials 
${\bf C}^*_{\Gamma_{0B}}$ and ${\bf C}^*_{\Gamma_{0C}}$ have disjoint
support, so that 
\begin{equation}
|\Gamma_{0D}\rangle \;=\; \delta\, {\bf C}^*_{\Gamma_{0B}}\,
{\bf C}^*_{\Gamma_{0C}}\, |0\rangle\;,
\label{v.30}
\end{equation}
where $\delta$ is a phase factor.  Hence, 
\begin{equation}
\epsilon (\Omega)  \;=\; \langle 0 |\,{\bf A}_D\,|0\rangle 
\label{aad}
\end{equation}
with
\begin{equation}
{\bf A}_D \;\bydef\; {\bf C}_{\Gamma_{0C}}\, {\bf C}_{\Gamma_{0B}}
\Phi_{\llD_{n+m}} \cdots\Phi_{\llD_1} 
\,{\bf C}^*_{\Gamma_{0B}}\,{\bf C}^*_{\Gamma_{0C}}\;.
\label{zetad}
\end{equation}
\item[{\bf (C2)}] 
The occurrence of a creation operator ${\bf c}^\dagger_x$ in
factors of ${\bf A}_B$ implies
that the site $x$ becomes, or continues to be, part of
the support of $\omega_B$ {\em at least until}\/ there is a further
occurrence of a destruction operator ${\bf c}_x$.  In
particular since $\omega_B$ and $\omega_C$ do not intersect, 
we have the following property:
\begin{equation}
\begin{minipage}{305pt}
{\em Between an occurrence of ${\bf c}_x$ in factors of ${\bf
A}_B$ and the preceding (ie.\ immediately to the right) occurrence of 
${\bf c}^\dagger_x$ in factors of ${\bf A}_B$, there cannot be an 
occurrence of $x$ in factors of ${\bf A}_C$.}
\label{nooc}
\end{minipage}
\end{equation}
\end{itemize}
Moreover, as the whole set $\Omega$ is periodic in the ``time''-direction, 
we have that
each occurrence of ${\bf c}_x$ in ${\bf A}_D$ must be
preceded by an occurrence of ${\bf c}^\dagger_x$.  Combining this
observation with {\bf (C2)} we obtain the last property needed:
\begin{itemize}
\item[{\bf (C3)}]  
Between an occurrence of ${\bf c}^\dagger_x$ in factors of ${\bf
A}_B$ and the immediately preceding occurrence of ${\bf c}_x$
in factors of ${\bf A}_B$, there is an even number of occurrences of
$x$ in factors of ${\bf A}_C$.  Of course, these occurrences
correspond to alternating creations and destructions.
The same property holds after the last occurrence of 
${\bf c}_x$ in factors of ${\bf A}_B$ and before the
first occurrence of ${\bf c}^\dagger_x$ in factors of ${\bf A}_B$.
\end{itemize}

{From} {\bf (C1)}--{\bf (C3)} we conclude that the ``collapse'' of ${\bf
A}_B$ gives {\em exactly}\/ the same factor as in the absence of
the component $\omega_C$.  Indeed, the last occurrence of a site 
$x$ in factors of ${\bf A}_B$ is a destruction operator, that we
can displace up to the previous occurrence to produce a factor  
${\bf 1} - {\bf n}_x$.  Between these two occurrences there is
{\em no}\/ occurrence of $x$ in factors of ${\bf A}_C$ because 
of {\bf (C2)}. Hence  ${\bf c}_x$ commutes with all the operators 
$\Phi_{\llC_i}$ encountered during the displacement (recall that
such operators are monomials of {\em even}\/ degree).  Thus, the
phase acquired during this displacement only depends on the operators
in ${\bf A}_B$ and hence it is the same phase 
$\alpha_x^{(1)}(\omega_B)$ obtained when collapsing the
component $\omega_B$ in the absence of any other circuit.
Moreover, by {\bf (C3)} the operators in ${\bf A}_D$ located to the
left of the factor ${\bf 1} - {\bf n}_x$, obtained in the above 
manner, involve an even number of creation and destruction 
operators supported in $x$.  Therefore, we can freely move this factor 
${\bf 1} - {\bf n}_x$ all the way to the left to obtain
\begin{equation}
{\bf A}_D \;=\; \alpha_x^{(1)}(\omega_B)\, ({\bf 1} - {\bf
n}_x)\, \widehat{\bf A}_D\;,
\label{ahatd}
\end{equation}
where $\widehat{\bf A}_D$ has two fewer occurrences of $x$ in
factors of ${\bf A}_B$ but otherwise satisfies {\bf (C1)}--{\bf (C3)}.
Iterating this process we collapse ${\bf A}_B$ exactly as done in
\reff{ahatb}--\reff{v.25}.  We obtain
\begin{equation}
{\bf A}_D\;=\; \epsilon (\omega_B)\, \prod_{x\in {\cal S}(\omega_B)} 
({\bf 1} - {\bf n}_x) \, {\bf A}_C\;.
\label{v.25d}
\end{equation}

Combining this expression with \reff{aad} and \reff{aac} we get the
desired factorization \reff{v.40}.  
\QED

\subsection*{A.3 Cluster expansion}

The logarithm of 
$$\intsum_\Omega 
\prod_{\omega\in \Omega} z(\omega)$$
can be developed in a cluster expansion.
Adapting Theorem 3.1 of \cite{pfister} to our case (in which one variable, 
the ``time'', 
is continuous) we have the following result:
\be
{\hbox{if}} \intsum_{\omega \in {\mathcal W}_j(S)} |z(\omega)| e^{|\omega|} 
\chi[{(x,\tau) \in \omega}] \leq C_j \hspace{5mm} {\rm{with }} 
\hspace{5mm} \sum_{j \geq 0} C_j < 1
\label{conv}
\ee
(where $\chi$ denotes the characteristic function)
then we have an absolutely convergent cluster expansion given by
\be
\intsum_\Omega \prod_{\omega \in \Omega} z(\omega) = \exp\Bigl\{ \sum_{n \geq 1} \frac1{n!} 
\intsum_{\omega_1 \ldots \omega_n} \varphi_n^T(\omega_1, \ldots, \omega_n) 
\prod_{k=1}^n z(\omega_k) \Bigr\}.
\ee

Here $\mathcal W_j(S)$ is the space of circuits compatible with the 
ion configuration $S$ and having $j$ jumps, and $\varphi_n^T(\omega_1, 
\dots, \omega_n)$ is a 
combinatoric function on families of circuits 
whose value is zero 
whenever $\{ \omega_1, \dots, \omega_n \}$ is not a {\em{cluster}} 
(i.e., whenever the support of the set of 
circuits $\{ \omega_1, \dots, \omega_n \}$
is not connected in ${\bf{R}}^{d+1}$).

Furthermore, we shall use the following bound, \reff{lemext}, which is an extension of Lemma 3.5 
of \cite{pfister}:
\be
\intsum_{\omega_1 \in {\mathcal W}_{j_1}(S)} \intsum_{\omega_2 \in {\mathcal W}_{j_2}(S)} 
\cdots \intsum_{\omega_n \in {\mathcal W}_{j_n}(S)}
\chi[{(x,\tau)\in \omega_1}] |\varphi_n^T(\omega_1, \dots, \omega_n)|  
\prod_{k=1}^n |z(\omega_k)| \leq (n-1)! \prod_{k=1}^n C_{j_k}.
\label{lemext}
\ee
We show that \reff{conv} holds in our case as follows.
For $j\ne 0$, (i.e., $j \ge 2$) we have
\bea
\intsum_{\omega \in {\mathcal W}_j(S)} \, |z(\omega)| e^{|\omega|} \chi[{(x,\tau)\in\omega}] 
&\le& \intsum_{\omega \in {\mathcal W}_j(S)}\,t^j e^{-(U-1)|\omega|} \chi[{(x,\tau)\in\omega}] 
\nonumber\\
&\le& (2dt)^j \bigl[ \int_0^\beta d\tau e^{-(U-1)\tau} \bigr]^j \nonumber\\
&\le& (2dt)^j \Bigl(\frac{1 - e^{-\beta(U-1)}}{U-1} \Bigr)^j
\label{Ularge}
\eea
For $U$ large enough, $U-1 \ge c\,U$, for some constant $c$ with $0<c<1$. Hence
\be
\mbox{RHS of } \ \reff{Ularge} \le C_j
\ee
where we define
\be
C_j := \Bigl(\frac{2dt}{cU}\Bigr)^j \quad \quad {\hbox{for $j\ge 2$}}.
\label{Cj}
\ee
For $j=0$ we find
\be
C_0 = e^{-\beta c U}
\label{C0}
\ee
{From} \reff{Cj} and \reff{C0} it follows that the bound \reff{conv} is 
satisfied for $U$ and $\beta$ large enough.

Hence we obtain an expression for the effective hamiltonian:
\be
\heff(\beta,S) = - U|\Lambda| - \frac1\beta \sum_{n\geq1} \frac1{n!} 
\intsum_{\omega_1 \ldots \omega_n} \varphi^T_n(\omega_1,\dots,\omega_n) 
\prod_{k=1}^n z(\omega_k).
\ee

We define the support, ${\rm{supp}}\,\omega$, of a circuit $\omega$, to be its 
orthogonal projection onto the plane $\tau=0$. Hence ${\rm{supp}}\,\omega 
\subset \Lambda$.

Let $A\subset \Lambda$. 
The potential $\Phi_A(\beta,S_A)$ is introduced as follows:
\be
\Phi_A(\beta,S_A) = -\frac1\beta \sum_{n\geq1} \frac1{n!} 
\intsum_{\omega_1 \ldots \omega_n} 
\chi[{ \bigcup_{k=1}^n {\rm{supp}}\,\omega_k = A}] 
\varphi^T_n(\omega_1,\dots,\omega_n) 
\prod_{k=1}^n z(\omega_k),
\ee
where $S_A$ denotes the restriction of the ion configuration 
$S$ to $A \subset \Lambda$. 
\begin{Lem}\label{expdec}
There exists positive constants $U_0 >> t$ and $\beta_0$, such that for all 
$U>U_0$ and $\beta >\beta_0$
\be
|\Phi_A(\beta,S_A)|\le c_1\, \frac{(c_2t)^{n(A)}}{U^{n(A)-1}} 
\hspace{3mm} {\hbox{for $|A| \geq 2$}},
\label{lemstate}
\ee
for some constants $c_1$ and $c_2$, with $n(A)$ being the minimum length
of a closed path which passes through all sites of $A$.
\end{Lem}

\begin{proof}
Let $j_1, \dots, j_n$ denote the number of jumps for the circuits 
$\omega_1, \dots, \omega_n$, such that $\{ \omega_1, \dots, \omega_n \}$ 
forms a cluster with support equal to $A$.
\smallskip

\noindent 
If $\bigcup_{k=1}^n \supp\,\omega_k = A$ then $\sum_{k=1}^n j_k \ge n(A)$
for any $\omega_1,\dots,\omega_n$, then
\bea
|\Phi_A(\beta,S_A)| &\le& \frac1\beta \sum_{n\ge 1} \frac{1}{n!} 
\sum_{{j_1, \ldots, j_n \ge 0}\atop{\sum_k j_k \ge n(A)}} 
\intsum_{\mathcal \omega_1\in W_{j_1}(S)} \cdots
\intsum_{\mathcal \omega_n\in W_{j_n}(S)} \nonumber\\
& &\chi[{ \bigcup_{k=1}^n {\rm{supp}}\omega_k = A}] 
|\varphi_n^T(\omega_1,\dots,\omega_n)| \prod_{k=1}^n |z(\omega_k)| .
\label{bound_a}
\eea
Now,
$$\sum_{x \in A} \int_0^\beta d\tau \, \chi[{(x,\tau)\in\omega_1}] 
\frac1{|\omega_1|}= 1,
$$ 
for any $\omega_1$ with ${\rm{supp}}\,\omega_1 \subset A$.
Introducing this identity into the above equation \reff{bound_a} yields
\bea
|\Phi_A(\beta,S_A)| &\le& \frac1\beta \sum_{x \in A} \sum_{n\ge 1} \frac1{n!} 
\sum_{{j_1, \ldots, j_n \ge 0}\atop{\sum_k j_k \ge n(A)}} 
\int_0^\beta d\tau 
\intsum_{\omega_1 \in {\mathcal W}_{j_1}(S)} 
\chi[{(x,t) \in \omega_1}] \frac1{|\omega_1|} \nonumber\\
&&\times\intsum_{\omega_2\in {\mathcal W}_{j_2}(S)} \ldots 
\intsum_{\omega_n\in {\mathcal W}_{j_n}(S)} 
|\varphi_n^T(\omega_1,\dots,\omega_n)| \prod_{k=1}^n |z(\omega_k)|.
\label{final}
\eea
Note that $|A|\ge 2$ implies that $j_i\ge 2$ for at least one 
$i\in\{1\ldots n\}$. We have assumed that $j_1 \ge 2$.

Let us define $z'(\omega,U)$ to be equal to the weight $z(\omega)$, but with 
explicit dependence on the coupling constant $U$. Let us define
\bea
F(U_1, \dots, U_n) &:=& \intsum_{\omega_1 \in {\mathcal W}_{j_1}(S)} 
\chi[{(x,t) \in \omega_1}] 
\frac{|z'(\omega_1,U_1)|}{|\omega_1|} 
\intsum_{\omega_2\in {\mathcal W}_{j_2}(S)} \ldots 
\intsum_{\omega_n\in {\mathcal W}_{j_n}(S)} 
\nonumber\\
&&\times |\varphi_n^T(\omega_1,\dots,\omega_n)|
\prod_{k=2}^n |z'(\omega_k,U_k)|
\eea
and obtain an upper bound for this function. Eventually, we shall set
$U_i=0$ for all $i=1\ldots n$.

Note that $F(\infty, \dots, U_n) = 0$. Moreover,
$$
\frac\partial{\partial U_1} |z'(\omega_1,U_1)|=-|\omega_1| |z'(\omega_1,U_1)|.
$$
Hence,
\bea
|\frac\partial{\partial U_1} F(U_1, \dots, U_n)| &\le& 
\intsum_{\omega_1 \in {\mathcal W}_{j_1}(S)} 
\chi[{(x,t) \in \omega_1}] 
\intsum_{\omega_2\in {\mathcal W}_{j_2}(S)} \ldots 
\intsum_{\omega_n\in {\mathcal W}_{j_n}(S)} 
\nonumber\\
&&\times |\varphi_n^T(\omega_1,\dots,\omega_n)|
\prod_{k=1}^n |z'(\omega_k,U_k)|
\eea

{From} \reff{lemext} we have that
\be
|\frac\partial{\partial U_1} f(U_1, \dots, U_n)| \leq (n-1)! 
\prod_{k=1}^n C'_{j_k} (U_k)
\ee
where $C'_{j_k} (U_k)$ is the same as $C_{j_k}$, except for the explicit 
dependence on $U_k$. Since 
$$
F(U_1, \dots, U_n) = -\int_{U_1}^\infty \frac\partial{\partial V} F(V, 
\dots, U_n) dV
$$
we have that 
\bea
F(U_1, \dots, U_n) &\le& \int_{U_1}^\infty |\frac\partial{\partial V} F(V, 
\dots, U_n)| dV\nonumber\\ 
&\le& (n-1)! \prod_{k=2}^n C'_{j_k}(U_k) 
\int_{U_1}^\infty C'_{j_1}(V) dV
\eea
{From} \reff{Cj} we have that
\bea
\int_{U_1}^\infty C'_{j_1}(V) dV &\le& (2d t)^{j_1} 
\int_{U_1}^\infty (cV)^{-j_1} \, dV \nonumber\\
&\le& \Bigl[\frac{2dt}{c}\Bigr]^{j_1}\,\frac{1}{U_1^{j_1 - 1}},
\eea
(since $j_1 \ge 2$).
Hence,
\be
|F(U_1, \dots, U_n)| \le (n-1)! 
\frac{2dt}{c} \Bigl[\frac{2d t}{cU_1} \Bigr]^{j_1-1}
 \prod_{{k=2}\atop{j_k \ne 0}}^n \Bigl[ \frac{2d t}{c U_k} \Bigr]^{j_k} 
\prod_{{k=2}\atop{j_k = 0}}^n C'_0(U_k).
\ee
Let $i$ be the number of circuits without any jump. From \reff{final} we 
we have that
\bea
|\Phi_A(\beta,S_A)| &\le& \frac{1}{\beta} \sum_{x\in A}  \sum_{n\geq 1}
\frac{1}{n!} \sum_{{j_1, \ldots, j_n \geq 0}\atop{\sum_k j_k \ge n(A)}} 
\int_0^\beta |F(U_1 \ldots U_n)|\nonumber\\
&\le& 
|A| \sum_{n\ge 1} \frac{1}{n} \sum_{i=0}^n \frac{n!}{(n-i)! i!} C_0^i 
\sum_{m=n(A)}^\infty 
\sum_{{j_1, \dots, j_{n-i} \ge 2}\atop{\sum_k j_k = m}} (2d t/c) \Bigl[
\frac{2d t}{cU} \Bigr]^{m-1}.
\label{theeq}
\eea
Further, we have that 
\be
\sum_{{j_1, \dots, j_r \ge 2}\atop{\sum_k j_k = m}} 1 \leq 2^{m-r}.
\label{estim}
\ee
It is true for all $j$ when $m=1$, and by induction
$$
\sum_{{j_1, \ldots, j_{r+1} \ge 2}\atop{\sum_k j_k = m}} 1 = \sum_{j_{r+1} 
\ge 2}\sum_{{j_1, \ldots, j_r \ge 2}\atop{\sum_k j_k = m-j_{r+1}}} 1 \leq 
2^{m-r}
\sum_{j_{r+1} \ge 2} 2^{-j_{r+1}} = 2^{m-(r+1)}.
$$
Using the bound \reff{estim} on the RHS of \reff{theeq}, we obtain
\bea
|\Phi_A(\beta,S_A)|
&\le& |A| \sum_{n\geq 1} \frac1n \sum_{i=0}^n \frac{n!}{(n-i)! i!} 
C_0^i 2^{-(n-i)}\sum_{m=n(A)}^\infty 2^m\,\frac{(2d t/c)^m}{U^{m-1}}
\nonumber\\
&=& |A|\Bigl[\sum_{n \geq 1} \frac1n \bigl( \frac12 + C_0 \bigr)^n\Bigr]\,
 \sum_{m \ge n(A)} \frac{(4d t/c)^m}{U^{m-1}}
\label{above}
\eea

Now, since $C_0 = e^{-\beta c U}$ [\reff{C0}], there exists positive constants
$U_0 >>t$ and $\beta_0$, such that for all $\beta> \beta_0$ and $U>U_0$ the
sums over $n$ and $m$ on RHS of \reff{above} converge. Thus we obtain the bound
\reff{lemstate} for some positive constants $c_1$ and $c_2$.
\end{proof}
\smallskip

\noindent
{\em{Remark}}: For $A \subset \Lambda$ we have that \reff{geebee} 
\be
g(A) = n(A) - 1
\ee
Set $t=1$. Then the bound \reff{lemstate} can be written as
\be
|\Phi_A(\beta,S_A)|
\le C_1 \Bigl(\frac{C_2}{U}\Bigr)^{g(A)},
\label{used_bound}\ee
for some positive constants $C_1$ and $C_2$, with $C_2 / U < 1$. 
This is our desired estimate. 

\section*{Appendix B: Proof of Lemma \ref{convergence}}
\label{appendixB}
\renewcommand{\thetheorem}{\mbox{A.\arabic{Thm}}}
\renewcommand{\theequation}{\mbox{B.\arabic{equation}}}
\setcounter{equation}{0}

{From} the standard results of cluster expansions \cite{pfister, seiler, 
BLPO,BLP,dobclu} it follows that a sufficient condition for the convergence 
of the series given in \reff{series2} is given by
\begin{equation}
\sum_{D \ni 0} W(D) e^{|D|} < 1, \label{cond}
\end{equation}
Hence, the task of proving Lemma \ref{convergence} amounts to proving
that the condition \reff{cond}  is satisfied. In order to do so, we consider
an auxiliary polymer system whose elements (the {\em{polymers}}) are denoted
by $\p$ and defined as follows.
\begin{equation}
\p := (\gamma, M_{\gamma})
\end{equation}
where $\gamma$ is a contour and $M_{\gamma}$ is a {\em{decoration}} of
$\gamma$. A decoration of $\gamma$ is a (possibly empty) set of bonds
$\{B_1,\ldots,B_j\}\subset {\cal{B}}$, such that each bond intersects
$\gamma$, i.e., $B_i\cap\gamma\neq\emptyset$ for $1 \le i \le j$. 
Hence a polymer $\p$ consists of a contour $\gamma$ and a 
finite set of bonds which intersect its support.
Let
${\cal{B}}_{\gamma}$ denote the set of {\em{all}} bonds which intersect  the
contour $\gamma$. Then $M_{\gamma} \subset {\cal{B}}_{\gamma}$. For a polymer
$\p=(\gamma,M_{\gamma})$, we define
\begin{equation}
|\p|:= |\gamma| + \sum_{B \in M_{\gamma}} g(B).
\label{pee}
\end{equation}
 The weight of a polymer is 
given by
\begin{equation}
w(\p) := e^{-\beta E(\gamma)} \prod_{B \in {{M}}_\gamma} 
(e^{-\beta G_B}-1),
\end{equation}
and satisfies the bound $|w(\p)| \le w_0(\p)$ where
\begin{equation}
w_0(\p) := e^{-\beta C_1 \lambda |\gamma|} 
\prod_{B \in {M}_{\gamma}} (e^{\beta C_2 \lambda^{g(B)}}-1).
\label{bound4}
\end{equation}
Each decorated contour $D$ can be considered to be the union of a finite 
number of intersecting polymers, i.e., a connected {\em{cluster of polymers}}.
The weight of a decorated contour can then be expressed in terms of the
weights of its constituent polymers.
The decomposition of a decorated contour into polymers 
is however not unique.  The condition \reff{cond} for the
model of decorated contours can be  transcribed into a condition for the
auxiliary polymer system, by making use of the ``tree-graph approximation''
used in cluster expansions  \cite{seiler, pfister}. We sketch the idea below,
following \cite{pfister}. We first bound the LHS of \reff{cond} in terms of a
sum over polymers:
\begin{equation}
\sum_{D \ni 0} W(D) e^{|D|} \le \sum_{n \ge 1} \frac{1}{n!} \sum_{{\p_1, 
\ldots , \p_n} \atop{\p_1 \cup \ldots \cup \p_n = D \atop {D \ni 0, 
{\rm{connected}}}}} \prod_{j=1}^n w(\p_j) e^{|\p_j|}.
\label{bound1}
\end{equation}
The fact that \reff{bound1} is not an equality but only a bound is due to the
non-uniqueness of decomposition of a decorated contour into polymers. A
decorated contour $D$ consisting of $n$ polymers, $\p_1, \ldots , \p_n$,  can
be represented by a connected, oriented graph, whose vertices are the 
polymers and the lines between pairs of vertices corresponding to 
intersecting polymers. The graph is oriented by introducing an ordering of
the  vertices. Let ${\cal{G}}_n$ be the corresponding  {\em{complete graph}},
i.e., the graph with $n$ vertices, with a line between each pair of vertices.
The sum over the polymers   on the RHS of \reff{bound1} can be bounded by a sum
over all tree graphs of  the corresponding complete graph. Hence, we can write
\begin{equation}
\sum_{D \ni 0} W(D) e^{|D|} \le \sum_{n \ge 1} \frac{1}{n!} 
\sum_{T \subset {\cal{G}}_n}
\sum_{\p_1 \ni 0} \cdots \sum_{\p_n}  \prod_{j=1}^n w(\p_j) e^{|\p_j|},
\end{equation}
where $T$ denotes a tree graph. 

{From} Lemma 3.5 of \cite{pfister} it follows that 
\begin{equation}
\sum_{T \subset {\cal{G}}_n}
\sum_{\p_1 \ni 0} \cdots \sum_{\p_n}  \prod_{j=1}^n w(\p_j) e^{|\p_j|}
\le (n-1)! C^n,
\label{pfist0}
\end{equation}
where
\begin{equation}
C:=  \sum_{\p \ni 0} w_0(\p) e^{2|\p|},
\end{equation}
with $w_0(\p)$ being defined through \reff{bound4}. 
Hence,
\begin{equation}
\sum_{D \ni 0} W(D) e^{|D|} \le \sum_{n\ge1} \frac{C^n}{n}.
\label{bound2}
\end{equation}
The series on the RHS of \reff{bound2} converges to $C/(1-C)$ if $C<1$, which
is $<1$ if  $C <1/2$. Hence the proof of \reff{cond}, and hence of
Lemma \ref{convergence}, is completed by proving the following lemma. 
\begin{Lem}
\label{poly}
For each $C'>0$, there exist constants $\lambda_0$ and $b_0$ such that for all
$\lambda < \lambda_0$ and $\beta \lambda > b_0$ one has the bound
\begin{equation}
\sum_{\p\ni 0} w_0(\p) e^{2\vert \p\vert} \leq C',
\end{equation}
where $w_0(\p)$ is given by \reff{bound4}.
\end{Lem}
\begin{proof} 
For convenience we consider the sum
\begin{equation}
\Zpol := \sum_{\p\ni 0} w_0(\p) e^{a\vert \p\vert} \leq C',
\end{equation}
for a constant $a>0$, and set $a=2$ at the end of the proof.  Let $k_0$ be the
smallest positive integer for which
\begin{equation}
C_2\beta ({\lambda c_d e^{a}})^{k_0} \leq 1,
\label{alpha1}
\end{equation}
where $c_d = 36$. 
We define
\begin{equation}
\alpha := C_2\beta ({\lambda c_d e^{a}})^{k_0} 
\label{alpha}
\end{equation}
Let $x=C_2 \beta \lambda^{k}$ and $x_0=C_2 \beta \lambda^{k_0}$. Then for
$k>k_0$ we use the bound $e^x -1 \le xe^x$ and for $k \le k_0$ we use the
bound $e^x -1 < e^x$.  It follows from \reff{pee}, \reff{bound4}, and the
above  bounds, that
\begin{eqnarray}
 w_0(\p) e^{a|\p|} &\le& e^{-\beta C_1 \lambda |\gamma|} e^{a|\gamma|}
\label{prefactor}
\times\left[\prod_{{B \in M_{\gamma}}\atop{g(B) \le k_0}} e^{\beta C_2 
\lambda^{g(B)}}\,e^{ag(B)}\right]\nonumber\\
&&\times\left[\prod_{{B \in M_{\gamma}}\atop{g(B) > k_0}} {\beta C_2 
\lambda^{g(B)}}e^{\beta C_2 
\lambda^{g(B)}}\,e^{ag(B)}\right]
\end{eqnarray}
We have that 
\begin{eqnarray}
\sum_{B \in M_{\gamma}} \lambda^{g(B)} &\le& \sum_{B\cap \gamma \ne 
\emptyset}\ \lambda^{g(B)} \le \sum_{x \in \gamma} \sum_{B \ni x} 
\frac{\lambda^{g(B)}}{|B|}
\le \sum_{x \in \gamma} \sum_{B \ni x}
\frac{\lambda^{g(B)}}{2}\nonumber\\
&\le& |\gamma|  \sum_{B}^* \lambda^{g(B)}
\le |\gamma| \sum_{k\ge 3} \sum_{B: \atop{g(B) =k}}^* \lambda^{k}
\le  |\gamma| \sum_{k\ge 3} c_d^k \lambda^{k},
\label{fixed}
\end{eqnarray}
The symbol ${\sum_{x \in \gamma}}$ denotes the sum over all sites $x \in
\Lambda$ for which at least one nearest neighbour bond of the lattice, which
contains the site $x$, is intersected by a face in $\gamma$. There are
$2|\gamma|$ such sites in the $\Lambda$.  The symbol ${\sum_{B}^*}$ denotes
the sum over all $B$'s containing a fixed point. For the last inequality we
used the K\"onigsberg Bridge lemma \cite[pp 464-465]{simon}, which gives 
$c_d=(2d)^2$, $d$ being the dimension of the lattice, i.e., $c_d = 36$. 

Hence, we obtain the uniform bound
\begin{equation}
\prod_{B \in M_{\gamma}} e^{\beta C_2 \lambda^{g(B)}}\le 
e^{\beta C_3 \lambda^{3}|\gamma|} \label{bound6}
\end{equation}
with 
\begin{equation}
C_3 := \frac{C_2\,c_d^3}{1 - c_d\, \lambda}
\end{equation}
{\em{provided}} 
\begin{equation}
c_d\,\lambda < 1.
\label{cond1}
\end{equation} 
For convenience we define
\begin{equation}
a_0 := \beta (C_1 \lambda  - C_3 \lambda^{3}) - a. 
\label{a0}
\end{equation}
Then, if \reff{cond1} is satisfied, we have that  
\begin{eqnarray}
 w_0(\p) e^{a|\p|} &\le& e^{-a_0|\gamma|}
\times\left[\prod_{{B \in M_{\gamma}}\atop{g(B) \le k_0}} e^{ag(B)}\right]
\label{factor1}\\
&&\times\left[\prod_{{B \in M_{\gamma}}\atop{g(B) > k_0}} {\beta C_2 
\lambda^{g(B)}}e^{\beta C_2 
\lambda^{g(B)}}\,e^{ag(B)}\right]\label{factor2}.
\end{eqnarray}
Hence,
\begin{eqnarray}
\sum_{\p \ni 0} w_0(\p) e^{a|\p|} &\le& \sum_{j \ge 0} \frac{1}{j!} 
\sum_{\gamma, B_1, \ldots, B_j \atop{\gamma\cup B_1 \cup \ldots \cup 
B_j \ni 0 \atop {B_i \cap \gamma \ne \emptyset}}} e^{-a_0 |\gamma|} 
\Bigl[\prod_{{1\le i \le j}\atop{g(B) \le k_0}}
e^{ag(B)}\Bigr]\nonumber\\
& &\Bigl[\prod_{{1\le i \le j}\atop{g(B) > k_0}} {\beta C_2 
\lambda^{g(B)}}e^{\beta C_2 
\lambda^{g(B)}}\,e^{ag(B)}\Bigr] \label{top}
\end{eqnarray}
The sum on the LHS of \reff{top} is over all polymers which contain  the
origin. For each term in the sum on the RHS of \reff{top}, the origin can be 
contained either in the contour of a polymer and/or in one or more of  the
bonds  intersecting it. This sum can be bound by a sum over all polymers for
which  the origin is contained in their respective contours. We refer to such
a polymer as a {\em{pinned polymer}}. More precisely, a pinned polymer $\p$ is
defined by the connected sequence $\gamma, B_1, \ldots, B_j$ such that $\gamma
\ni 0$. The contribution of each pinned polymer must be multiplied by the
number of translations containing the origin. Since  for all $i, B_i\cap\gamma
\neq\emptyset$, this number is bounded by 
\begin{equation}
|\cup_{1\leq i\leq j} B_i \setminus \gamma|\leq \sum_{1\leq i\leq j} \vert B_i
\vert -1\leq \sum_{1\leq i\leq j} \Bigl(g(B_i) -1\Bigr) \end{equation} since
$|B| \le g(B)$. The factors of ``$-1$'' arise from the fact that each  bond $B$
must intersect the contour $\gamma$. Starting from a given contour $\gamma$
containing the origin, a pinned polymer can be constructed by  successively
adding bonds which intersect it. For a given $\gamma$, the sum over
decorations can be split into a sum over decorations with `small' bonds, i.e.,
$B$ with $g(B)\leq k_0$,  and a sum over decorations with `large' bonds, i.e.,
$B$ with $g(B)> k_0$.  This gives the following estimate: 
\bea
\Zpol&\leq&\sum_{\gamma\ni 0} e^{-a_0 \vert\gamma\vert}\!\!\!\!
\sum_{j\geq 0, \{B_1,\ldots,B_j\}\subset\B_\gamma\atop g(B_i)\leq k_0}
\sum_{j^\prime\geq 0, \{B^\prime_1,\ldots,B^\prime_{j^\prime}\}
\subset\B_\gamma\atop g(B^\prime_i)> k_0} \Bigl[\sum_{1\leq i\leq j}
(g(B_i)-1) + \sum_{1\leq i\leq j^\prime} (g(B^\prime_i)-1)\Bigr]\nonumber\\
\label{first}\\
& &\times\left[\prod_{1\leq i\leq j} e^{ag(B_i)}\right]
\left[\prod_{1\leq i\leq j^\prime} \beta C_2 \lambda^{g(B^\prime_i)}
e^{ag(B^\prime_i)}\right].\label{second}
\eea
If both $j\geq 1$ and $j^\prime\geq 1$ then (since $g(B)\geq 3$) the
sum of the two sums in paranthesis  in \reff{first}  can be bounded by the
product of the two sums, such that the first sum can be combined with the
first factor of \reff{second} and the second sum with the second factor. The
result is
\bea
\Zpol&\leq&\sum_{\gamma\ni 0} e^{-a_0\vert\gamma\vert}
\left\{1+\sum_{j\geq 1, \{B_1,\ldots,B_j\}\subset\B_\gamma\atop 
g(B_i)\leq k_0}\Bigl(\sum_{1\leq i\leq j} g(B_i)-1\Bigr)
\left[\prod_{1\leq i\leq j} e^{ag(B_i)}\right]\right\}\label{smalldec}\\
&&\times\left\{1+\sum_{j^\prime\geq 1, \{B^\prime_1,\ldots,
B^\prime_{j^\prime}\}\subset\B_\gamma\atop g(B^\prime_i)> k_0} 
\Bigl(\sum_{1\leq i\leq j^\prime} g(B^\prime_i)-1\Bigr)
\left[\prod_{1\leq i\leq j^\prime} \beta C_2 \lambda^{g(B^\prime_i)}
e^{ag(B^\prime_i)}\right]\right\}\nonumber\\
\label{largedec}
\eea
The cases with $j=0$ or $j^\prime=0$ have been incorporated by  adding $1$ to
each factor. Next, we estimate the sums in \reff{smalldec}.

The sum in \reff{smalldec} can be treated with a `reverse' high-temperature
expansion, i.e., resummation, as follows. 

\begin{eqnarray}
&&\sum_{j\geq 1, \{B_1,\ldots,B_j\}\subset\B_\gamma
\atop g(B_i)\leq k_0}(\sum_{1\leq i\leq j} g(B_i)-1)
\left[\prod_{1\leq i\leq j} e^{ag(B_i)}\right]\\
&\leq&\sum_{j\geq 1, \{B_1,\ldots,B_j\}\subset\B_\gamma
\atop g(B_i)\leq k_0}(\sum_{1\leq i\leq j} g(B_i))
\left[\prod_{1\leq i\leq j} e^{ag(B_i)}\right]\\
&=&\frac{d}{da}\sum_{j\geq 0, \{B_1,\ldots,B_j\}\subset\B_\gamma
\atop g(B_i)\leq k_0}\left[\prod_{1\leq i\leq j} e^{ag(B_i)}\right]\\
&=&\frac{d}{da}\left[\prod_{B\in\B_\gamma \atop g(B)\leq k_0}
(e^{a g(B)}+1)\right]\\
&=&\sum_{B\in\B_\gamma \atop g(B)\leq k_0}g(B)
\frac{e^{ag(B)}}{e^{ag(B)}+1}\prod_{B^\prime\in\B_\gamma \atop g(B^\prime)
\leq k_0}(e^{a g(B^\prime)}+1)\\
&\leq& \Bigl[\sum_{B\in\B_\gamma \atop g(B)\leq k_0}g(B)\Bigr]
\prod_{B\in\B_\gamma \atop g(B)\leq k_0}e^{a^\prime g(B)}\\
&=&\bigl[\sum_{B\in\B_\gamma \atop g(B)\leq k_0}g(B)\bigr]
\exp\Bigl(a^\prime \sum_{B\in\B_\gamma \atop g(B)\leq k_0}g(B)\Bigr)
\end{eqnarray}
with $a^\prime=a+(\log 2)/3$, guaranteeing that for all $g(B)\geq 3$,
$\exp(ag(B)) + 1 \leq \exp(a^\prime g(B))$. 
The estimate of the
sum over `small' decorations can now be completed by using the bound
\be
\sum_{B\in\B_\gamma \atop g(B)\leq k_0}g(B)
=\sum_{x \in \gamma} \sum_{B \ni x \atop g(B)\leq k_0} \frac{g(B)}{|B|} 
\le \vert\gamma\vert \sum_{k=3}^{k_0} 
 \sum_{B \atop g(B) = k}^* k
\le   \vert\gamma\vert \sum_{k=3}^{k_0} c_d^k k
\le C_4 \vert\gamma\vert
\ee
with
\be
C_4 = (k_0+1)(c_d)^{k_0+1}.
\label{c5}
\ee
At this point we have:
\bea
\Zpol&\leq&\sum_{\gamma\ni 0} e^{-a_0\vert\gamma\vert}
\left[ 1 + C_4 \vert\gamma\vert e^{C_4 a^\prime \vert\gamma\vert}\right]\\
&&\times\left[1+\left(\sum_{j^\prime\geq 1, \{B^\prime_1,\ldots,
B^\prime_{j^\prime}\}\subset\B_\gamma\atop g(B^\prime_i)> k_0} 
(\sum_{1\leq i\leq j^\prime} g(B^\prime_i)-1)
\prod_{1\leq i\leq j^\prime} \beta C_2 U^{-g(B^\prime_i)}
e^{ag(B^\prime_i)}\right)\right]\nonumber\\
\label{part}
\eea

The second term in \reff{part} represents a sum over 
the `large' decorations. Let us denote this term by $L$, i.e., 
\begin{equation}
L:= \sum_{j^\prime\geq 1, \{B^\prime_1,\ldots,
B^\prime_{j^\prime}\}\subset\B_\gamma\atop g(B^\prime_i)> k_0} 
(\sum_{1\leq i\leq j^\prime} g(B^\prime_i)-1)
\prod_{1\leq i\leq j^\prime} \beta C_2 U^{-g(B^\prime_i)}
e^{ag(B^\prime_i)}.
\end{equation}
and find an upper bound for it. We first make some simplifying estimates.
 Using $k_0\geq 3$ and $\vert\gamma\vert\geq 6$ one can see that
\be
1+C_4\vert\gamma\vert e^{C_4 a^\prime \vert\gamma\vert}
\leq C_4\vert\gamma\vert e^{C_4 a^\pp \vert\gamma\vert}
\label{simp}
\ee
with, 
\begin{equation}
a^\pp =a + 1/4.
\label{app} 
\end{equation}
For convenience we introduce
the notations
\be
\B_{\gamma,>k_0}=\{B\in\B\mid B\cap\gamma\neq\emptyset, g(B)> k_0\}
\ee
and
\be
\tilde{\beta}=\beta C_2, \quad z= \lambda {e^a}
\ee
Using the trivial observation
\be
\sum_{\{B_1,\ldots,B_j\}\subset\B_{\gamma,>k_0}} (\cdot )
\leq \frac{1}{j!}\sum_{B_1,\ldots,B_j \in \B_{\gamma,>k_0}} (\cdot)
\ee
we obtain
\bea
L &\leq&  \sum_{j\geq 1} \frac{1}{j!}\sum_{B_1,\ldots,B_j\in \B_{\gamma,>k_0}
}
(\sum_{1\leq i\leq j} g(B_i)-1)
\prod_{1\leq i\leq j} \tilde{\beta} z^{g(B_i)}\nonumber\\
&\leq& \sum_{j\geq 1} \sum_{B_1 \in \B_{\gamma,>k_0}} \ldots 
\sum_{B_{j-1} \in \B_{\gamma,>k_0}} 
\bigl[\prod_{1\leq i\leq j} \tilde{\beta} z^{g(B_i)}\bigr] \nonumber\\
& &\times \sum_{B_{j} \in \B_{\gamma,>k_0}} (\sum_{1\leq i\leq j} g(B_i)-1)
 \tilde{\beta} z^{g(B_j)}.
\label{ell}
\eea
We have that
\be
\sum_{B_{j} \in \B_{\gamma,>k_0}} (\sum_{1\leq i\leq j} g(B_i)-1)
 \tilde{\beta} z^{g(B_j)} \nonumber\\
\le \sum_{x \in \gamma} \sum_{B_j \ni x \atop{g(B_j) > k_0}}
(\sum_{1\leq i\leq j} g(B_i)-1) \tilde{\beta} z^{g(B_j)} 
\label{ellpart}
\ee
The sum over $B_j$ is independent of the choice of the point $x$. Hence
\bea
{\hbox{RHS of \reff{ellpart}}} &\le&  2\vert \gamma \vert \sum_{k_j \ge k_0}
\left[ g(B_1) -1 + \ldots + g(B_{j-1}) - 1 + k_j -1 \right]\nonumber\\
& & \times \sum_{B_j\atop{g(B_j)=k_j}}^*  \tilde{\beta} z^{k_j} 
\label{ellpart2}
\eea
Iterating the above steps, we obtain the following upper bound to the 
contribution of the large decorations:
\be
L \le   \sum_{j\geq 1} \frac{{\vert \gamma \vert}^j}{j!} \sum_{k_1>k_0} \ldots 
\sum_{k_j>k_0} \left[ \bigl(\sum_{i=1}^j k_i -1\bigr)\prod_{i=1}^j \bigl(
\sum_{B_i\atop{g(B_i)=k_i}}^* \tilde{\beta} z^{k_i}\bigr) \right]
\ee
Since, $k_i>k_0 \ge  3$, we have that 
\be
\sum_{i=1}^j  (k_i  -  1)  \le  \prod_{i-1}^j (k_i  -1)
\ee
Each factor of $(k_i -1)$ can be inserted into the sum over  $k_i$. This
yields
\bea
L&\le&  \sum_{j\geq 1} \frac{{\vert \gamma \vert}^j}{j!}\prod_{i=1}^j \bigl(
 \sum_{k_i>k_0} ( k_i  -  1)  \sum_{B_i \atop{g(B_i)=k_i}}^*  
\tilde{\beta} z^{k_i}\bigr)\nonumber\\
&\le&  \sum_{j\geq 1} \frac{{\vert \gamma \vert}^j}{j!}\prod_{i=1}^j 
 \sum_{k_i>k_0} ( k_i  -  1) c_d^{k_i}\tilde{\beta} z^{k_i}\nonumber\\
&\le&  \sum_{j\geq 1} \frac{{\vert \gamma \vert}^j}{j!}
\tilde{\beta} k_0 (c_d  z)^{k_0} \left[\sum_{k \ge 1}  k (c_d  z)^k 
\right]^j\nonumber\\
&\le&  \sum_{j\geq 1} \frac{{\vert \gamma \vert}^j}{j!} k_0 
[\tilde{\beta}(c_d  z)^{k_0}] \left[\frac{c_d  z}{(1-c_d  z)^2}\right]^j,
\label{part3}
\eea
{\em{provided}}
\be
c_d z \equiv c_d e^a \lambda < 1
\label{cond2}
\ee
By definition (\reff{alpha1}, \reff{alpha}) we have that 
\be
\tilde{\beta} (c_d z)^{k_0} \equiv \beta C_2 (c_d e^a \lambda)^{k_0}= 
\alpha < 1
\ee
Hence, {\em{if}} \reff{cond2} {\em{holds}}, then 
\bea
L&\le&  \sum_{j\geq 1} \frac{{\vert \gamma \vert}^j}{j!} 
\left(\frac{k_0 c_d  z}{(1-c_d  z)^2}\right)^j \nonumber\\
&\le& {\rm{exp}}\left[\frac{\vert \gamma \vert k_0{c_d  z}}{(1-c_d  z)^2}
\right]
\nonumber\\
&:=&  {\rm{exp}}(a_1 \vert \gamma \vert)
\label{last}
\eea
with
\be
a_1 := \frac{k_0{c_d  z}}{(1-c_d  z)^2}
\label{a1}
\ee
Hence from \reff{part}, \reff{simp} and \reff{last} it follows that 
\bea
\Zpol &\leq& \sum_{\gamma \ni 0} e^{-a_0 \vert \gamma \vert} C_4 
\vert \gamma \vert e^{C_4 a^\pp \vert \gamma \vert} \left[ 1 + 
e^{a_1 \vert \gamma \vert} \right] \nonumber\\
&\leq& \sum_{\gamma \ni 0} e^{-a_0 \vert \gamma \vert} 2 C_4 
\vert \gamma \vert e^{C_4 a^\pp \vert \gamma \vert} 
e^{a_1 \vert \gamma \vert},
\eea
since $a_1 > 0$. Defining
\be
q:= a_0 - {\rm{log}}\,c_d - a_1 -  a^\pp C_4
\label{que}
\ee
we have that 
\bea
\Zpol &\leq&  \sum_{k \ge 6} 2 C_4 k e^{-kq} \\ \label{sum}
 &\leq&  2 C_4 \frac{e^{-q}}{(1 - e^{-q})^2},
\label{result}
\eea
{\em{provided}}
\be
q > 0 \label{cond4}.
\ee

Let us inspect the condition \reff{cond4} in more detail. From the 
definitions, \reff{a0}, \reff{a1}, \reff{app} and \reff{c5}, of
 $a_0$, $a_1$, $a^\pp$ and $C_4$, and \reff{cond4}, it
follows that \reff{result} holds provided
\begin{equation}
\lambda \beta \, X(\lambda) > a + {\rm{log}}\,c_d + k_0\, 
\frac{c_d\,\lambda e^a}
{(1 - c_d\, \lambda e^a)^2} + (a+\frac{1}{4})\,(k_0 + 1) c_d^{k_0+ 1},
\label{ineq}
\end{equation}
where
\begin{equation}
X(\lambda) := C_1 - \frac{C_2 c_d^3 \lambda^2}{1 - c_d \lambda}
\end{equation}
and $k_0$ satisfies the bound \reff{alpha1}.  Before proceeding further,
let us recall the conditions which have been  imposed on $\lambda$ in order to
arrive at the form \reff{sum}. These are given by \reff{cond1} and
\reff{cond2}. Moreover, for \reff{ineq} to hold, it is necessary that
$X(\lambda)>0$. The inequalities \reff{cond1} and \reff{cond2} are satisfied
if 
\begin{equation}
\lambda < \lambda_1 := (c_d e^a)^{-1}.
\label{u1}
\end{equation}
Since $X(\lambda)$ is monotone decreasing, $X(\lambda) > 0$ 
can be satisfied by requiring $\lambda$ to be smaller than the solution of 
$X(\lambda) = 0$, i.e., 
\begin{equation}
\lambda < \lambda_2 := \frac{2}{c_d 
\Bigl(1 + \sqrt{ 1+ \frac{4 c_d C_2}{C_1}} \Bigr)}.
\label{u2}
\end{equation}
In order to satisfy \reff{u1} and \reff{u2} we choose $\lambda >
\lambda_0$ 
with 
\begin{equation}
\lambda_0 := (B\, c_d^2\, e^a)^{-1} 
\label{lamb0}
\end{equation}
where 
\begin{equation}
B := \frac{1}{2}\Bigl(1 + \sqrt{1 + \frac{4 c_d C_2}{C_1}}\Bigr).
\label{bigB}
\end{equation}
Note that $B > 1$ since $C_1, C_2$ and $c_d$ are positive. 

The quantity $q$ defined through \reff{que} can be expressed as a  function of
$\lambda$ and $b:=\beta \lambda$. Next, let us determine $b_0$ such that  for
all $b >  b_0$ and $\lambda < \lambda_0$, \reff{cond4} is satisfied.  On the
RHS of \reff{ineq}, $k_0$ (expressed as a function of $\lambda$ and $b$) 
has to satisfy
\begin{equation}
k_0 = k_0 (\lambda,b) > 1 - \frac{{\rm{log}}\,\bigl(C_2 c_d e^a b \bigr)}
{{\rm{log}}\,\bigl( \lambda c_d e^a\bigr)}.
\label{k00}
\end{equation}
This follows from the defining relation \reff{alpha1} for $k_0$.
For all $\lambda < \lambda_0$ this can be achieved by choosing
\begin{eqnarray}
k_0 = \Bar{k}_0 (b) &=&   1 - \frac{{\rm{log}}\,\bigl(C_2 c_d e^a b \bigr)}
{{\rm{log}}\,\bigl( \lambda_0 c_d e^a\bigr)}\nonumber\\
&=&  1 + \frac{{\rm{log}}\,\bigl(C_2 c_d e^a b \bigr)}
{{\rm{log}}\,\bigl(B c_d\bigr)},
\label{bark0}
\end{eqnarray}
where $B$ is defined through \reff{bigB}. Let us denote the RHS of
\reff{ineq}, with the above choice $\Bar{k}_0$, by $A(\lambda,b)$, i.e.,
\begin{equation}
A(\lambda,b) = a + {\rm{log}}\,c_d + \frac{\Bar{k}_0\lambda
e^a}{(1 - c_d \lambda e^a)^2}
+ c_d (a + \frac{1}{4}) (\Bar{k}_0 + 1) c_d^{\Bar{k}_0}.
\label{aub}
\end{equation}
Recall that $X(\lambda)$ is monotone decreasing, and note that $A(\lambda,b)$,
for a  fixed value of $b$, is monotone increasing in $\lambda$. Hence, for all
$\lambda < \lambda_0$  we can satisfy \reff{ineq} by requiring that 
\begin{equation}
b X(\lambda_0) > A(\lambda_0, b).
\label{strong}
\end{equation}
That \reff{strong} is satisfied for all $b > b_0$, for some $b_0 > 0$, follows
from the fact that $A(\lambda_0, b)$ increases strictly less than linearly  in
$b$. More precisely,
\begin{equation}
A(\lambda_0, b) = A_0 + A_1 {\rm{log}}\,(d_0\,b) + A_2 (d_0\, b)^r + 
A_3 (d_0\, b)^r \, {\rm{log}}\,(d_0\,b),
\end{equation}
where $A_0$, $A_1$, $A_2$, $A_3$ and $d_0$ depend on $C_1$, $C_2$, $e^a$
and $c_d$, and $r$ is given by
\begin{equation}
r = \frac{{\rm{log}}\,c_d}{{\rm{log}}\,B\,c_d}.
\end{equation}
In particular, 
\begin{equation}
d_0 = C_2 c_d e^a 
\label{c0}
\end{equation}
Since $B > 1$ (see \reff{bigB}), we have that $r < 1$. This proves the
convergence of the series for $\Zpol$ and the bound \reff{result}. 

It is now easy to see that, in fact, the bound $C'$ for $\Zpol$ as in the
statement of the lemma, can be made arbitrarily small by choosing $b_0$ 
arbitrarily large. With $k_0 = \Bar{k}_0$, $C'$ is given by \begin{equation}
C'(b) = 2\,c_d^2 \Bigl[ 2 + \frac{{\rm{log}}(C_1 b)}{{\rm{log}}(B c_d)}
\Bigr]\, (C_1 b)^r  \times \frac{e^{- q(b)}}{(1 - e^{- q(b)})^2},
\end{equation}
where we have used the definitions of $C_4$ [\reff{c5}] and $\Bar{k}_0$
[\reff{bark0}]. The constants $d_0$ and $B$ are defined  in \reff{c0} and
\reff{bigB}. The function $q(b)$ is bounded below by $b X(\lambda_0) -
A(\lambda_0, b)$, which increases linearly in $b$. Then  again since $r<1$,
$C'(b) \longrightarrow 0$ as $b \longrightarrow \infty$.
\end{proof}
\vspace{10pt}\noindent {\large\bf Acknowledgements}

\noindent
B.N. was partially supported by the U.S. National Science Foundation under
Grant No. PHY9513072, and DMS-9706599. B.N. and N.D.  would like to thank
the Centre de Physique Th\'eorique, Luminy, where part of this work was done,
for hospitality. N.D. is grateful to the Dublin Institute of Advanced Studies,
Dublin and the University of Strathclyde, Glasgow for financial support. She
would also like to thank C.E. Pfister  and S. Sen for helpful discussions. 

\vspace{.5truecm}

{
\def\thebibliography#1{\noindent
{\large\bf References} \par \list
{\arabic{enumi}.}{\settowidth\labelwidth{[#1]} \leftmargin\labelwidth
\advance\leftmargin\labelsep \itemsep=0pt \usecounter{enumi}}
\def\newblock{\hskip .11em plus .33em minus -.07em} \sloppy
\sfcode`\.=1000\relax}
}
\end{document}